\pdfoutput=1

 \documentclass[apj]{emulateapj}

\usepackage{epsfig}
\usepackage{graphicx}
\usepackage{amsmath}

\usepackage{natbib}
\usepackage{aas_macros}
\usepackage{hyperref}
\usepackage[3D]{movie15}

\def\OII{[{\ion{O}{2}}]}
\def\4959_5007{[\ion{O}{3}]~$\lambda \lambda$4959,5007}
\def\OIII49595007{[\ion{O}{3}]~$\lambda \lambda 4959,5007$}
\def\ratioR23{([\ion{O}{2}]~$\lambda \lambda$3727,9 + [\ion{O}{3}]~$\lambda\lambda$4959,5007)/H$\beta$}
\def\R23{${\rm R}_{23}$}
\def\dS23{${\rm S}_{23}$}

\def\NII{[{\ion{N}{2}}]}
\def\OIIIOII{[\ion{O}{3}]/[\ion{O}{2}]}
\def\OIIISII{[\ion{O}{3}]/[\ion{S}{2}]}

\def\NIIOII{[\ion{N}{2}]/[\ion{O}{2}]}

\def\ratioS23{([\ion{S}{2}]~$\lambda \lambda$6717,31 +[\ion{S}{3}]~$\lambda\lambda$9069,9532)/H$\beta$}
\def\NIIHa{[\ion{N}{2}]/H$\alpha$}
\def\SIIHa{[\ion{S}{2}]/H$\alpha$}
\def\OIHa{[\ion{O}{1}]/H$\alpha$}

\def\SII{[{\ion{S}{2}}]}

\def\O4363{[{\ion{O}{3}}]~$\lambda$4363}
\def\OIII{[{\ion{O}{3}}]}
\def\OIIIHb{[\ion{O}{3}]/H$\beta$}

\shorttitle{Three-dimensional Line Ratio Diagrams}
\shortauthors{Vogt et al.}

\begin{document}

\title{Galaxy emission line classification using 3D line ratio diagrams}

\author{Fr\'ed\'eric P.A. Vogt\altaffilmark{1,2}, Michael A. Dopita\altaffilmark{1,3,4}, Lisa J. Kewley\altaffilmark{1,4}, Ralph S. Sutherland\altaffilmark{1},\\
Julia Scharw\"achter\altaffilmark{5}, Hassan M. Basurah \altaffilmark{3}, Alaa Ali\altaffilmark{3,6} \& Morsi A. Amer\altaffilmark{3,6}}
\email{frederic.vogt@anu.edu.au}

\altaffiltext{1}{Research School of Astronomy and Astrophysics, Australian National University, Canberra, ACT 2611, Australia.}
\altaffiltext{2}{Department of Physics and Astronomy, Johns Hopkins University, 3400 N. Charles Street, Baltimore, MD 21218, USA.}
\altaffiltext{3}{Astronomy Department, King Abdulaziz University, P.O. Box 80203, Jeddah, Saudi Arabia.}
\altaffiltext{4}{Institute for Astronomy, University of Hawaii, 2680 Woodlawn Drive, Honolulu, HI 96822, USA.}
\altaffiltext{5}{Observatoire de Paris, LERMA (CNRS: UMR8112), 61 Av. de l'Observatoire, 75014 Paris, France.}
\altaffiltext{6}{Astronomy Department, Faculty of Science, Cairo University, Egypt.}

\begin{abstract}
Two-dimensional (2D) line ratio diagnostic diagrams have become a key tool in understanding the excitation mechanisms of galaxies. The curves used to separate the different regions - \ion{H}{2}-like or else excited by an active galactic nucleus (AGN) -  have been refined over time but the core technique has not evolved significantly. However, the classification of galaxies based on their emission line ratios really is a multi-dimensional problem. Here we exploit recent software developments to explore the potential of three-dimensional (3D) line ratio diagnostic diagrams. We introduce a specific set of 3D diagrams, the $\cal{ZQE}$ diagrams, which separate the oxygen abundance and the ionisation parameter of \ion{H}{2} region-like spectra, and which also enable us to probe the excitation mechanism of the gas. By examining these new 3D spaces interactively, we define a new set of 2D diagnostics, the $\cal{ZE}$ diagnostics, which can provide the metallicity of objects excited by hot young stars, and which cleanly separate  \ion{H}{2} region-like objects from the different classes of AGNs. We show that these $\cal{ZE}$ diagnostics are consistent with the key $\log${\NIIHa} \emph{vs.} $\log${\OIIIHb} diagnostic currently used by the community. They also have the advantage of attaching a probability that a given object belongs to one class or to the other. Finally, we discuss briefly why $\cal{ZQE}$ diagrams can provide a new way to differentiate and study the different classes of AGNs in anticipation of a dedicated follow-up study. \\
\end{abstract}

\keywords{galaxies: abundances, galaxies: starburst, galaxies: Seyfert, galaxies: general, ISM: lines and bands, \ion{H}{2} regions}

\section{Introduction}\label{sec:intro}

The use of specific line ratios to distinguish line emission regions depending on their gas excitation mechanism was pioneered by \cite{Baldwin81} and extended by \cite{Veilleux87}. The line ratios most frequently used,  specifically designed to be insensitive to reddening, are:
\begin{enumerate}
\item  $\log${\NIIHa} \emph{vs.} $\log${\OIIIHb}, 
\item  $\log${\SIIHa} \emph{vs.} $\log${\OIIIHb} and
\item  $\log${\OIHa} \emph{vs.} $\log${\OIIIHb}. 
\end{enumerate}
Theoretical progress has since allowed the placement of different diagnostic lines separating the different excitation mechanisms in these diagrams: regions photoionised by hot stars giving H\,{\sc ii}-like spectra, regions excited by an Active Galactic Nucleus (AGN), either Seyferts, or the low ionization nuclear emission-line regions (LINERs).  Currently, the \emph{maximum starburst lines} from \cite{Kewley01a,Kewley01b}, the empirical starburst line from \cite{Kauffmann03b}, and the LINER-Seyfert lines from \cite{Kewley06} are commonly used. Other, similar diagnostics include \cite{ Heckman80, Osterbrock85, Veilleux87, Tresse96, Ho97, Dopita00, Stasinska06}. 

The maximum starburst lines as defined by \cite{Kewley01b} are based on theoretical modelling of starburst galaxies.  Specifically, the wrap-round of theoretical model grids inside these optical  line ratio diagnostic diagrams justifies the definition of a theoretical upper bound of emission line ratios from gas photoionised by hot young stars. \cite{Kauffmann03b} used the large number statistics of the Sloan Digital Sky Survey \citep[SDSS,][]{York00} to set an observational lower bound to the maximum starburst line in the $\log${\NIIHa} \emph{vs.} $\log${\OIIIHb} diagram. The region between these two starburst lines is known as the ``composite" region. Recently, several objects in the composite region have been recognised as being (at least in part) excited by shocks \citep{Farage10, Rich11, Rich13}, although these do not rule out a \emph{mixed} excitation mechanism (starburst+AGN) for other composite objects \citep[e.g.][]{Scharwachter11,Davies14,Dopita14b}. \cite{Kewley06} also exploited the large number statistics from SDSS to define the separation lines between the LINER and the Seyfert branches on the AGN side of the $\log${\SIIHa} \emph{vs.} $\log${\OIIIHb} and $\log${\OIHa} \emph{vs.} $\log${\OIIIHb} diagrams. 

The classical optical line ratio diagnostic diagrams have proved to be useful and resilient, ever since their introduction. Recently, their usage has been extended as new IR surveys of galaxies measure the key line ratios for galaxies at intermediate and high redshifts. The key instruments are MOSFIRE on Keck \citep{McLean10}, FMOS on Subaru \citep{Kimura10}, MMIRS on Magellan \citep{McLeod04}, FLAMINGOS II on Gemini \citep{Eikenberry08} and LUCI at the Large Binocular Telescope \citep{Buschkamp12}. The sensitivity of the optical line ratio diagnostics to metallicity and other factors which could influence the diagnostics at high-redshift was investigated by \citet{Kewley13a}, and this insight was applied to actual samples of high-redshift galaxies by \citet{Kewley13b}.

A further stimulus to the use of optical line ratio diagnostic diagrams has been the advent of integral field spectrographs (IFS). These instruments provide spectral information for individual spectral pixels (commonly referred to as  \emph{spaxels}). With this approach, it is possible to reveal the presence of metallicity gradients across the entire spatial extent of galaxies \citep{Rich12}, explore the trends in the local excitation pressure \citep{Dopita14}, or study the AGN zone of influence \citep{Scharwachter11}. These analyses, in turn, rely on an accurate classification scheme. 

In reality, the full set of line ratios forms a multi-dimensional space, the topology of which needs to be understood before a final classification can be set. Progress towards this goal can be made by looking at alternative line ratio diagrams. For the case of \ion{H}{2} regions, \citet{Dopita13a} made a comprehensive study of the utility of alternative diagnostic diagrams, discussing previously used ones, as well as introducing some new ones. Few of these diagrams separate the AGN branch from the stellar excited objects as well as the traditional $\log${\NIIHa} \emph{vs.} $\log${\OIIIHb} diagram. Notable exceptions are provided by the $\log${\NIIHa} \emph{vs.} $\log${\OIIIOII} diagram and the $\log${\NIIOII}  \emph{vs.} $\log${\OIIIHb} diagram.

In this paper, we revisit the concept of optical line ratio diagram itself, and introduce new 3D line ratio diagrams. These diagrams, combining three different and complimentary line ratios, are a first step towards a better understanding of the distribution of galaxies in their multi-dimensional line ratio space. This article is organized as follows. We first describe the observational datasets that we employ in our analysis in Section~\ref{sec:datasets}, and  the theoretical models we use in Section~\ref{sec:HIImodels}. We introduce the new 3D line ratio diagrams derived from these datasets in Section~\ref{sec:3d-diag}. In Section~\ref{sec:3d-2d}, we use specific 3D line ratio diagrams to generate a new and consistent set of line ratio diagnostics to separate \ion{H}{2}-like galaxies from the AGN-like objects. In Section~\ref{sec:discussion}, we compare these new diagnostics with the standard optical line ratio diagnostic diagrams, and discuss their compatibility with intermediate and high redshifts spectroscopic observations. Finally, we highlight the potential of these new 3D line ratio diagrams to investigate the different AGN families in Section~\ref{sec:AGN}, and summarise our conclusions in Section~\ref{sec:conclusion}.

\section{Observational datasets}\label{sec:datasets}

\subsection{SDSS galaxies}

We construct our sample of emission line galaxies from the Sloan Digital Sky Survey (SDSS) data release (DR) 8 \citep{Aihara11a,Aihara11b,Eisenstein11}. Specifically, we exploit the ``galSpec''  Value Added Catalogue from the Max Planck Institute for Astronomy and Johns Hopkins University (MPA-JHU) group. The data is in fact identical to that associated with the SDSS DR7 \citep{Abazajian09}, but was first made accessible via the general SDSS data release in DR8. This dataset has been freely accessible since the SDSS DR4 \citep[][]{Adelman06}, and the associated fitting procedure for the stellar continuum and emission lines are described in detail in \cite{Kauffmann03a, Brinchmann04,Tremonti04}. Each spectrum is corrected for the foreground Galactic extinction using the \cite{Odonnell94} extinction curve. The stellar continuum is fitted with a linear combination of ten single-age stellar population models based on a new version of the \texttt{GALEXEV} code of \cite{Bruzual03, Bruzual11} plus an additional parameter accounting for internal dust attenuation \citep{Charlot00}. The different emission lines are fitted with single Gaussians. Balmer lines share a unique rest-frame velocity and velocity dispersion (accounting for the instrumental resolution), and so do the forbidden lines. 

From the 1843200 objects provided in DR8, we extract a sub-sample of high-quality spectra with reliable galSpec fit parameters, following the methodology of \cite{Kewley06}. Our detailed selection criteria (including the explicit SDSS \texttt{keywords} expressions) are:

\begin{enumerate}
\item an existing galSpec fit (i.e. [\texttt{PLATEID};\texttt{FIBERID};\texttt{MJD}] $\neq$ -1),
\item the galSpec fit is flagged as ``reliable'' by the MPA-JHU group (i.e. \texttt{RELIABLE} = 1),
\item a reliable redshift measurement (i.e. \texttt{Z\_WARNING} = 0),
\item a redshift between 0.04 and 0.1 (i.e. $0.04<\mathtt{Z}<0.1$),
\item a signal-to-noise $\geq3$ in the following strong lines: [\ion{O}{2}]$\lambda$3726, [\ion{O}{2}]$\lambda$3729, H$\beta$, [\ion{O}{3}]$\lambda$5007, H$\alpha$, [N\,{\sc ii}]$\lambda$6584, [\ion{S}{2}]$\lambda$6717 and [\ion{S}{2}]$\lambda$6731,
\item a signal-to-noise $\geq3$ for the continuum measurement around H$\beta$, and
\item H$\alpha$/H$\beta_\text{corr} \geq 2.86$. 
\end{enumerate}

The redshift selection is identical to \cite{Kewley06}: the lower limit ensures that at least 20\% of the galaxy is covered by the 3 arcseconds fiber of the SDSS spectrograph, so that the spectra is representative of the global properties of the galaxy \citep{Kewley05}. The higher redshift bound is designed to ensure the completeness of the LINER class, comparatively dimmer than Seyferts. We calculated the S/N of each emission lines from the line flux and its associated error scaled by the amount suggested by \cite{Juneau14} (see Table~\ref{table:sdss_error}). These correction factors have been obtained by comparing the different duplicate observations in the dataset, and are lower than the values recommended by the MPA-JHU group for their DR4 Value Added Catalogue. Following the recommendation of \cite{Groves12}, we also add 0.35{\AA} to the equivalent width of H$\beta$ (with H$\beta_\text{corr}$ the corrected line flux), which was found to be underestimated because of an error in the 2008 version of the \texttt{GALEXEV} code \citep{Bruzual11}. Hence, we require the continuum level around H$\beta$ to have S/N$\geq3$ to ensure a reliable correction. The median correction for our sample is $\sim$6\% of the original H$\beta$ flux.

\begin{table}[htb!]
\caption{Corrections applied to the errors of emission line fluxes.}\label{table:sdss_error}
\center
\vspace{-10pt}
\begin{tabular}{c c}	
\hline
\hline
Line & Correction \\
\hline \\[-1.5ex]
 [\ion{O}{2}]$\lambda$3726 & 1.33 \\ [2ex]
 [\ion{O}{2}]$\lambda$3729 & 1.33 \\ [2ex]
 H$\beta$ & 1.29\\ [2ex]
 [\ion{O}{3}]$\lambda$5007 & 1.33 \\ [2ex] 
 [\ion{O}{1}]$\lambda$6300 & 1.02 \\ [2ex]
 H$\alpha$ & 2.06 \\ [2ex]
 [\ion{N}{2}]$\lambda$6584 & 1.44 \\ [2ex]
 [\ion{S}{2}]$\lambda$6717 & 1.36 \\ [2ex]
 [\ion{S}{2}]$\lambda$6731 & 1.36 \\ [2ex]
\hline
\end{tabular}
\footnotetext[0]{ Note: although S/N([\ion{O}{1}]$\lambda$6300) is not used in our sample selection, the associated error scaling correction is included here for completeness.}
\end{table}

We have removed duplicate observations in the sample using our own \texttt{Python} routine. For every galaxy, we look for all other objects located within 3 arcseconds (with no restriction on the redshift), and remove them all from our sample except for the one with the largest S/N(H$\alpha$). Our final sample is comprised of 105070 galaxies.

\subsubsection{Extragalactic reddening correction}
We correct the emission line fluxes for extragalactic reddening based on the Balmer decrement $R_{\alpha\beta}$, using the extinction law from \cite{Fischera05} for R$_V^A$=4.5 (and A$_V$=1). This extinction law is very close to that of \cite{Calzetti00} for starburst galaxies, and \cite{Wijesinghe11} have shown that it provides very good agreement between different SFR indicators ([\ion{O}{2}],H$\alpha$, near-UV, far-UV) for the GAMA galaxies \citep{Driver09}. Specifically, we follow the procedure described in detail in Appendix A of \cite{Vogt13a}, with:
\begin{eqnarray}
\frac{E_{\lambda-V}}{E_{B-V}} &=& -4.61777 + 1.41612 \cdot \lambda^{-1} + 1.52077 \cdot \lambda^{-2} \nonumber\\ 
& & - 0.63269 \cdot \lambda^{-3} + 0.07386 \cdot \lambda^{-4}
\end{eqnarray}
where $\lambda$ is in $\mu$m, and the actual reddening correction is given by :
\begin{equation}\label{eq:red_corr}
F_{\lambda,0} = F_{\lambda}\cdot \left( \frac{F_{H\alpha}/F_{H\beta}}{R_{\alpha\beta}}\right) ^{-\frac{ \frac{E_{\lambda-V}}{E_{B-V}}+R_V^A}{\frac{E_{H\alpha-V}}{E_{B-V}}-\frac{E_{H\beta-V}}{E_{B-V}}}}
\end{equation}
with $F_{\lambda,0}$ the intrinsic emission line flux, $F_{\lambda}$ the measured emission line flux, $\lambda$ the rest-frame emission line wavelength, and in our case, $R_V^A=4.5$. 

We adopt an intrinsic Balmer ratio $R_{\alpha\beta}=2.86$ corresponding to Case B recombination for \emph{every} object in our sample, irrespective of their classification. This value is appropriate for star-forming galaxies, but the presence of an AGN can result in an higher intrinsic Balmer ratio \citep[i.e.  $R_{\alpha\beta}\cong3.1$, e.g.][]{Osterbrock89, Kewley06}. However, it is unclear what intrinsic ratio should be applied for ``composite'' objects possibly containing a mix of star-formation and AGN. Hence, we use an intrinsic Balmer ratio of 2.86 to ensure a uniform sample without artificial separation. For consistency, we will indicate visually in all line ratio diagrams throughout this article the spatial displacement $\zeta$ associated with a intrinsic Balmer decrement $R_{\alpha\beta}$=3.1 instead of 2.86. Analytically, for an observed line ratio $F_{\lambda_{1}}/F_{\lambda_{2}}$, we can write using Eq.~\ref{eq:red_corr} : 

\begin{equation}
\frac{F_{\lambda_1,0}}{F_{\lambda_{2},0}} = \frac{F_{\lambda_1}}{F_{\lambda_{2}}} \cdot \left( \frac{2.86}{R_{\alpha\beta}} \right)^\frac{\tau_{\lambda_{2}}-\tau_{\lambda_{1}}}{\tau_{H\alpha}-\tau_{H\beta}} \cdot \left(\frac{F_{H\alpha}/F_{H\beta}}{2.86}\right)^\frac{\tau_{\lambda_{2}}-\tau_{\lambda_{1}}}{\tau_{H\alpha}-\tau_{H\beta}},
\end{equation}
where
\begin{equation}
\tau_\lambda =  \frac{E_{\lambda-V}}{E_{B-V}}+R_V^A,
\end{equation}
so that
\begin{equation}\label{eq:zeta}
\zeta(\lambda_1,\lambda_{2}) =  \left( \frac{2.86}{R_{\alpha\beta}} \right)^\frac{\tau_{\lambda_{2}}-\tau_{\lambda_{1}}}{\tau_{H\alpha}-\tau_{H\beta}}.
\end{equation}

As we will discuss in the next Sections, $\zeta$ is small enough so that the choice of $R_{\alpha\beta}$ is not critical to our analysis. Especially, as we will focus on the separation between AGN-dominated and star-forming galaxies, the objects located close-to or on the classification diagnostic lines (i.e. with very little AGN influence) can be expected to have $R_{\alpha\beta}\cong 2.86$. As mentioned above, we have removed $\sim$200 galaxies with measured H$\alpha$/H$\beta<2.86$ from our sample, under the assumptions that these low ratios are indicative of observational and/or fitting issues. 

After correcting the emission line fluxes for extragalactic reddening, our sample contains 88933 (84.6\%) galaxies classified as star-forming, 11447 (10.9\%) classified as composites and 4690 (4.5\%) classified as AGN-dominated, based \emph{solely} on their position in the $\log${\NIIHa} \emph{vs.} $\log${\OIIIHb} diagram.

\subsection{\ion{H}{2} region spectra}
Since the SDSS spectra represent nuclear spectra of whole galaxies, it is important for classification purposes that we also have a set of well-observed isolated \ion{H}{2} regions covering a wide range of chemical abundances. For this purpose we adopt the excellent homogeneous dataset from \citet{vanZee98}. 

This dataset is somewhat deficient in the most metal rich objects, so we have supplemented the \citet{vanZee98}  \ion{H}{2} regions with our own data on the \ion{H}{2} regions in the Seyfert galaxy NGC\,5427. These bright \ion{H}{2} regions are unaffected by the weak Seyfert 2 nucleus, and their abundances range up to three times solar. We refer the reader to \cite{Dopita14b} for more details on the observations, data reduction and emission line flux measurements for these \ion{H}{2} regions.

\section{The theoretical \ion{H}{2} region models}\label{sec:HIImodels}

Throughout this article, we rely on the grids of line intensities for \ion{H}{2} regions derived from the modelling code \emph{MAPPINGS IV} by \cite{Dopita13a}. These grids cover a wide range of abundances ($5 - 0.05~Z_{\odot}$) and ionisation parameters ($6.5 \lesssim \log q \lesssim 8.5$). 

\emph{MAPPINGS IV} is the latest evolution of the \emph{MAPPINGS} code \citep{Dopita82,Binette82,Binette85,Sutherland93, Groves04, Allen08}, that (among other updates) can now account for the possible non-Maxwellian energy distribution of electrons in astrophysical plasmas. The idea that the energy distribution of electrons in planetary nebulae and \ion{H}{2} regions may depart from a standard Maxwell-Boltzmann distribution to resemble a $\kappa$-distribution, characterised by a high-energy tail, was recently suggested by \cite{Nicholls12}. The consequences of a $\kappa$-distribution of electron energies on temperature and abundance measurements in \ion{H}{2} regions have been discussed in detail by  \cite{Nicholls13} and in respect of the effect on strong line intensities by \citet{Dopita13a}. For these, the effect of a $\kappa$-distribution is only minor, and does not significantly affect our analysis. Throughout this paper, we have adopted $\kappa=20$.

\section{Creating 3D line ratio diagrams}\label{sec:3d-diag}
There is \emph{a priori} no reason to restrict line ratio diagrams to 2 dimensions, other than the evident practicality of visualisation. Here, we exploit recent software developments to explore the potential of 3D line ratio diagrams. The basic concept is as follows. As a starting point we use the 2D diagnostics from \cite{Dopita13a} which cleanly separate the ionisation parameter, $q$, and the oxygen abundance, $12+\log$(O/H). We then couple them with an additional line ratio, chosen specifically to help differentiate  \ion{H}{2}-like objects from AGNs. This third ratio ought to be more sensitive to the hardness of the radiation field. In Table~\ref{table:keys}, we list the different line ratios used for each of the three categories;
\begin{itemize}
\item Category I: abundance sensitive ratios,
\item Category II: $q$- sensitive ratios, and
\item Category III: radiation hardness-sensitive ratios.
\end{itemize}
In practice, of course, the separation is not as clean as implied by this list, since each ratio is in some part sensitive to all three parameters we are trying to dissociate. Nonetheless the exercise remains useful as a means of teasing out these parameters.
\begin{table}[htb!]
\caption{Line ratios and associated keys.}\label{table:keys}
\center
\vspace{-10pt}
\begin{tabular}{c c c c c c}	
\hline
\hline
\multicolumn{2}{c}{Category I } & \multicolumn{2}{c}{Category II } & \multicolumn{2}{c}{Category III }\\
Key & Ratio & Key & Ratio & Key & Ratio\\
\hline \\[-1.5ex]
a: & $\log\frac{[\text{N\,{\sc ii}}]}{[\text{O\,{\sc ii}]}}$ & c: &  $\log\frac{[\text{O\,{\sc iii}}]}{[\text{O\,{\sc ii}}]}$ & f: & $\log\frac{[\text{O\,{\sc iii}}]}{\text{H}\beta}$\\ [2ex]
b: & $\log\frac{[\text{N\,{\sc ii}}]}{[\text{S\,{\sc ii}}]}$ & d: & $\log\frac{[\text{O\,{\sc iii}}]}{[\text{S\,{\sc ii}}]}$ & g: & $\log\frac{[\text{N\,{\sc ii}}]}{\text{H}\alpha}$\\ [2ex]
& & e: & $\log\frac{[\text{O\,{\sc iii}}]}{[\text{N\,{\sc ii}}]}$ & h: & $\log\frac{[\text{S\,{\sc ii}}]}{\text{H}\alpha}$\\ [2ex]
& & & & i: & $\log\frac{[\text{O\,{\sc i}}]}{\text{H}\alpha}$\\ [2ex]
\hline
\end{tabular}
\footnotetext[0]{Note: throughout this paper and unless noted otherwise, when we refer to specific emission lines we mean [N\,{\sc ii}]$\equiv$[N\,{\sc ii}$]\lambda$6583, [S\,{\sc ii}]$\equiv$[S\,{\sc ii}$]\lambda$6717+$\lambda$6731, [O\,{\sc ii}]$\equiv$[O\,{\sc ii}$]\lambda$3727+$\lambda$3729, [O\,{\sc iii}]$\equiv$[O\,{\sc iii}$]\lambda$5007 and [O\,{\sc i}]$\equiv$[O\,{\sc i}$]\lambda$6300.}
\end{table}

We restrict ourselves to ratios involving (usually) intense emission lines commonly observed in both \ion{H}{2} regions and AGN-dominated objects. To each ratio we associate a ``key", defined in Table~\ref{table:keys}, to unambiguously identify them throughout this paper. 

\begin{figure*}[htb!]
\centerline{\includemovie[toolbar, 3Dviews2 = 3dviews_movie15.txt, label=my_movie15, text={\includegraphics[scale=0.5]{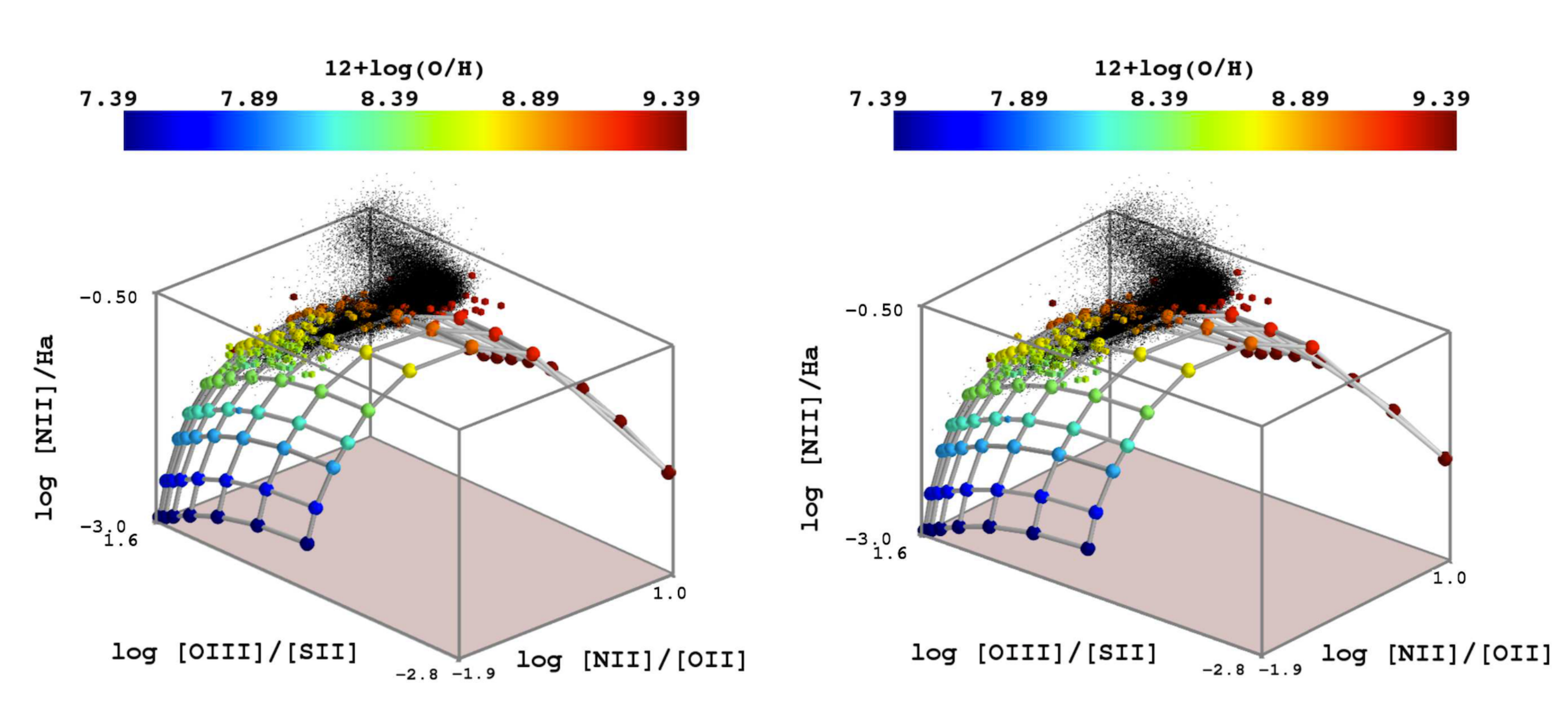}} ]{1\linewidth}{ 0.6\linewidth}{./fig1.u3d}} 
\caption{Example of a 3D line ratio diagram, labelled $\cal{ZQE}_\text{adg}$ (a.k.a. $\log${\NIIOII} \emph{vs.} $\log${\OIIISII} vs $\log${\NIIHa}), in the form of a cross-eyed stereo pair. The plane of \emph{MAPPINGS IV} simulations of H\,{\sc ii} regions is represented by the coloured spheres connected by the grey rods, where each sphere corresponds to one distinct simulation. The colour indicates the oxygen abundance in terms of $12+\log$(O/H). Individual cubes correspond the \cite{vanZee98} data points, and the small cones to the NCG\,5472 measurements of individual \ion{H}{2} regions, also coloured as a function of their metallicity. Detailed instructions to view this cross-eyed stereo pair can be found in \cite{Vogt12}. An interactive version of this Figure, that allows the reader to freely rotate and/or zoom in and out, can be accessed by using Adobe Acrobat Reader v9.0 or above. In the interactive model, the red, green and blue axes correspond to the $\log$\NIIOII, $\log${\OIIISII} and $\log${\NIIHa} directions, respectively. }\label{fig:3D_example}
\end{figure*}

One example of a 3D line ratio diagram ($\log${\NIIOII} \emph{vs.} $\log${\OIIISII} \emph{vs.} $\log${\NIIHa}) is shown in Figure~\ref{fig:3D_example}. Figure~\ref{fig:3D_example} is interactive, and allows the reader to freely rotate, zoom in/out and/or fly through the 3D diagram.\footnote{Accessing the interactive model requires Adobe Acrobat Reader v9.0 or above, which is freely accessible. This figure, which follows a concept described by \cite{Barnes08}, was created using the \texttt{Python} module \texttt{Mayavi2} \citep{Ram11}, and the commercial software \texttt{PDF3D}, similarly to the interactive counterpart of Figure 9 in \cite{Vogt13b}.} Figure~\ref{fig:3D_example} is also 3D printable using the STL file provided as supplementary material (see Appendix~\ref{app:3d_print} for more details).

We refer to these new 3D line ratio diagrams as $\cal{ZQE}$ diagrams, following the categorisation of the line ratios involved. To uniquely identify all possible $\cal{ZQE}$ diagrams, we attach the key of the three line ratios involved (in the order defining a right-handed orthogonal base), in the form of $\cal{ZQE}$$_{\text{x}_1\text{x}_2\text{x}_3}$, where $\text{x}_1$,$\text{x}_2$ and $\text{x}_3$ corresponds to keys of line ratios in the Category I, II and III defined in Table~\ref{table:keys}. For example, the 3D line ratio diagram shown in Figure~\ref{fig:3D_example} is $\cal{ZQE}_{\text{adg}}$.

In this new 3D diagram, the spatial structure of the cloud of points of SDSS galaxies resemble that of a \emph{nudibranch}. \ion{H}{2}-like objects are located on, or close to, the photoionization model grid and can be associated with the sea slug's body. This sequence is clearly separated from the AGN sequence, which extends away from the \ion{H}{2} region model grid (and which can be regarded as the ``feelers'' of the nudibranch). These AGN-dominated regions also display a clear substructure in the spatial density of galaxies, best revealed in the interactive version of Figure~\ref{fig:3D_example}. 

From the line ratios listed in Table~\ref{table:keys}, it is possible to construct $2\times3\times4=24$ different 3D spaces combining one ratio of each category, and we list them all in Table~\ref{table:all_3d} with their $\cal{ZQE}$ denomination. 

\begin{table}[htb!]
\caption{All possible $\cal{ZQE}$ diagrams combining one line ratio from each Category I, II and III defined in Table~\ref{table:keys}, and existence of an associated $\cal{ZE}$ diagnostic.}\label{table:all_3d}
\center
\vspace{-10pt}
\begin{tabular}{c c c c c}
\hline\hline
Name & Ratio 1& Ratio 2 & Ratio 3 & $\cal{ZE}$ diagnostic ? \\
\hline \\ [-1.5ex]
$\cal{ZQE}_\text{acf}$&$\log\frac{[\text{N\,{\sc ii}}]}{[\text{O\,{\sc ii}]}}$ & $\log\frac{[\text{O\,{\sc iii}}]}{[\text{O\,{\sc ii}}]}$ & $\log\frac{[\text{O\,{\sc iii}}]}{[\text{H}\beta]}$ & yes\\ [2ex]
$\cal{ZQE}_\text{acg}$&$\log\frac{[\text{N\,{\sc ii}}]}{[\text{O\,{\sc ii}]}}$ & $\log\frac{[\text{O\,{\sc iii}}]}{[\text{O\,{\sc ii}}]}$ & $\log\frac{[\text{N\,{\sc ii}}]}{[\text{H}\alpha]}$ & yes \\ [2ex]
$\cal{ZQE}_\text{ach}$&$\log\frac{[\text{N\,{\sc ii}}]}{[\text{O\,{\sc ii}]}}$ & $\log\frac{[\text{O\,{\sc iii}}]}{[\text{O\,{\sc ii}}]}$ & $\log\frac{[\text{S\,{\sc ii}}]}{[\text{H}\alpha]}$ & no \\ [2ex]
$\cal{ZQE}_\text{aci}$&$\log\frac{[\text{N\,{\sc ii}}]}{[\text{O\,{\sc ii}]}}$ & $\log\frac{[\text{O\,{\sc iii}}]}{[\text{O\,{\sc ii}}]}$ & $\log\frac{[\text{O\,{\sc i}}]}{[\text{H}\alpha]}$ & no \\ [2ex]
$\cal{ZQE}_\text{adf}$&$\log\frac{[\text{N\,{\sc ii}}]}{[\text{O\,{\sc ii}]}}$ & $\log\frac{[\text{O\,{\sc iii}}]}{[\text{S\,{\sc ii}}]}$ & $\log\frac{[\text{O\,{\sc iii}}]}{[\text{H}\beta]}$ &no\\ [2ex]
$\cal{ZQE}_\text{adg}$&$\log\frac{[\text{N\,{\sc ii}}]}{[\text{O\,{\sc ii}]}}$ & $\log\frac{[\text{O\,{\sc iii}}]}{[\text{S\,{\sc ii}}]}$ & $\log\frac{[\text{N\,{\sc ii}}]}{[\text{H}\alpha]}$ &  yes \\ [2ex]
$\cal{ZQE}_\text{adh}$&$\log\frac{[\text{N\,{\sc ii}}]}{[\text{O\,{\sc ii}]}}$ & $\log\frac{[\text{O\,{\sc iii}}]}{[\text{S\,{\sc ii}}]}$ & $\log\frac{[\text{S\,{\sc ii}}]}{[\text{H}\alpha]}$ & no \\ [2ex]
$\cal{ZQE}_\text{adi}$&$\log\frac{[\text{N\,{\sc ii}}]}{[\text{O\,{\sc ii}]}}$ & $\log\frac{[\text{O\,{\sc iii}}]}{[\text{S\,{\sc ii}}]}$ & $\log\frac{[\text{O\,{\sc i}}]}{[\text{H}\alpha]}$ & no \\ [2ex]
$\cal{ZQE}_\text{aef}$&$\log\frac{[\text{N\,{\sc ii}}]}{[\text{O\,{\sc ii}]}}$ & $\log\frac{[\text{O\,{\sc iii}}]}{[\text{N\,{\sc ii}}]}$ & $\log\frac{[\text{O\,{\sc iii}}]}{[\text{H}\beta]}$ & yes \\ [2ex]
$\cal{ZQE}_\text{aeg}$&$\log\frac{[\text{N\,{\sc ii}}]}{[\text{O\,{\sc ii}]}}$ & $\log\frac{[\text{O\,{\sc iii}}]}{[\text{N\,{\sc ii}}]}$ & $\log\frac{[\text{N\,{\sc ii}}]}{[\text{H}\alpha]}$ & yes \\ [2ex]
$\cal{ZQE}_\text{aeh}$&$\log\frac{[\text{N\,{\sc ii}}]}{[\text{O\,{\sc ii}]}}$ & $\log\frac{[\text{O\,{\sc iii}}]}{[\text{N\,{\sc ii}}]}$ & $\log\frac{[\text{S\,{\sc ii}}]}{[\text{H}\alpha]}$ & no \\ [2ex]
$\cal{ZQE}_\text{aei}$&$\log\frac{[\text{N\,{\sc ii}}]}{[\text{O\,{\sc ii}]}}$ & $\log\frac{[\text{O\,{\sc iii}}]}{[\text{N\,{\sc ii}}]}$ & $\log\frac{[\text{O\,{\sc i}}]}{[\text{H}\alpha]}$ & no \\ [2ex]
$\cal{ZQE}_\text{bcf}$&$\log\frac{[\text{N\,{\sc ii}}]}{[\text{S\,{\sc ii}]}}$ & $\log\frac{[\text{O\,{\sc iii}}]}{[\text{O\,{\sc ii}}]}$ & $\log\frac{[\text{O\,{\sc iii}}]}{[\text{H}\beta]}$ & no \\ [2ex]
$\cal{ZQE}_\text{bcg}$&$\log\frac{[\text{N\,{\sc ii}}]}{[\text{S\,{\sc ii}]}}$ & $\log\frac{[\text{O\,{\sc iii}}]}{[\text{O\,{\sc ii}}]}$ & $\log\frac{[\text{N\,{\sc ii}}]}{[\text{H}\alpha]}$ & yes  \\ [2ex]
$\cal{ZQE}_\text{bch}$&$\log\frac{[\text{N\,{\sc ii}}]}{[\text{S\,{\sc ii}]}}$ & $\log\frac{[\text{O\,{\sc iii}}]}{[\text{O\,{\sc ii}}]}$ & $\log\frac{[\text{S\,{\sc ii}}]}{[\text{H}\alpha]}$ & yes \\ [2ex]
$\cal{ZQE}_\text{bci}$&$\log\frac{[\text{N\,{\sc ii}}]}{[\text{S\,{\sc ii}]}}$ & $\log\frac{[\text{O\,{\sc iii}}]}{[\text{O\,{\sc ii}}]}$ & $\log\frac{[\text{O\,{\sc i}}]}{[\text{H}\alpha]}$ & no \\ [2ex]
$\cal{ZQE}_\text{bdf}$&$\log\frac{[\text{N\,{\sc ii}}]}{[\text{S\,{\sc ii}]}}$ & $\log\frac{[\text{O\,{\sc iii}}]}{[\text{S\,{\sc ii}}]}$ & $\log\frac{[\text{O\,{\sc iii}}]}{[\text{H}\beta]}$ & yes \\ [2ex]
$\cal{ZQE}_\text{bdg}$&$\log\frac{[\text{N\,{\sc ii}}]}{[\text{S\,{\sc ii}]}}$ & $\log\frac{[\text{O\,{\sc iii}}]}{[\text{S\,{\sc ii}}]}$ & $\log\frac{[\text{N\,{\sc ii}}]}{[\text{H}\alpha]}$ & yes \\ [2ex]
$\cal{ZQE}_\text{bdh}$&$\log\frac{[\text{N\,{\sc ii}}]}{[\text{S\,{\sc ii}]}}$ & $\log\frac{[\text{O\,{\sc iii}}]}{[\text{S\,{\sc ii}}]}$ & $\log\frac{[\text{S\,{\sc ii}}]}{[\text{H}\alpha]}$ & yes \\ [2ex]
$\cal{ZQE}_\text{bdi}$&$\log\frac{[\text{N\,{\sc ii}}]}{[\text{S\,{\sc ii}]}}$ & $\log\frac{[\text{O\,{\sc iii}}]}{[\text{S\,{\sc ii}}]}$ & $\log\frac{[\text{O\,{\sc i}}]}{[\text{H}\alpha]}$ & no \\ [2ex]
$\cal{ZQE}_\text{bef}$&$\log\frac{[\text{N\,{\sc ii}}]}{[\text{S\,{\sc ii}]}}$ & $\log\frac{[\text{O\,{\sc iii}}]}{[\text{N\,{\sc ii}}]}$ & $\log\frac{[\text{O\,{\sc iii}}]}{[\text{H}\beta]}$ & yes \\ [2ex]
$\cal{ZQE}_\text{beg}$&$\log\frac{[\text{N\,{\sc ii}}]}{[\text{S\,{\sc ii}]}}$ & $\log\frac{[\text{O\,{\sc iii}}]}{[\text{N\,{\sc ii}}]}$ & $\log\frac{[\text{N\,{\sc ii}}]}{[\text{H}\alpha]}$ & yes \\ [2ex]
$\cal{ZQE}_\text{beh}$&$\log\frac{[\text{N\,{\sc ii}}]}{[\text{S\,{\sc ii}]}}$ & $\log\frac{[\text{O\,{\sc iii}}]}{[\text{N\,{\sc ii}}]}$ & $\log\frac{[\text{S\,{\sc ii}}]}{[\text{H}\alpha]}$ & yes \\ [2ex]
$\cal{ZQE}_\text{bei}$&$\log\frac{[\text{N\,{\sc ii}}]}{[\text{S\,{\sc ii}]}}$ & $\log\frac{[\text{O\,{\sc iii}}]}{[\text{N\,{\sc ii}}]}$ & $\log\frac{[\text{O\,{\sc i}}]}{[\text{H}\alpha]}$ & no \\ [2ex]
\hline
\end{tabular}
\end{table} 

\subsection{Exploiting $\cal{ZQE}$ diagrams}

One of the key advantages of $\cal{ZQE}$ diagrams is the ability to inspect them interactively (in a similar manner to the interactive counterpart of Figure~\ref{fig:3D_example}). Following this approach, it is possible to identify new points of view of interest on the multi-dimensional space of galaxy line ratios. Working interactively with 3D line ratio diagrams may seem (at first) cumbersome. As we will argue here, it really is not the case anymore. We rely on the \texttt{Python} module \texttt{Mayavi2} to create our interactive 3D diagrams \citep{Ram11}. We refer the reader to the full package documentation available online\footnote{\url{http://docs.enthought.com/mayavi/mayavi/}, accessed on October 29th, 2013.}. \texttt{Mayavi2} is a module dedicated to ``3D scientific data visualization and plotting in \texttt{Python}''. It is in some ways reminiscent of the \texttt{Matplotlib} module dedicated to 2D plotting \citep{Hunter07}. We stress here that unlike dedicated computer-assisted design (CAD) softwares, using \texttt{Mayavi2} does not require any specific knowledge \emph{a priori}. The module syntax is relatively intuitive, as illustrated by the basic examples available online\footnote{\url{http://docs.enthought.com/mayavi/mayavi/auto/examples.html}}. Similarly to other \texttt{Python} modules, \texttt{Mayavi2} can be integrated seamlessly in any given \texttt{Python} script, and within a few lines, allows the creation of an interactive 3D model, for example a $\cal{ZQE}$ diagram. We note that in addition to a ''cursor-based'' approach, the interactive diagrams generated with \texttt{Mayavi2} can also be manipulated from a \texttt{Python} shell and scripts. Readers with practical questions regarding the implementations of interactive 3D diagrams with \texttt{Python} are welcome to contact us. 

At the time of publication of this article, the interactive 3D models generated by \texttt{Mayavi2} cannot be directly integrated in documents in a \emph{Portable Document Format} (PDF). While \texttt{Mayavi2} can save 3D models in different dedicated file formats (e.g. .VRML, .OBJ, .IV), an additional step is required to transform these in the U3D format, compatible for inclusion in PDF documents. For this article, we relied on the commercial software \texttt{PDF3DReportGen} to transform the VRML file generated by \texttt{Mayavi} into a .U3D file, and included it in this article with \texttt{pdftex} and the \texttt{media9} package in \LaTeX.
$ $\\

\section{From 3D $\cal{ZQE}$ diagrams to new 2D line ratios diagnostics}\label{sec:3d-2d}

Having introduced the interactive $\cal{ZQE}$ 3D line ratio diagrams as a new \emph{tool} to study the multi-dimensional galaxy emission line space, we turn our attention to one possible application: the definition of new diagnostic diagrams to separate \ion{H}{2}-like and AGN-like objects independently of the standard line ratio diagrams. To that end, we have visually and interactively inspected all twenty-four $\cal{ZQE}$ diagrams, and selected a subsample of thirteen in which the starburst sequence and the AGN sequence are best separated. The other eleven diagrams (that do no have a $\cal{ZE}$ diagnostic associated to in Table~\ref{table:all_3d}) do not show an evident separation between the AGN and starburst sequences. Hence, these diagrams are less suitable for the kind of analysis presented here. As we discuss in Section~\ref{sec:AGN}, these $\cal{ZQE}$ diagrams may become of interest in a different type of application, for example when looking at the inherent structure of the AGN branch in the multi-dimensional line ratio space of galaxy spectra, which will be explored in a separate article.

\subsection{ The $\cal{ZE}$$_{\text{x}_1\text{x}_2\text{x}_3}(\phi;\theta)$ diagrams}\label{sec:3d-2d_def}

The original \emph{MAPPINGS IV} simulation grids created by \cite{Dopita13a} define a set of surfaces in the $\cal{ZQE}$ line ratio spaces (see Figure~\ref{fig:3D_example}). For some of the grids, the intrinsic curvature in the third dimension is small, such that it is possible to find a specific point-of-view from which the grid collapses onto itself, with a thickness $\lesssim$0.3 dex. By identifying these specific viewpoints, we effectively identify new (composite) 2D line ratio diagrams - the $\cal{ZE}$$_{\text{x}_1\text{x}_2\text{x}_3}(\phi;\theta)$ diagrams -  which rely on the combination of three different line ratios, and in which \ion{H}{2}-like objects are degenerate and constrained to a small region in the diagram.

\begin{figure}[htb!]
\centerline{\includegraphics[scale=0.35]{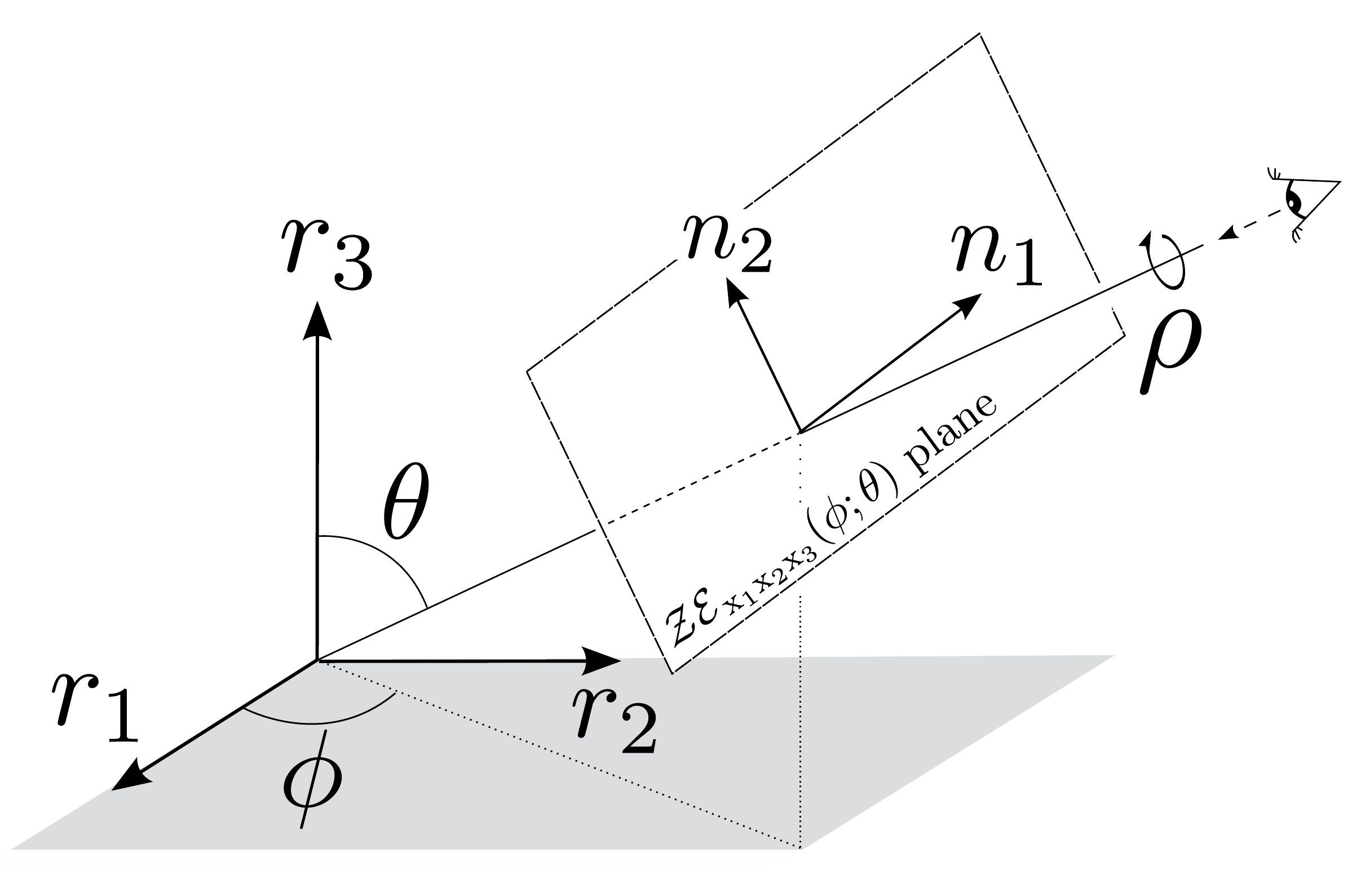}}
\caption{Schematic illustrating the concept of the $\cal{ZE}$$_{\text{x}_1\text{x}_2\text{x}_3}(\phi;\theta)$ diagram, and the associated notation defined in this article.}\label{fig:drawing}
\end{figure}

These new 2D line ratio diagrams are uniquely defined by 
\begin{enumerate} 
\item the three line ratios involved, and 
\item the angles $\phi$ and $\theta$ defining the viewing angle in the $\cal{ZQE}$ space defined by the ratios.
\end{enumerate}
Here $\phi$ and $\theta$ are defined following the standard spherical coordinates convention (see Figure~\ref{fig:drawing}). We adopt the convention that the roll angle $\rho=0$ to ensure the uniqueness of each diagram. For reasons highlighted below, we refer to these 2D line ratio diagrams constructed from the projection of a $\cal{ZQE}$ space as Metallicity-Excitation ($\cal{ZE}$) diagrams. With this naming convention, we avoid the confusion which may arise through the use of the standard optical line ratio diagnostic diagrams, or with the more recent Mass-Excitation (MEx) diagram from \cite{Juneau11, Juneau14}. 

We introduce the following notation (illustrated in Figure~\ref{fig:drawing}), that allows to uniquely identify any $\cal{ZE}$ diagram (always provided that $\rho=0$);

\begin{equation}\label{eq:def_ze}
\mathcal{ZE}_{\text{x}_1\text{x}_2\text{x}_3}(\phi;\theta)\quad \text{with} \quad \phi\in[0,180[\text{ }; \theta\in[0,180[
\end{equation}
where $\text{x}_1$,x$_2$ and x$_3$ are the three line ratio keys involved (as defined in Table~\ref{table:keys}). We limit $\phi$ to 180 degrees to avoid a \emph{mirror} version of each diagram. 

For any given triplet of line ratio values $(r_1;r_2;r_3)$, the $\cal{ZE}$$_{\text{x}_1\text{x}_2\text{x}_3}(\phi;\theta)$ diagram associates a unique doublet of composite line ratios $(n_1;n_2)$, defined by :

 \begin{eqnarray}
n_1 &=& -r_1\sin\phi + r_2\cos\phi\label{eq:n1}\\
n_2 &=& -r_1\cos\phi\cos\theta  - r_2\sin\phi\cos\theta + r_3\sin\theta\label{eq:n2}
\end{eqnarray}

One should note that because we adopted the convention of $\rho=0$, the composite line ratio $n_1$ is simply a combination of the first two ratios $r_1$ and $r_2$.

\subsection{The $\cal{ZE}$$_{\text{x}_1\text{x}_2\text{x}_3}(\phi^*;\theta^*)$ diagnostics}\label{sec:me_diag}
From the twenty-four initial $\cal{ZQE}$ diagrams listed in Table~\ref{table:all_3d}, we have identified thirteen for which:
\begin{enumerate}
\item  we could find a $\cal{ZE}$ plane in which the \ion{H}{2} regions collapse onto a line with a thickness $\lesssim$ 0.3 dex, and for which
\item the starburst branch of the SDSS galaxies is well separated from the AGN-like objects.
\end{enumerate}

Hence, we can construct thirteen new composite line ratio diagnostic diagrams to classify galaxies as \ion{H}{2}-like or AGN-like. In the next two subsections, we describe in details how we determine the specific angles $\phi^*$, $\theta^*$ and the diagnostic line parameters for each of the $\cal{ZE}$ diagrams.

\subsubsection{Identifying $\phi^*$ and $\theta^*$: manual vs. automated approach}\label{sec:star}

\begin{figure}[htb!]
\centerline{ \includegraphics[scale=0.32]{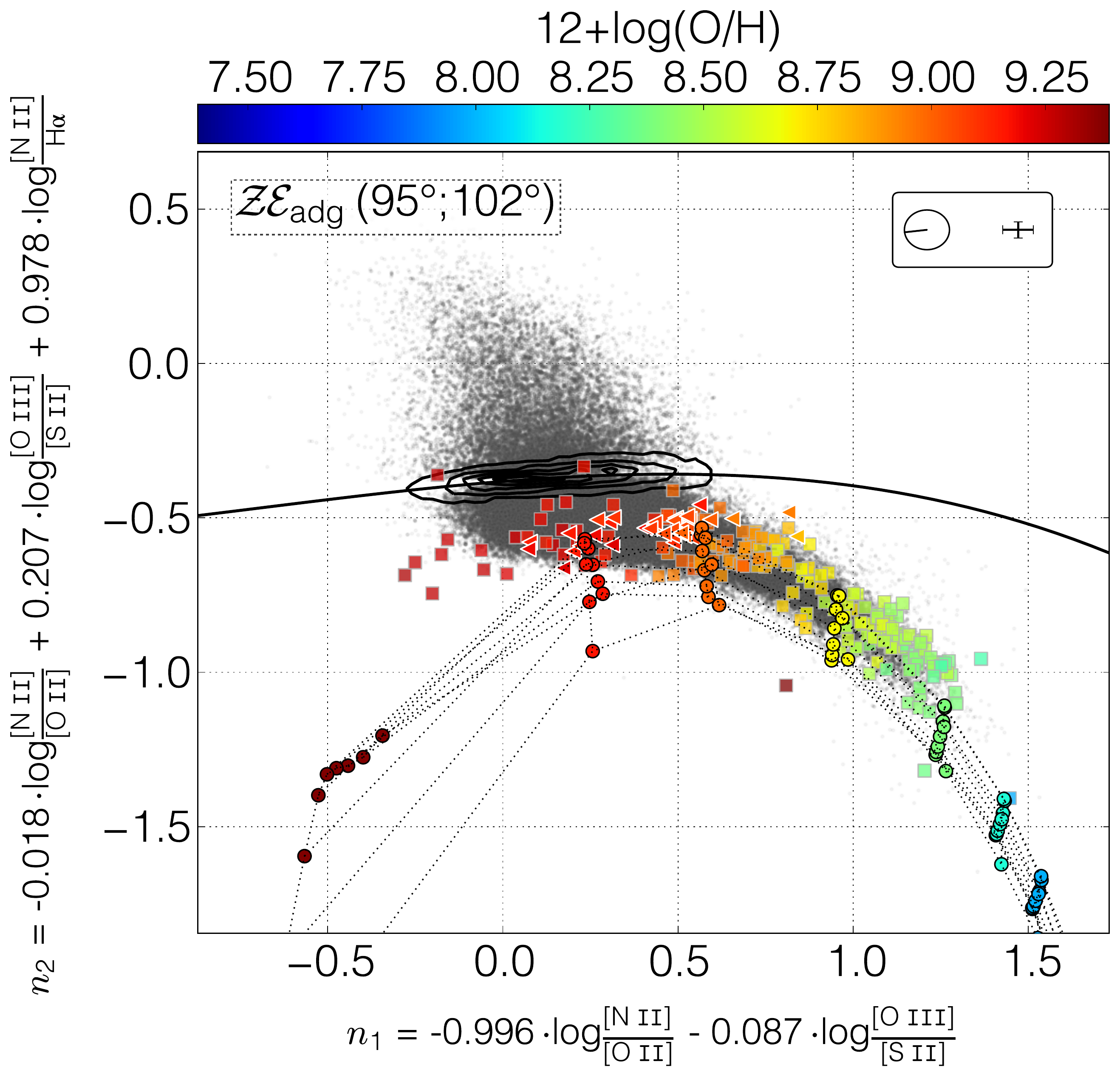}}
\caption{The $\cal{ZE}$$_{\text{adg}}$(95$^{\circ}$;102$^{\circ}$) diagram. The diagram name and associated values of $(\phi^*,\theta^*)$ are shown in the top-left corner for completeness. \ion{H}{2}-like and AGN-like SDSS galaxies are in grey. Uncertain galaxies (based on all $\cal{ZE}$ diagnostics) are represented by density contours (5\%, 20\%, 40\% and 80\% of the maximum density). The round coloured dots (connected by the dotted lines) correspond to the \emph{MAPPINGS IV} models from \cite{Dopita13a}. These provide guidance about the theoretical shape of the H\,{\sc ii} regions space. The \cite{vanZee98} points are represented by small squares with 75\% opacity, and the measurements from NGC\,5427 are marked with small triangles.  All measured \ion{H}{2} regions are color-coded according to their oxygen abundance.  The black thick line traces our diagnostic line separating the H\,{\sc ii}-like objects from the AGN-like ones, for which we adopt a 3rd degree polynomial functional form. The black cross (top right) indicates the median error associated with the given combination of SDSS flux ratios, and the circle-and-bar symbol marks the intensity and direction (taken from the circle center outwards) of the displacement associated with $R_{\alpha\beta}$=3.1 instead of 2.86.}\label{fig:me_1}
\end{figure}

\begin{figure*}[htb!]
\centerline{ \includegraphics[scale=0.27]{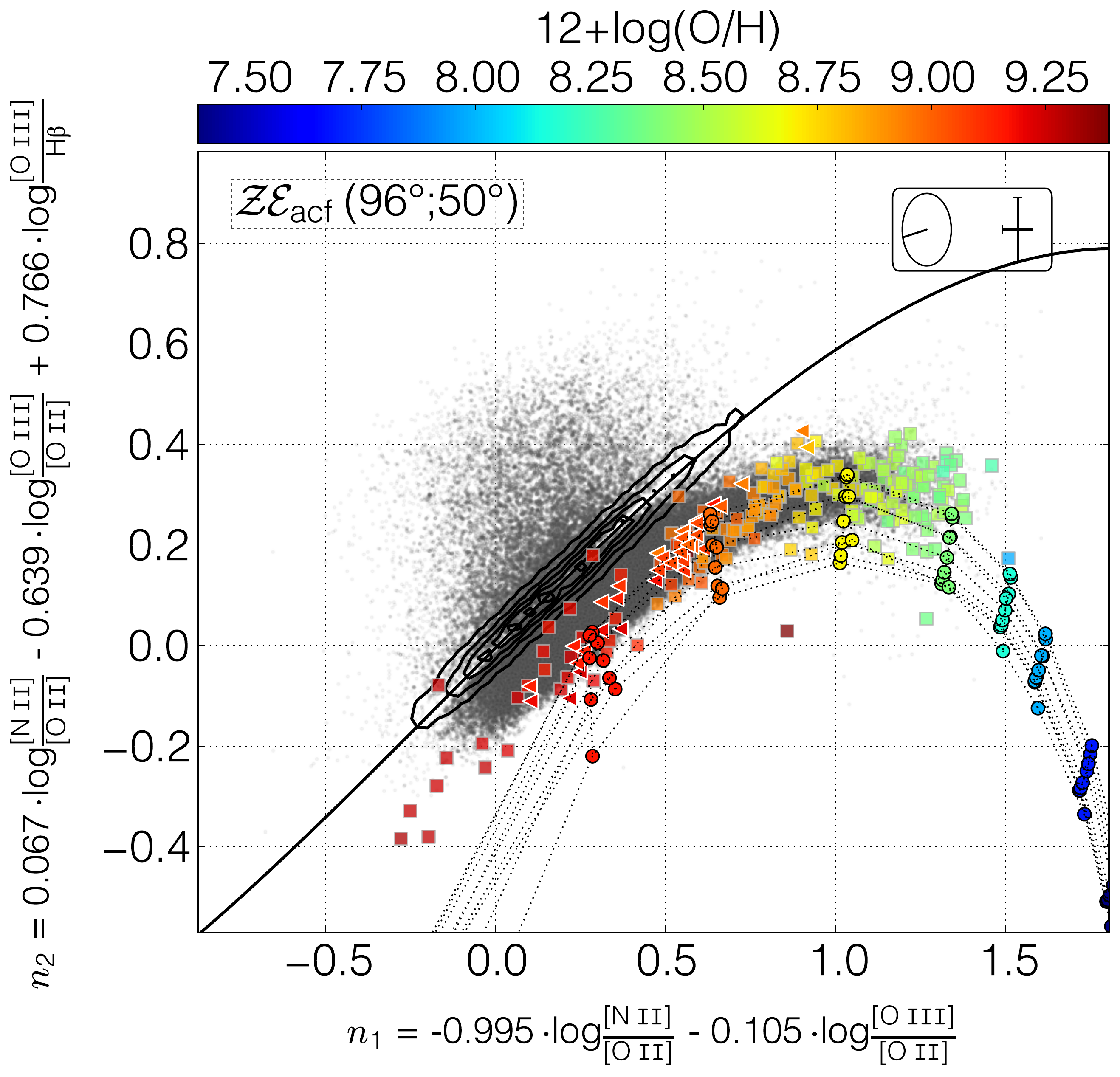}\qquad\qquad \includegraphics[scale=0.27]{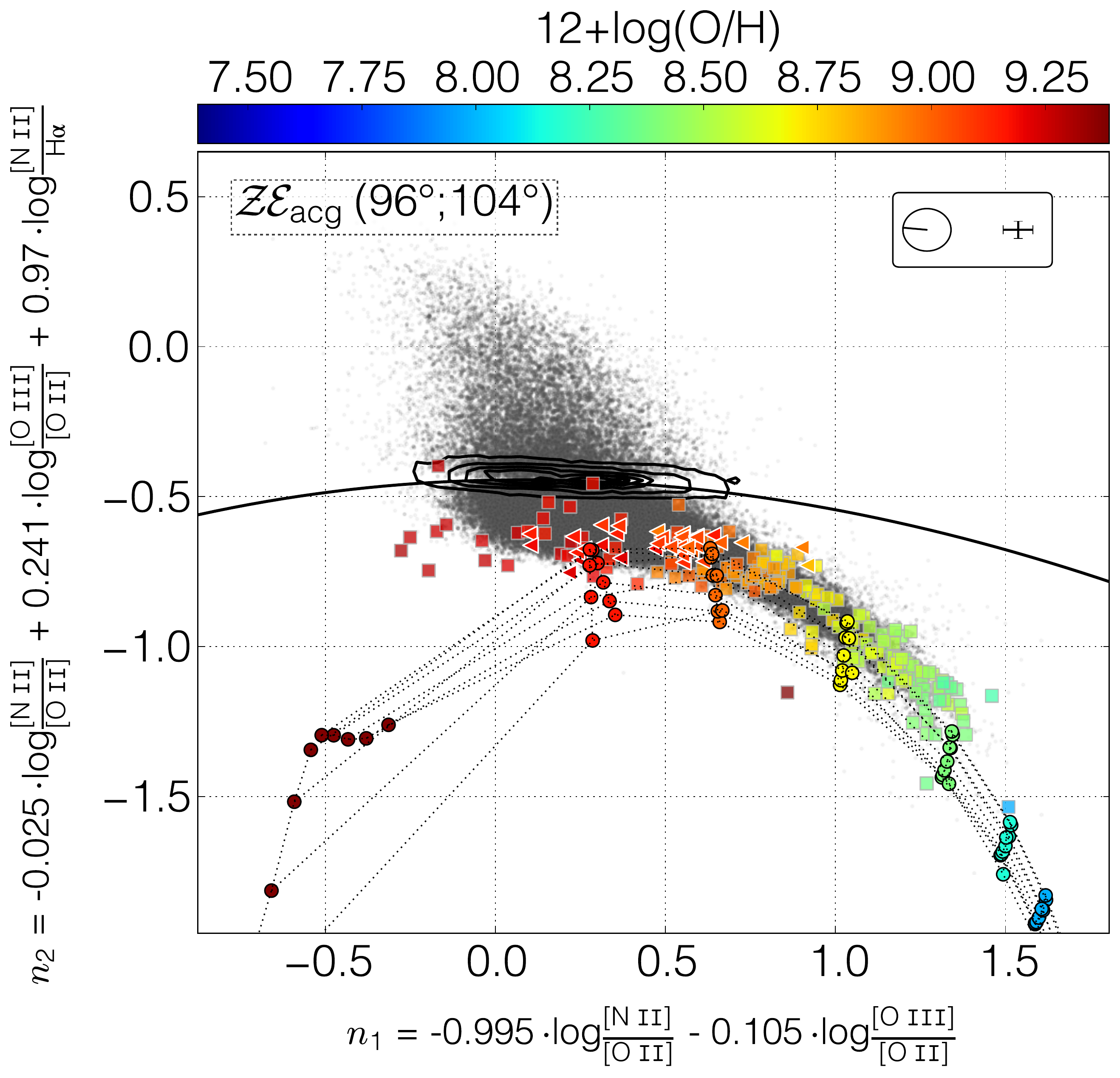}}\centerline{  \includegraphics[scale=0.27]{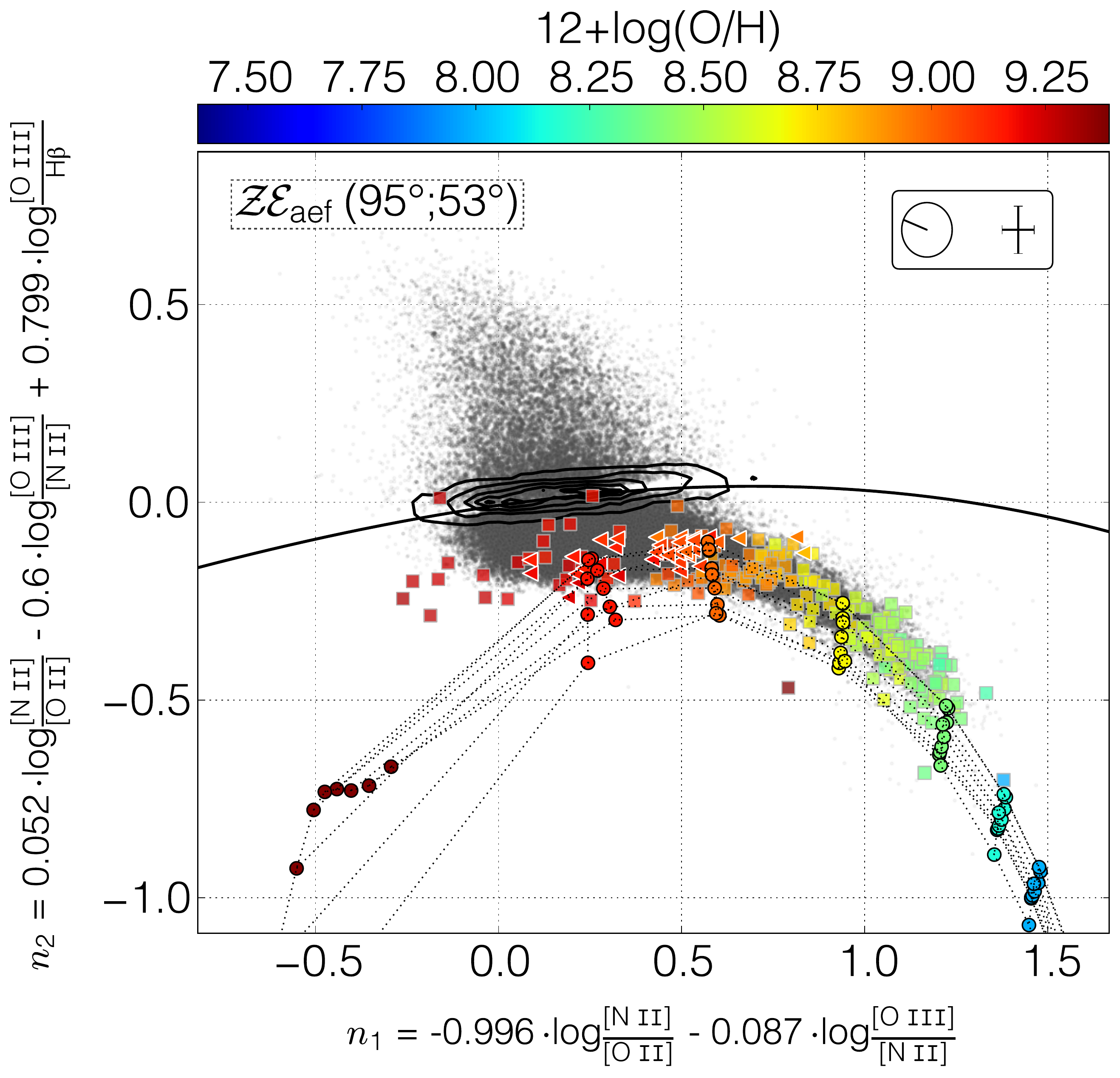}\qquad\qquad \includegraphics[scale=0.27]{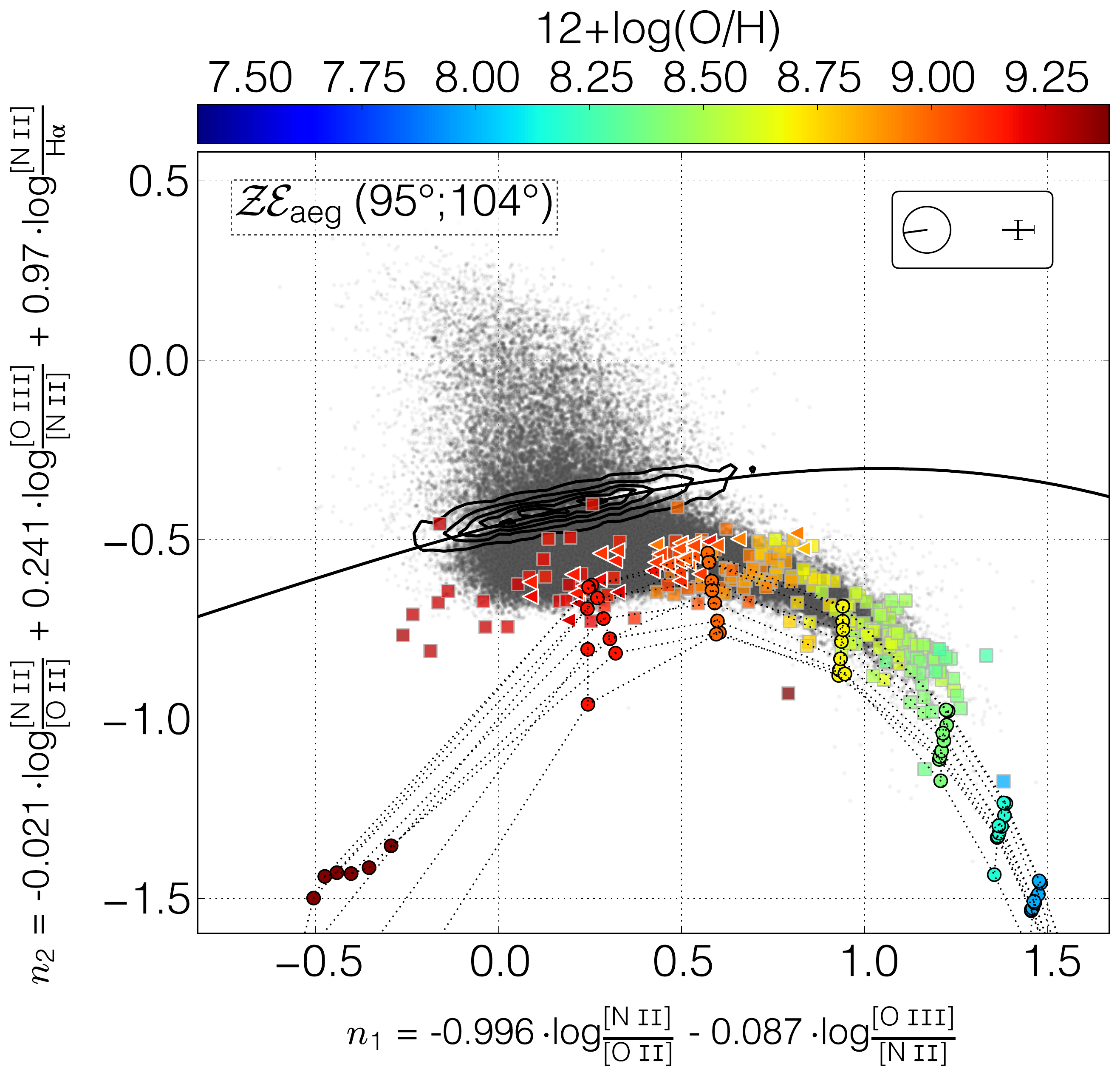}}
\centerline{  \includegraphics[scale=0.27]{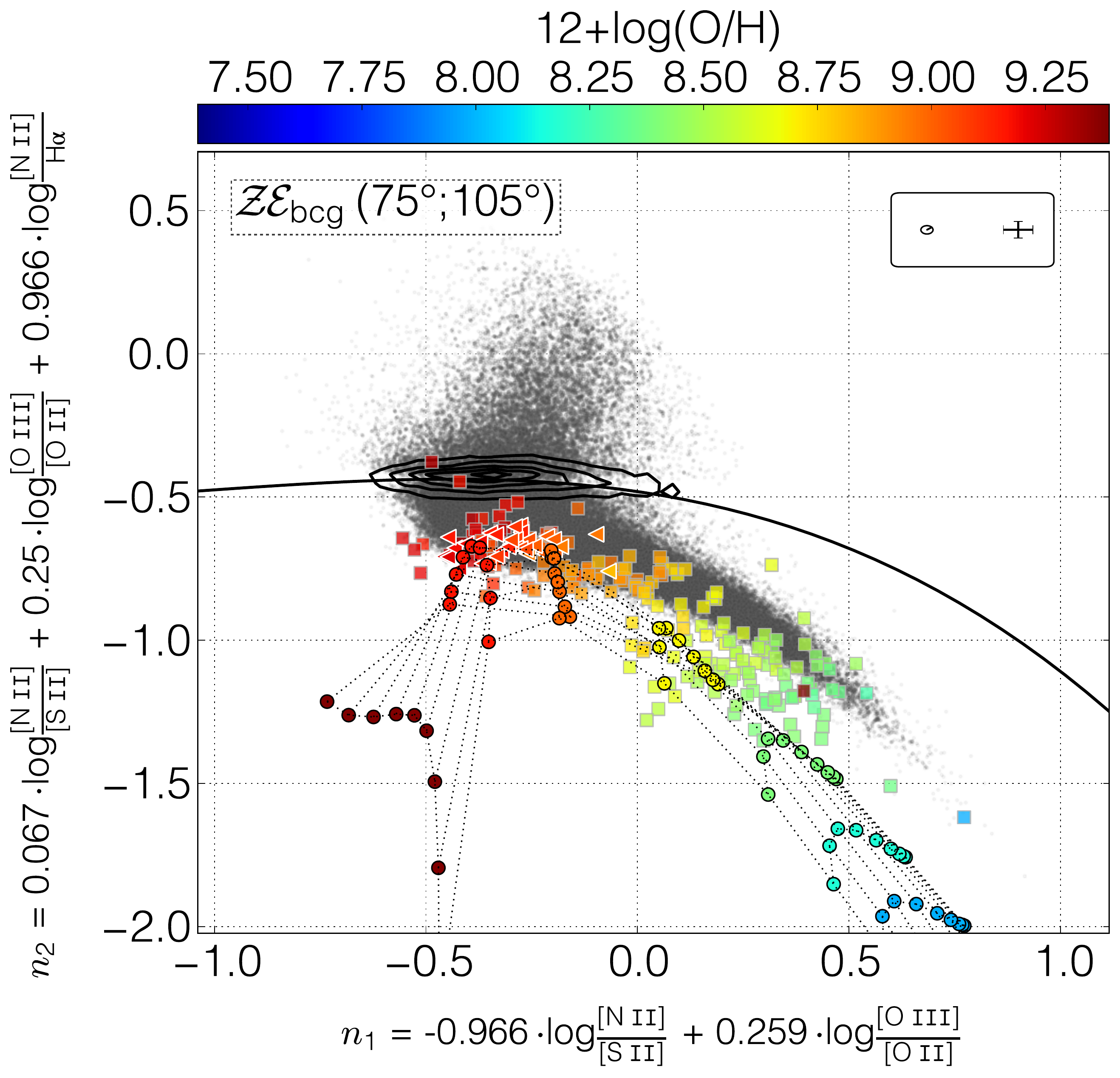}\qquad\qquad \includegraphics[scale=0.27]{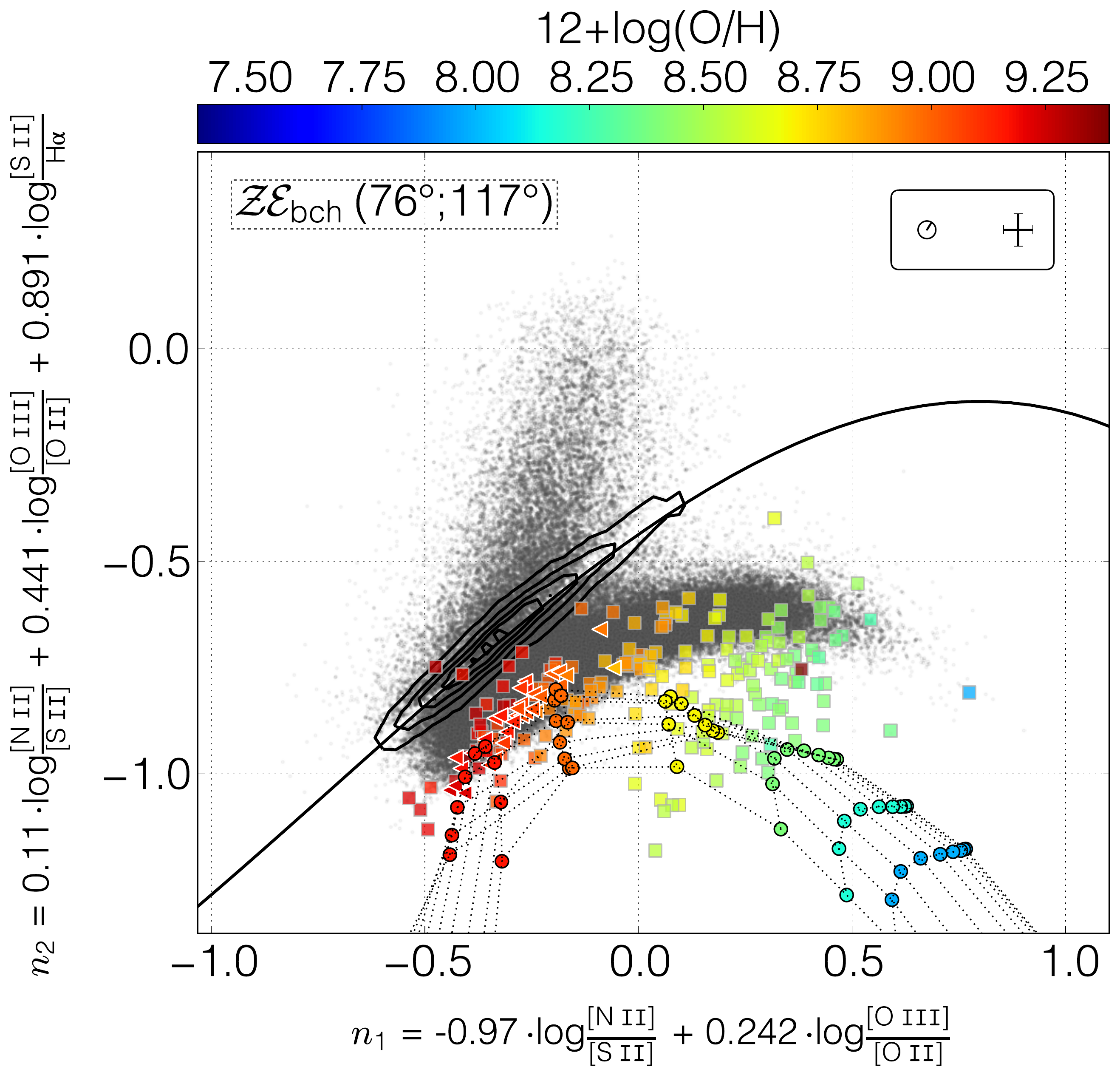}}
\caption{Same as Figure~\ref{fig:me_1}, for the other $\cal{ZE}$ diagnostics involving [O\,{\sc ii}].}\label{fig:me_2}
\end{figure*}

\begin{figure*}[htb!]
\centerline{ \includegraphics[scale=0.27]{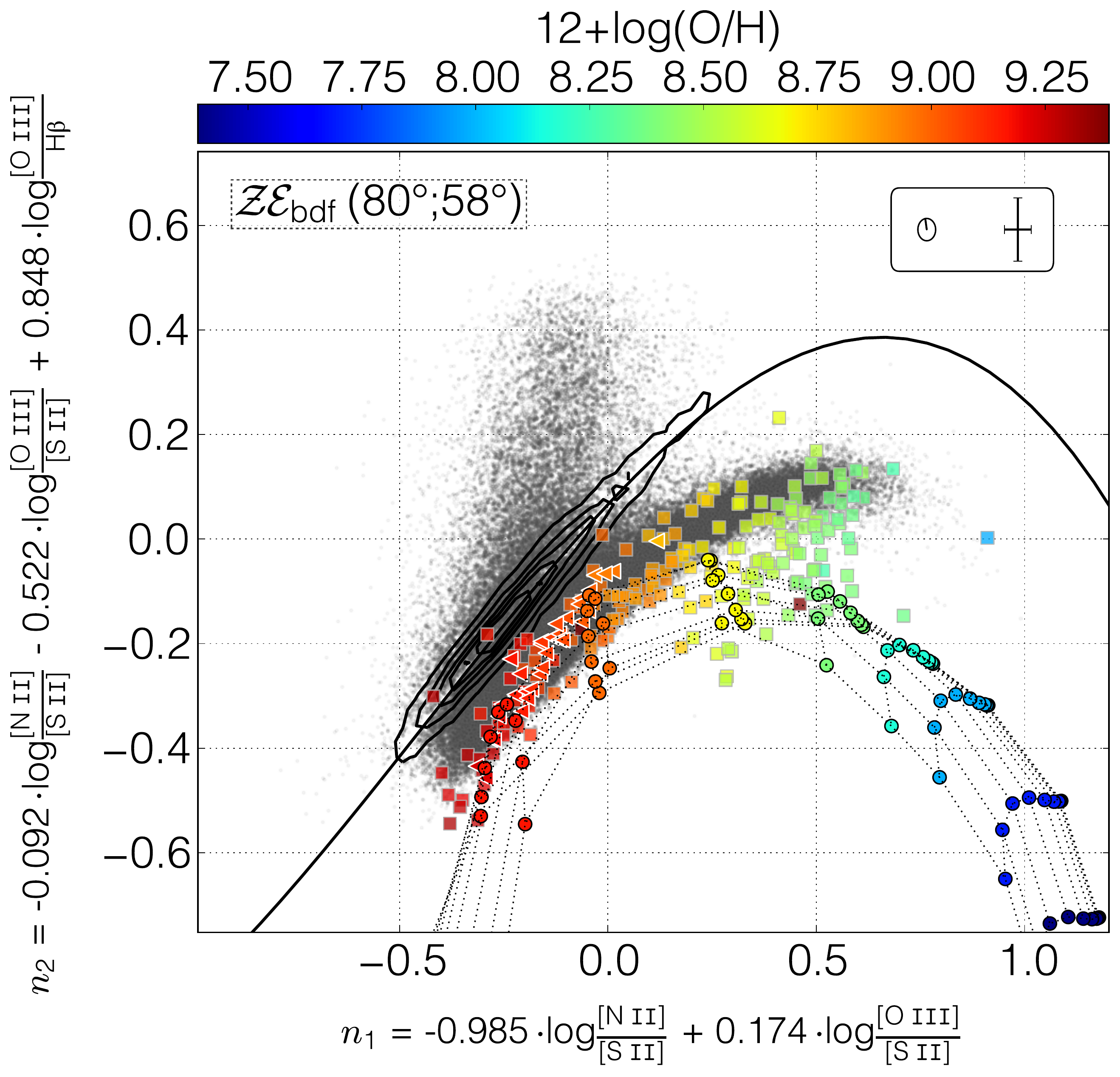}\qquad\qquad \includegraphics[scale=0.27]{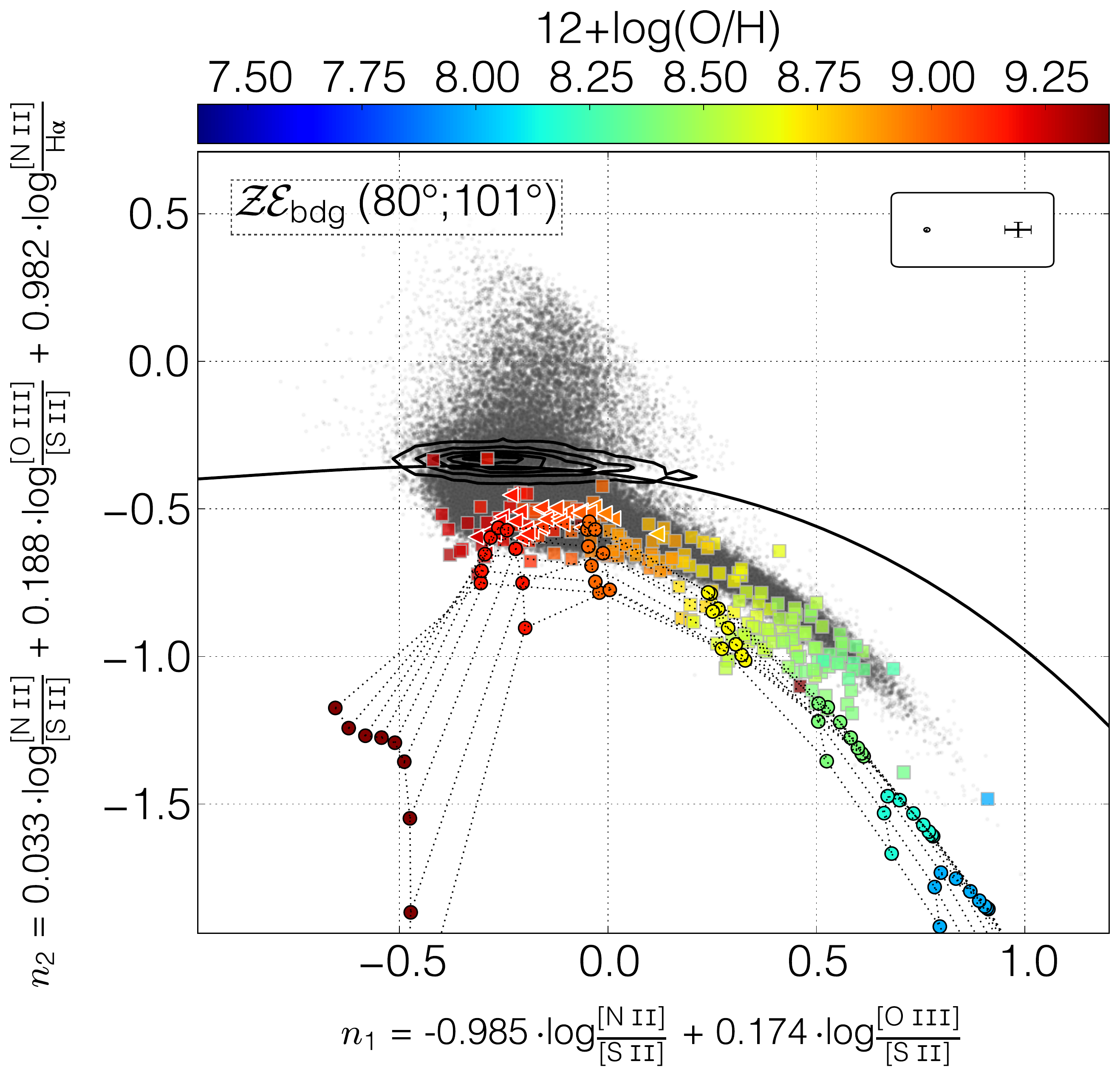}}\centerline{  \includegraphics[scale=0.27]{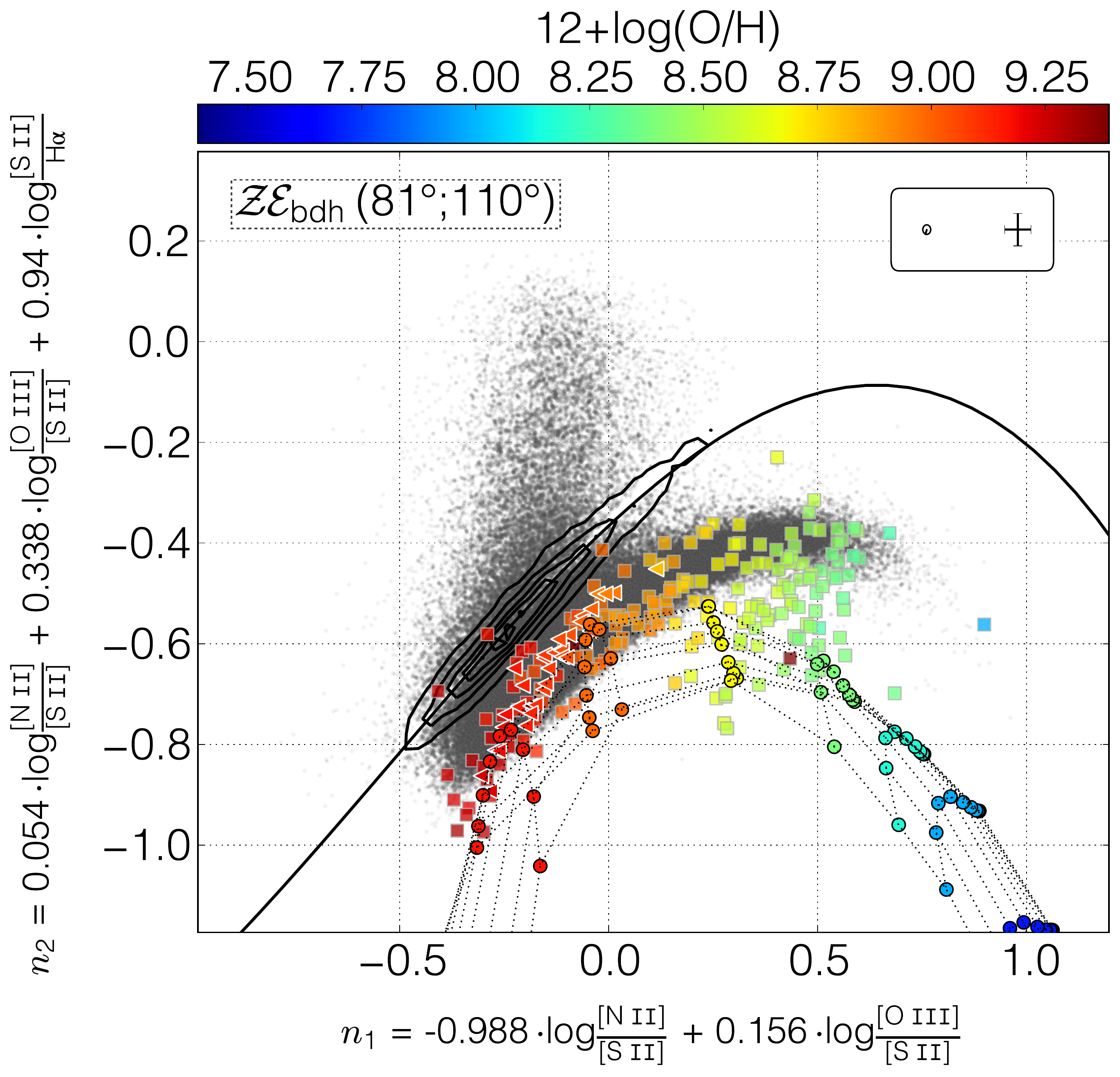}\qquad\qquad \includegraphics[scale=0.27]{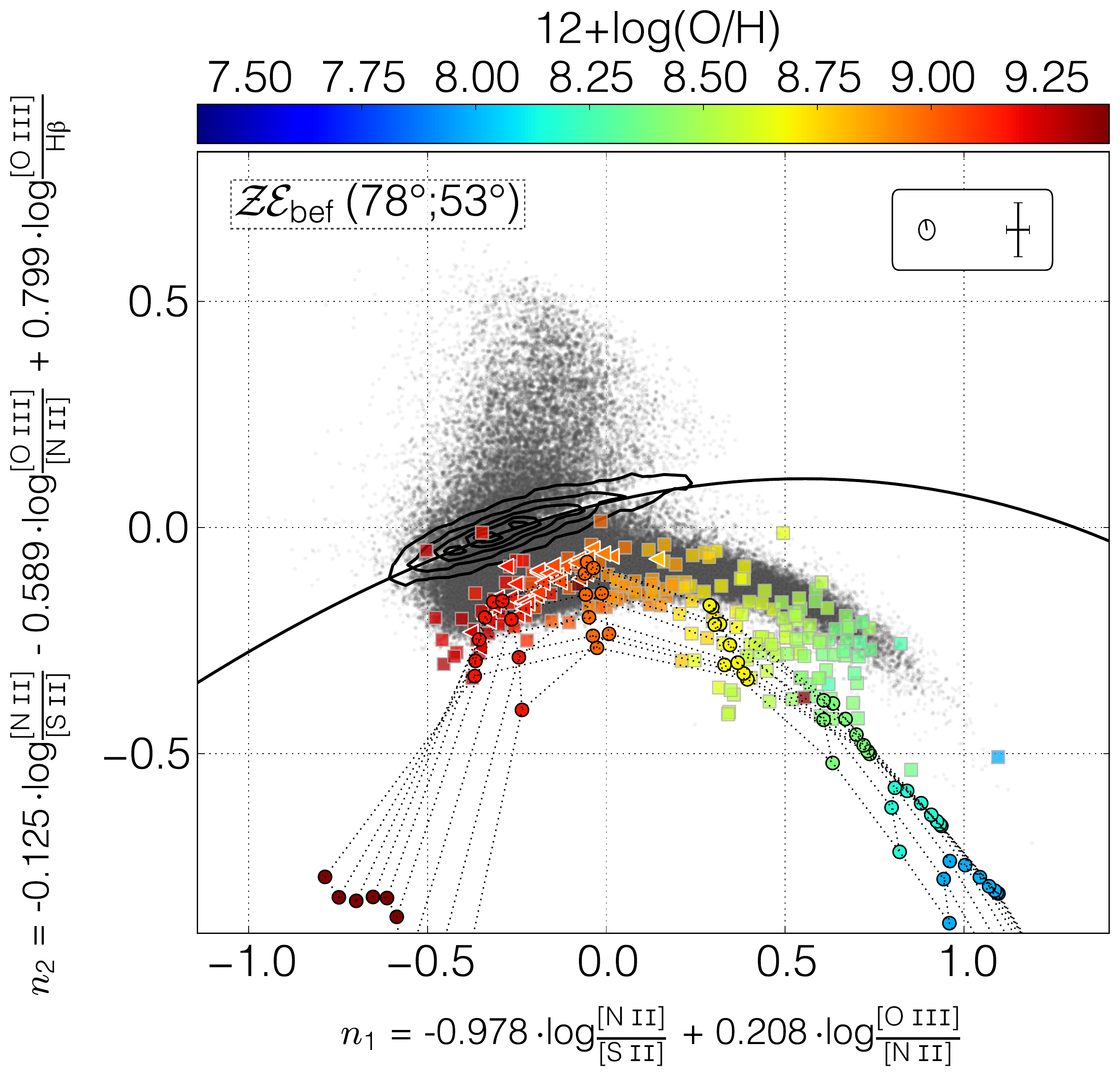}}
\centerline{  \includegraphics[scale=0.27]{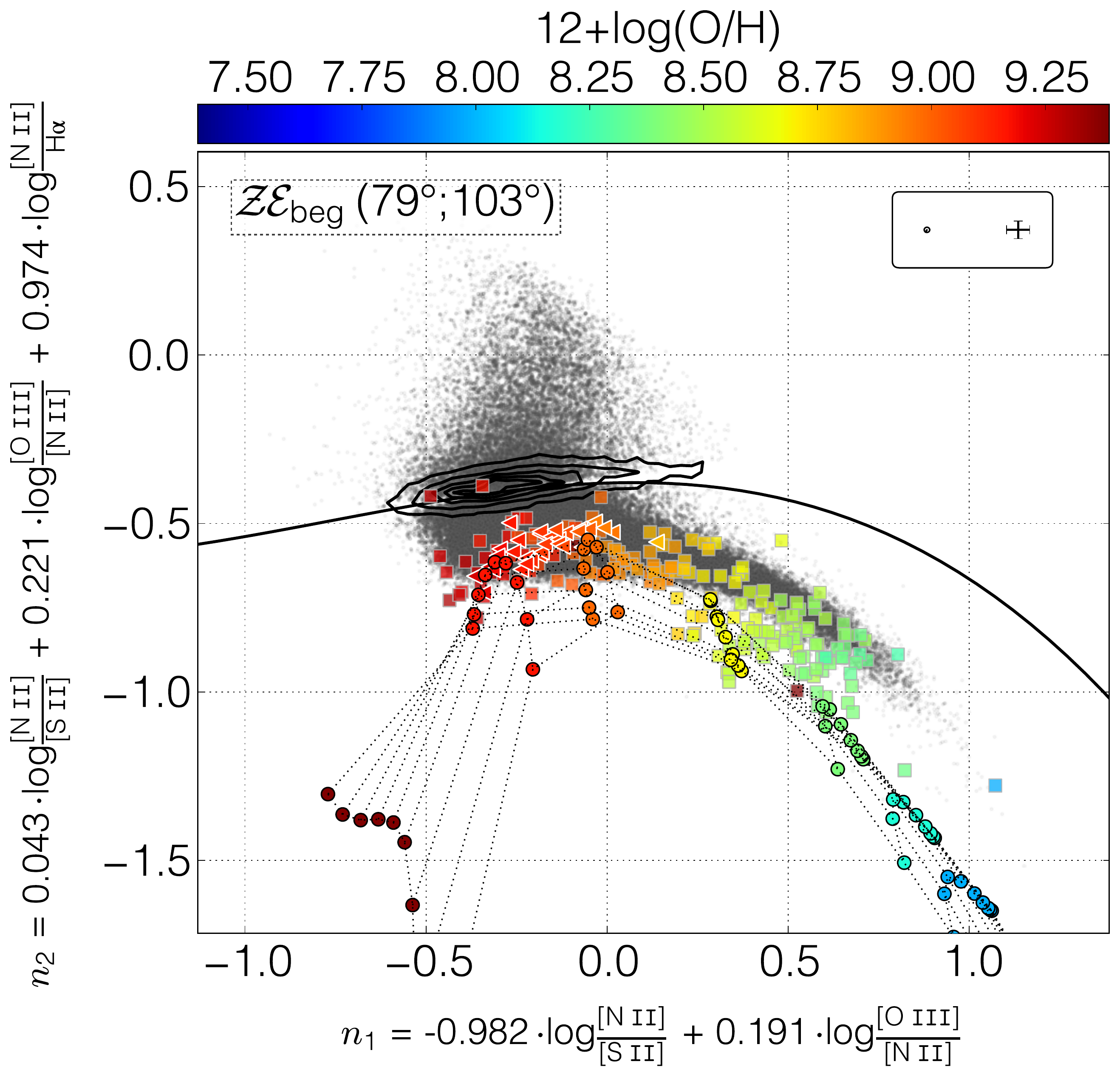}\qquad\qquad \includegraphics[scale=0.27]{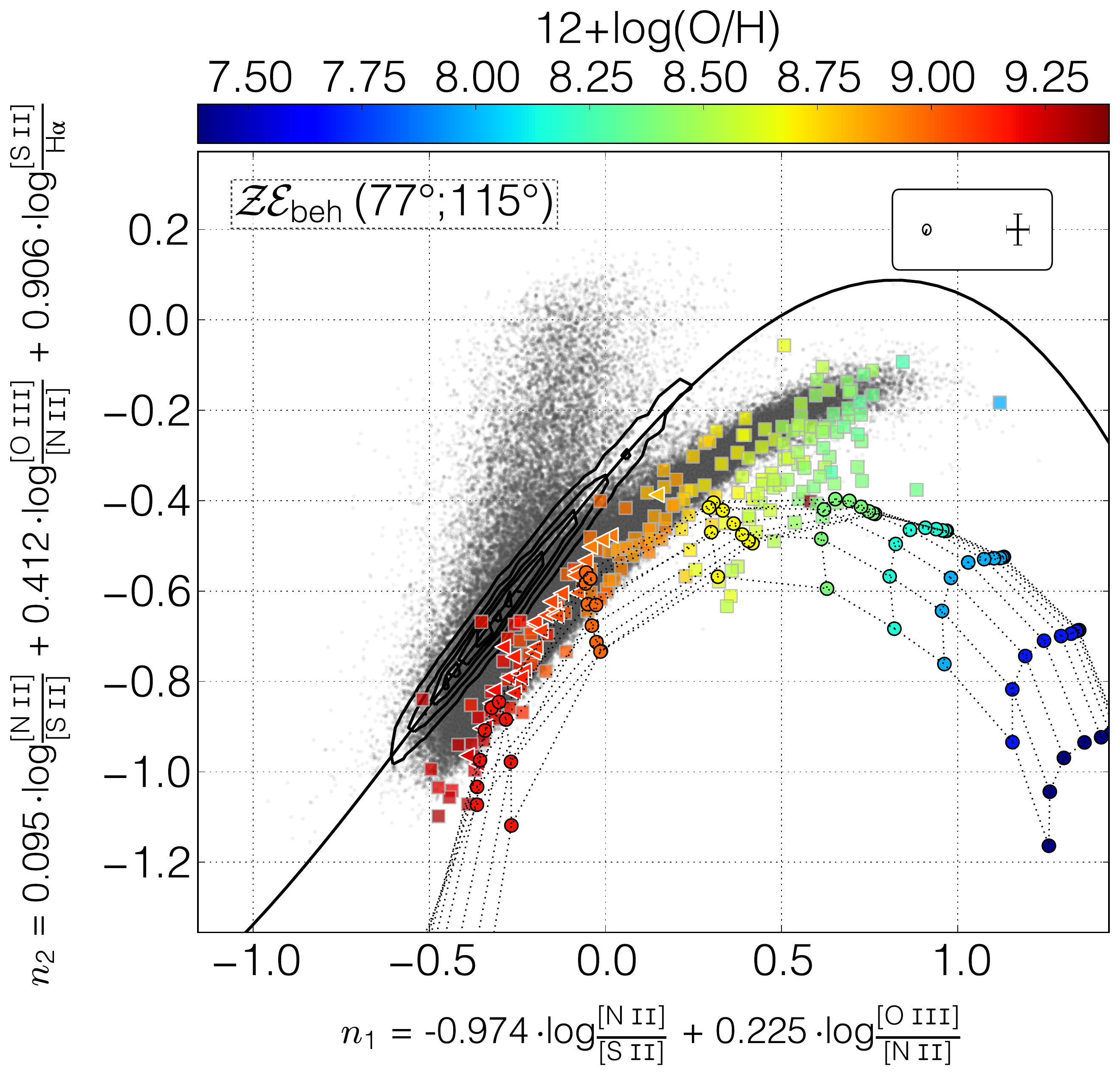}}
\caption{Same as Figure~\ref{fig:me_1}, but for the $\cal{ZE}$ diagnostics not involving [O\,{\sc ii}].}\label{fig:me_3}
\end{figure*}

We define ($\theta^*;\phi^*$) the specific values of $\phi$ and $\theta$ which define the thirteen $\cal{ZE}$ diagrams suitable to classify galaxies as \ion{H}{2}-like or AGN-like. Each $\cal{ZE}$$_{\text{x}_1\text{x}_2\text{x}_3}(\phi^*,\theta^*)$ diagram is shown in Figure~\ref{fig:me_1} , \ref{fig:me_2} and \ref{fig:me_3}. The corresponding $\cal{ZE}$$_{\text{x}_1\text{x}_2\text{x}_3}(\phi^*;\theta^*)$ denomination is shown in the top left corner of each diagram. For clarity, the $x$ and $y$ axes are labelled with the complete $n_1$ and $n_2$ composite line ratio equations, derived from Eqs.~\ref{eq:n1} and \ref{eq:n2}. All the parameters of the thirteen $\cal{ZE}$ diagrams are also summarised in Table~\ref{table:all_me}. In each diagram, we show to the top right the median error associated with the SDSS data points, given the mix of line ratios involved. \cite{Juneau14} observed (in a set of duplicate observations extracted from SDSS DR7) that the error associated with line ratios are comparatively smaller than those associated with individual line fluxes. Hence, our median errors (computed from the individual line errors) reported in the different panels of Figure~\ref{fig:me_1}, \ref{fig:me_2} and \ref{fig:me_3} can be regarded as upper bounds on the \emph{real} errors of the composite line ratios. These median errors can be compared to the theoretical displacement that $R_{\alpha\beta}$=3.1 (instead of 2.86) would imprint on the data. The circle-and-bar traces the intensity and direction (from the circle center outwards) of the total $\zeta$ spatial shift (see Eq.~\ref{eq:zeta}). The $\zeta$ shift is always similar to or smaller than the median measurement errors, and largely influenced by the $\log${\NIIOII} ratio.

The values of $\phi^*$ and $\theta^*$ have been found by interactive inspection of the $\cal{ZQE}$ diagrams\footnote{The capability to handle 3D models and structures interactively is an intrinsic characteristic of \texttt{Mayavi2}.}. It should be noted here that in all cases, $\phi^*$ and $\theta^*$ are not tightly constrained. Typically, a variation of $\pm$2 degrees will not significantly modify the general appearance of the projection, so that we restricted our choice to integer values of $\phi^*$ and $\theta^*$. We find that the theoretical grids created with MAPPINGS IV have a slightly different curvature depending on the chosen value of $\kappa$; for most $\cal{ZE}$ diagrams shown in Figure~\ref{fig:me_1}, \ref{fig:me_2} and \ref{fig:me_3}, a different value of $\kappa$ could influence the choice of the angles $\phi^*$ and $\theta^*$ by $\pm$2 degrees. The values quoted in Table~\ref{table:all_me} are our favoured ones for $\kappa=20$. 

Our choices of $\phi^*$ and $\theta^*$ were guided jointly by the appearance of the theoretical models, individual \ion{H}{2} regions, and the SDSS starburst branch. Specifically, we first used the model grids to identify a ``first-order'' point of view from which the grids collapse onto themselves. We then turned our attention to the cloud of SDSS galaxies, and specifically to the starburst branch, to fine-tune the final choice of $\phi^*$ and $\theta^*$ so that the observational data appears at its thinnest. For all the $\cal{ZE}$ diagrams but two, the \emph{MAPPINGS IV} simulation grid (marked by filled circles coloured as a function of the corresponding oxygen abundance of the model) is narrow and degenerate, mostly in the $q$ direction. Hence, the $x$ axis of these $\cal{ZE}$ diagrams can be associated with a metallicity ($\cal{Z}$) direction. By contrast, most of the differentiation between starburst-like and AGN-like objects is achieved in the $y$ direction, which can therefore be seen as the excitation or $\cal{E}$ direction, which is the basis of our chosen nomenclature. For the particular case of the $\cal{ZE}_\text{beh}$ and $\cal{ZE}_\text{bch}$ diagrams, a two dimensional twist inherent to the simulation grid makes it impossible to find a point-of-view from which the grid collapses for the entire metallicity range. In that case, we chose $\phi^*$ and $\theta^*$ so that the \ion{H}{2} space is most degenerate in the area of largest confusion between \ion{H}{2}-like and AGN-like objects. 

Identifying a specific viewpoint on the 3D distribution of SDSS observational data points is somewhat reminiscent of the notion of the Fundamental Plane (FP) for early-type galaxies \citep[][]{Dressler87,Djorgovski87}. In that situation, the identification of the parameters of the best-fit FP is often performed automatically, for example by computing the direction of smallest scatter in the data \citep[e.g.][]{Jorgensen96}, or with similar but more sophisticated approaches \citep[e.g.][]{Bernardi03, Saulder13}. While it is in principle not impossible to perform an analytical identification of $\phi^*$ and $\theta^*$, it is in practice less straightforward than our adopted manual solution. First, the structure of the 3D distribution of SDSS galaxies in the $\cal{ZQE}$ diagrams is significantly more complex than that of a plane. Second, the dataset contains both \ion{H}{2}-like and AGN-like objects, but in the present case one only is interested in collapsing the starburst branch onto itself - not the entire cloud of data points. If it is possible to identify and track the location of the starburst branch ``by eye'', it is significantly more complex to do so analytically and without any prior knowledge of the classification of the different objects. 

As we have found manually, any choice of $\phi^*$ and $\theta^*$ is not tightly constrained, and could vary by $\pm2$ degrees without significantly affecting the structure of the $\cal{ZE}$ diagram. Under this circumstance and at this point in time, our manual identification of $\phi^*$ and $\theta^*$ appears as satisfactory and useful as any analytical approach. Especially, analytical determinations of $\phi^*$ and $\theta^*$ would still depend on the underlying dataset and the chosen methodology, and would therefore not be ``unique'' \citep[as is the case for the FP parameters, see][]{Bernardi03}. The implementation of an automated routine to identify $\phi^*$ and $\theta^*$ is outside the scope of this paper, but ought to be explored in the future as the quality of the observational data points and theoretical datasets improves further. For example, a spaxel-based analysis relying on ongoing or upcoming IFU surveys such as Califa \citep{Sanchez12}, SAMI \citep{Croom12} or MANGA could better differentiate between the core and the outskirts of galaxies, and possibly reduce the inherent confusion at the interface between star-formation dominated and AGN-dominated objects \citep{Maragkoudakis14, Davies14}.

Principal Component Analysis (PCA) is a statistical technique which can identify \emph{directions of interest} in multi-dimensional datasets by calculating the successive normal directions of maximum variance \citep[see e.g.][for a brief introduction]{Francis99}. When performing a PCA analysis, the main challenge resides in interpreting these directions of interest, and connecting them to the physical world. The approach we adopt for creating the $\cal{ZE}$ diagrams (and associated diagnostics) follows the opposite path. Here, we use direct physical insight to separate line ratios into three complementary categories, and only then, once we have constructed the corresponding $\cal{ZQE}$ space, inspect it interactively to find point-of-views of interest. The interactive aspect of our approach is especially useful in allowing us to compare at the same time the grids of theoretical models, the individual measurements of \ion{H}{2} regions, and SDSS galaxies. Of course, the prime advantage of PCA is that it is not restricted to three-dimensional spaces. That is, a PCA analysis could be applied to the entire multi-dimensional line ratio space of galaxies, unlike the $\cal{ZQE}$ diagrams approach, which for obvious reasons cannot probe beyond three dimensions. Hence, while a detailed comparison is outside the scope of this article, the PCA approach and the $\cal{ZQE}$ approach are (conceptually) very complimentary.

In any situation, to avoid misunderstandings and ensure repeatability, we strongly advise any use of the $\cal{ZE}$ diagrams to clearly state the values of $\phi$ and $\theta$ employed, along with the line ratios involved, which are required to uniquely define any $\cal{ZE}$ diagram (see Section~\ref{sec:3d-2d_def} and Eq.~\ref{eq:def_ze}). 

\begin{table*}[htb!]
\caption{$\cal{ZE}$ diagrams, and their associated parameters.}\label{table:all_me}
\center
\vspace{-10pt}
\begin{tabular}{c c c c c}
\hline\hline
Name & $\phi^*$ & $\theta^*$ & $n_1$ & $n_2$ \\
\hline \\ [-1.5ex]
$\cal{ZE}_\text{acf}$ & 96 & 50 &
	$-0.995\cdot\log\frac{[\text{N\,{\sc ii}}]}{[\text{O\,{\sc ii}]}}-0.105\cdot\log\frac{[\text{O\,{\sc iii}}]}{[\text{O\,{\sc ii}}]}$&
	$+0.067\cdot\log\frac{[\text{N\,{\sc ii}}]}{[\text{O\,{\sc ii}]}}-0.639\cdot\log\frac{[\text{O\,{\sc iii}}]}{[\text{O\,{\sc ii}}]}+0.766\cdot\log\frac{[\text{O\,{\sc iii}}]}{[\text{H}\beta]}$  \\ [2ex]
$\cal{ZE}_\text{acg}$ & 96 & 104 &
	$-0.995\cdot\log\frac{[\text{N\,{\sc ii}}]}{[\text{O\,{\sc ii}]}}-0.105\cdot\log\frac{[\text{O\,{\sc iii}}]}{[\text{O\,{\sc ii}}]}$&
	$-0.025\cdot\log\frac{[\text{N\,{\sc ii}}]}{[\text{O\,{\sc ii}]}}+0.241\cdot\log\frac{[\text{O\,{\sc iii}}]}{[\text{O\,{\sc ii}}]}+0.970\cdot\log\frac{[\text{N\,{\sc ii}}]}{[\text{H}\alpha]}$ \\ [2ex]
$\cal{ZE}_\text{adg}$ & 95 & 102 &
	$-0.996\cdot\log\frac{[\text{N\,{\sc ii}}]}{[\text{O\,{\sc ii}]}}-0.087\cdot\log\frac{[\text{O\,{\sc iii}}]}{[\text{O\,{\sc ii}}]}$&
	$-0.018\cdot\log\frac{[\text{N\,{\sc ii}}]}{[\text{O\,{\sc ii}]}}+0.207\cdot\log\frac{[\text{O\,{\sc iii}}]}{[\text{O\,{\sc ii}}]}+0.978\cdot\log\frac{[\text{N\,{\sc ii}}]}{[\text{H}\alpha]}$ \\ [2ex]
$\cal{ZE}_\text{aef}$ & 95 & 53 &
	$-0.996\cdot\log\frac{[\text{N\,{\sc ii}}]}{[\text{O\,{\sc ii}]}}-0.087\cdot\log\frac{[\text{O\,{\sc iii}}]}{[\text{N\,{\sc ii}}]}$&
	$+0.052\cdot\log\frac{[\text{N\,{\sc ii}}]}{[\text{O\,{\sc ii}]}}-0.600\cdot\log\frac{[\text{O\,{\sc iii}}]}{[\text{N\,{\sc ii}}]}+0.799\cdot\log\frac{[\text{O\,{\sc iii}}]}{[\text{H}\beta]}$ \\ [2ex]
$\cal{ZE}_\text{aeg}$ & 95 & 104 &
	$-0.996\cdot\log\frac{[\text{N\,{\sc ii}}]}{[\text{O\,{\sc ii}]}}-0.087\cdot\log\frac{[\text{O\,{\sc iii}}]}{[\text{N\,{\sc ii}}]}$&
	$-0.021\cdot\log\frac{[\text{N\,{\sc ii}}]}{[\text{O\,{\sc ii}]}}+0.241\cdot\log\frac{[\text{O\,{\sc iii}}]}{[\text{N\,{\sc ii}}]}+0.970\cdot\log\frac{[\text{N\,{\sc ii}}]}{[\text{H}\alpha]}$ \\ [2ex]
$\cal{ZE}_\text{bcg}$ & 75 & 105 &
	$-0.966\cdot\log\frac{[\text{N\,{\sc ii}}]}{[\text{S\,{\sc ii}]}}+0.259\cdot\log\frac{[\text{O\,{\sc iii}}]}{[\text{O\,{\sc ii}}]}$&
	$+0.067\cdot\log\frac{[\text{N\,{\sc ii}}]}{[\text{S\,{\sc ii}]}}+0.250\cdot\log\frac{[\text{O\,{\sc iii}}]}{[\text{O\,{\sc ii}}]}+0.966\cdot\log\frac{[\text{N\,{\sc ii}}]}{[\text{H}\alpha]}$ \\ [2ex]
$\cal{ZE}_\text{bch}$ & 76 & 117 &
	$-0.970\cdot\log\frac{[\text{N\,{\sc ii}}]}{[\text{S\,{\sc ii}]}}+0.242\cdot\log\frac{[\text{O\,{\sc iii}}]}{[\text{O\,{\sc ii}}]}$&
	$+0.110\cdot\log\frac{[\text{N\,{\sc ii}}]}{[\text{S\,{\sc ii}]}}+0.441\cdot\log\frac{[\text{O\,{\sc iii}}]}{[\text{O\,{\sc ii}}]}+0.891\cdot\log\frac{[\text{N\,{\sc ii}}]}{[\text{H}\alpha]}$ \\ [2ex]
$\cal{ZE}_\text{bdf}$ & 80 & 58 &
	$-0.985\cdot\log\frac{[\text{N\,{\sc ii}}]}{[\text{S\,{\sc ii}]}}+0.174\cdot\log\frac{[\text{O\,{\sc iii}}]}{[\text{S\,{\sc ii}}]}$&
	$-0.092\cdot\log\frac{[\text{N\,{\sc ii}}]}{[\text{S\,{\sc ii}]}}-0.522\cdot\log\frac{[\text{O\,{\sc iii}}]}{[\text{S\,{\sc ii}}]}+0.848\cdot\log\frac{[\text{O\,{\sc iii}}]}{[\text{H}\beta]}$\\ [2ex]
$\cal{ZE}_\text{bdg}$ & 80 & 101&
	$-0.985\cdot\log\frac{[\text{N\,{\sc ii}}]}{[\text{S\,{\sc ii}]}}+0.174\cdot\log\frac{[\text{O\,{\sc iii}}]}{[\text{S\,{\sc ii}}]}$&
	$+0.033\cdot\log\frac{[\text{N\,{\sc ii}}]}{[\text{S\,{\sc ii}]}}+0.188\cdot\log\frac{[\text{O\,{\sc iii}}]}{[\text{S\,{\sc ii}}]}+0.982\cdot\log\frac{[\text{N\,{\sc ii}}]}{[\text{H}\alpha]}$ \\ [2ex]
$\cal{ZE}_\text{bdh}$ & 81 & 110 &
	$-0.988\cdot\log\frac{[\text{N\,{\sc ii}}]}{[\text{S\,{\sc ii}]}}+0.156\cdot\log\frac{[\text{O\,{\sc iii}}]}{[\text{S\,{\sc ii}}]}$&
	$+0.054\cdot\log\frac{[\text{N\,{\sc ii}}]}{[\text{S\,{\sc ii}]}}+0.338\cdot\log\frac{[\text{O\,{\sc iii}}]}{[\text{S\,{\sc ii}}]}+0.940\cdot\log\frac{[\text{S\,{\sc ii}}]}{[\text{H}\alpha]}$ \\ [2ex]
$\cal{ZE}_\text{bef}$ & 78 & 53 &
	$-0.978\cdot\log\frac{[\text{N\,{\sc ii}}]}{[\text{S\,{\sc ii}]}}+0.208\cdot\log\frac{[\text{O\,{\sc iii}}]}{[\text{N\,{\sc ii}}]}$&
	$-0.125\cdot\log\frac{[\text{N\,{\sc ii}}]}{[\text{S\,{\sc ii}]}}-0.589\cdot\log\frac{[\text{O\,{\sc iii}}]}{[\text{N\,{\sc ii}}]}+0.799\cdot\log\frac{[\text{O\,{\sc iii}}]}{[\text{H}\beta]}$\\ [2ex]
$\cal{ZE}_\text{beg}$ & 79 & 103 &
	$-0.982\cdot\log\frac{[\text{N\,{\sc ii}}]}{[\text{S\,{\sc ii}]}}+0.191\cdot\log\frac{[\text{O\,{\sc iii}}]}{[\text{N\,{\sc ii}}]}$&
	$+0.043\cdot\log\frac{[\text{N\,{\sc ii}}]}{[\text{S\,{\sc ii}]}}+0.221\cdot\log\frac{[\text{O\,{\sc iii}}]}{[\text{N\,{\sc ii}}]}+0.974\cdot\log\frac{[\text{N\,{\sc ii}}]}{[\text{H}\alpha]}$ \\ [2ex]
$\cal{ZE}_\text{beh}$ & 77 & 115 &
	$-0.974\cdot\log\frac{[\text{N\,{\sc ii}}]}{[\text{S\,{\sc ii}]}}+0.225\cdot\log\frac{[\text{O\,{\sc iii}}]}{[\text{N\,{\sc ii}}]}$&
	$+0.095\cdot\log\frac{[\text{N\,{\sc ii}}]}{[\text{S\,{\sc ii}]}}+0.412\cdot\log\frac{[\text{O\,{\sc iii}}]}{[\text{N\,{\sc ii}}]}+0.906\cdot\log\frac{[\text{S\,{\sc ii}}]}{[\text{H}\alpha]}$\\ [2ex]
\hline
\end{tabular}
\end{table*}

\subsubsection{Defining the diagnostic lines}\label{sec:def_line}

For each $\cal{ZE}$$_{\text{x}_1\text{x}_2\text{x}_3}(\phi^*;\theta^*)$ diagram illustrated in Figure~\ref{fig:me_1}, \ref{fig:me_2} and \ref{fig:me_3}, we define a 3rd order polynomial that separates the \ion{H}{2}-like objects (below the line) and the AGN-like objects (above the line). The semi-empirical polynomial coefficients $\alpha,\beta,\gamma$ and $\delta$ are summarised in Table~\ref{table:all_params}, where the line equation is defined by
\begin{equation}
y=f(n_1)=\alpha (n_1)^3+\beta (n_1)^2+\gamma (n_1) +\delta.
\end{equation}
This approach is similar to that used by \cite{Kewley01b} and \cite{Kauffmann03b} to define diagnostic lines for the classical optical line ratio diagnostic diagrams, although the chosen functional forms are different. 

The theoretical grids do not match the envelope of the observations of \ion{H}{2} regions perfectly (see Figures~\ref{fig:me_1}, \ref{fig:me_2} and \ref{fig:me_3}). This is especially true for $\cal{ZE}$ diagrams involving the {\SII} lines, which as noted by \cite{Dopita13a} appear to be 0.1 dex too weak in the models. There exist several possible origins for the theoretical mismatch. At the low abundance end in particular, some \ion{H}{2} regions may possibly have a higher electron density than expected (up to n$_e\backsimeq100$ cm$^3$) (Nicholls et al., in preparation). At the high-abundance end, all lines become very sensitive to the electron temperature, which varies very rapidly through the models. Thus, small changes in the geometry of the ionised gas (assumed to be spherically symmetric in the models) can make large differences in the predicted emission line spectrum. The underlying stellar synthesis models may also be largely responsible for the offset between the theoretical grids and the SDSS galaxies (especially for line ratios involving the [\ion{S}{2}] lines) if these theoretical models do not produce enough far-UV ionizing photons, as suggested by \cite{Kewley01b} and \cite{Levesque10}. Lastly, we note that the spacing between the two highest abundance set of simulations are $\sim$2-3 times larger than the spacing between the other abundance sets. As a result, linearly interpolating (as traced by the dotted lines in Figures~\ref{fig:me_1}, \ref{fig:me_2} and \ref{fig:me_3}) can be a poorer estimation and result in a larger mismatch between the theoretical grid and the observations.

Given the mismatch between the shape of the model grids and the observational data points in some of the $\cal{ZE}$ diagrams, we use the theoretical models as a general guide, but choose the final coefficients $\alpha$, $\beta$, $\gamma$ and $\delta$ so that the diagnostic lines trace the full extent of the starburst sequence of the SDSS galaxies in all cases. Hence, keeping in mind that we indirectly rely on the theoretical models in the manual determination procedure for the values of $\phi^*$ and $\theta^*$ (see Section~\ref{sec:star}), the different $\cal{ZE}$ diagnostics do not depend explicitly on the \emph{MAPPINGS IV} grids.

In practice, the diagnostic line coefficients are identified as follows. We first choose manually a series of five-to-seven positions in the $\cal{ZE}$ diagram, spaced by  0.2-0.5 dex along the x-direction, defining a first-order separation between \ion{H}{2}-like and AGN-like objects. We subsequently obtain the corresponding polynomial coefficients by performing a least-square minimisation of a 3rd order polynomial to these data-points using the \texttt{Python} implementation of the IDL\footnote{Interactive Data Language} non-linear least-square minimization routine \texttt{mpfit} \citep[][]{Markwardt09}. Since we set these diagnostic lines manually and independently for each diagram, using the theoretical grid for guidance only, and given that each diagnostic is subject to both observational errors and theoretical uncertainties, it is possible that an SDSS galaxy classified as \ion{H}{2}-like by one diagnostic will be classified as AGN-like by others. However, because we now have thirteen diagnostics, we can combine them to ensure consistency and reduce the classification uncertainty. 

To that end, we separate all SDSS galaxies into three groups:
\begin{itemize}
\item {\bf \ion{H}{2}-like}: galaxies classified as \ion{H}{2}-like by all thirteen $\cal{ZE}$ diagnostics,
\item {\bf AGN-like}: galaxies classified as AGN-like by all thirteen $\cal{ZE}$ diagnostics, and
\item {\bf uncertain}: galaxies for which the thirteen $\cal{ZE}$ diagnostics are inconsistent.
\end{itemize}

In Figure~\ref{fig:me_1}, \ref{fig:me_2} and \ref{fig:me_3}, density contours delineate the location of SDSS galaxies having an uncertain classification. The contours have been obtained by distributing all the galaxies with uncertain classification in a regular grid with resolution of 0.03 dex, with the subsequent smoothing of the grid with a symmetric gaussian filter of 0.15 dex in radius (5 grid elements). As can be expected, the uncertain galaxies are clustered around each of the diagnostic lines, with the 20\% contour within $\pm$0.1 dex of the diagnostic line. Using a manual and iterative approach, we have adapted the parameters of each of the diagnostic equations to minimise the number of uncertain galaxies. Following this approach, we reduced the number of galaxies with uncertain classification to 2636 (2.5\% of a total of 105070 objects). We have 88918 \ion{H}{2}-like galaxies  (84.6\%) and 13516 galaxies classified as AGN-like (12.9\%), as classified by the $\cal{ZE}$ diagnostics.

Improving the overall agreement between the different diagnostics required in some cases to alter the shape of the diagnostic lines, especially for high metallicities. Because the SDSS points are not distributed uniformly, very small modifications of the diagnostic line in denser regions can strongly influence the overall agreement of the different diagnostics. Since we rely on 3rd order polynomials, the inner-most regions of the diagnostic lines are very much influenced by the slope at higher (and lower) metallicities. In other words, the lack of observations make it impossible to tightly constrain the position of the diagnostic line  in the outer-most region of the different $\cal{ZQE}$ diagrams.

\begin{table}[htb!]
\caption{Starburst diagnostic line parameters for each $\cal{ZE}$ diagram.}\label{table:all_params}
\center
\vspace{-10pt}
\begin{tabular}{c c c c c c c c c}
\hline\hline
Name &$\alpha$ &$\beta$ &$\gamma$ & $\delta$\\
\hline \\ [-1.5ex]
$\cal{ZE}_\text{acf}$ &
	$-0.059$&$-0.024$&$+0.676$&$-0.005$ \\ [2ex]
$\cal{ZE}_\text{acg}$ &
	$+0.005$&$-0.124$&$+0.020$&$-0.445$ \\ [2ex]
$\cal{ZE}_\text{adg}$ &
	$-0.034$&$-0.071$&$+0.091$&$-0.382$ \\ [2ex]
$\cal{ZE}_\text{aef}$ &
	$-0.013$&$-0.082$&$+0.133$&$-0.008$ \\ [2ex]
$\cal{ZE}_\text{aeg}$ &
	$-0.032$&$-0.079$&$+0.268$&$-0.459$ \\ [2ex]
$\cal{ZE}_\text{bcg}$ &
	$-0.101$&$-0.311$&$-0.216$&$-0.481$ \\ [2ex]
$\cal{ZE}_\text{bch}$ &
	$-0.132$&$-0.280$&$+0.700$&$-0.437$ \\ [2ex]	
$\cal{ZE}_\text{bdf}$ &
	$-0.283$&$-0.368$&$+0.851$&$+0.066$\\ [2ex]
$\cal{ZE}_\text{bdg}$ &
	$-0.118$&$-0.307$&$-0.171$&$-0.382$\\ [2ex]
$\cal{ZE}_\text{bdh}$ &
	$-0.238$&$-0.374$&$+0.774$&$-0.366$\\ [2ex]
$\cal{ZE}_\text{bef}$ &
	$-0.013$&$-0.157$&$+0.186$&$+0.055$\\ [2ex]
$\cal{ZE}_\text{beg}$ &
	$-0.097$&$-0.222$&$+0.034$&$-0.380$\\ [2ex]
$\cal{ZE}_\text{beh}$ &
	$-0.221$&$-0.289$&$+0.920$&$-0.350$\\ [2ex]
\hline
\end{tabular}
\end{table}

We examine the consistency of each of these $\cal{ZE}$ diagnostics in more detail in the next subsection. This is not to be confused with the validity of the final classification itself, which we examine in Section~\ref{sec:comp_bpt}. In the Appendix, we show for completeness the $\cal{ZE}$ diagrams that best collapse the grid of photoionization models in the eleven $\cal{ZQE}$ spaces for which we did not derive any $\cal{ZE}$ diagnostic. These diagrams were not selected as reliable diagnostics because of the high confusion between the starburst and AGN branch of the SDSS galaxies. In Figures~\ref{fig:ZQE_Oi} and \ref{fig:ZQE_drop}, the confusion is emphasized by showing the density contours of galaxies with uncertain classification. Although the final $\cal{ZE}$ classification and hence the density contours of uncertain galaxies were derived ``after'' the visual selection of $\cal{ZQE}$ diagrams with a clean separation between the starburst and AGN branch of SDSS galaxies, these contours act as an \emph{a posteriori} confirmation of the initial selection. In every $\cal{ZE}$ diagram shown in Figure~\ref{fig:ZQE_Oi} and \ref{fig:ZQE_drop} the uncertain galaxies spread out over large areas ($>$0.2 dex), unlike in the $\cal{ZE}$ diagnostic diagrams listed in Table~\ref{table:all_me}.

\subsection{Consistency of the $\cal{ZE}$ diagnostics}\label{sec:consistency}

\begin{figure}[htb!]
\centerline{\includegraphics[scale=0.3]{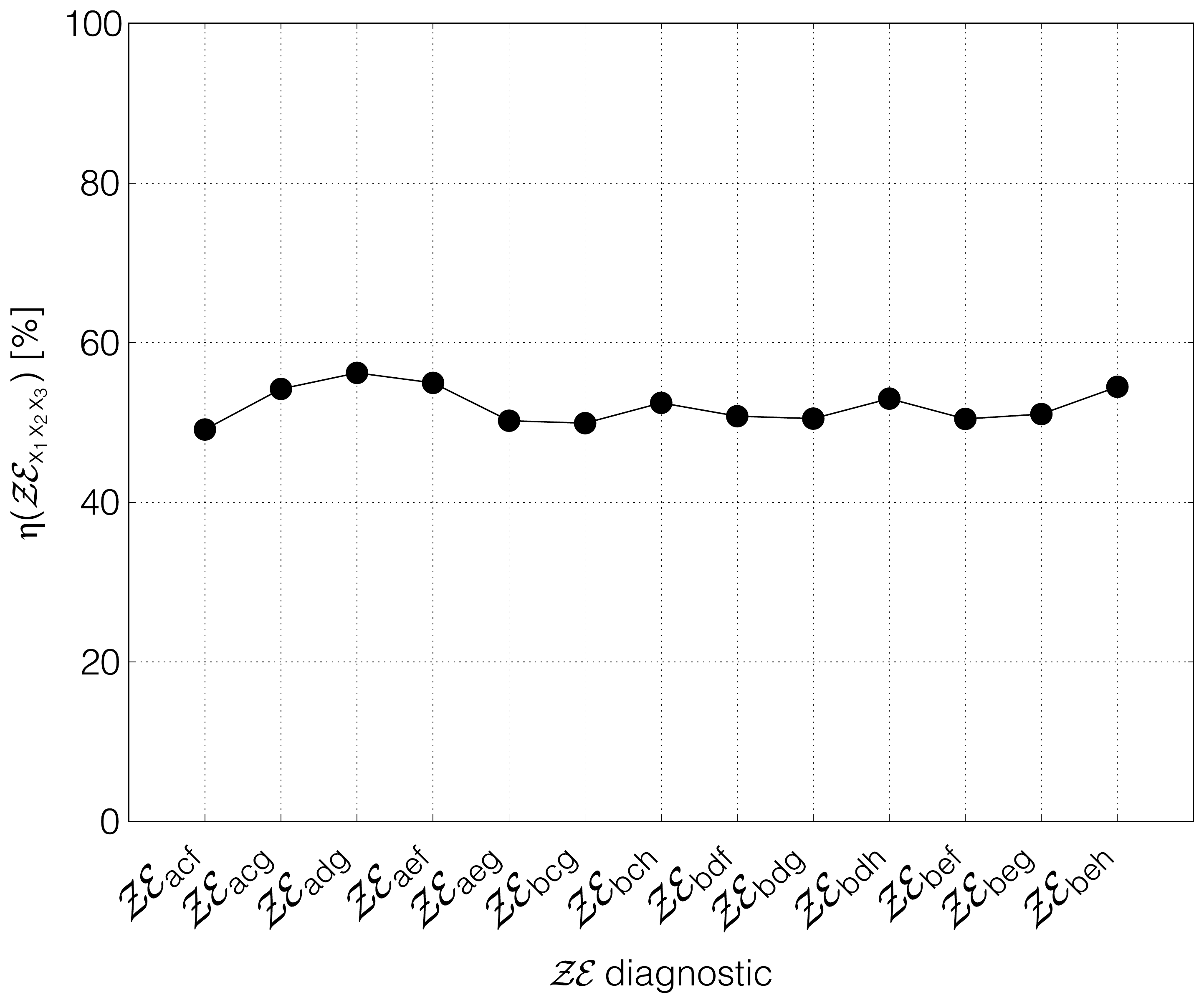}}
\caption{$\eta(\cal{ZE}$$_{\text{x}_1\text{x}_2\text{x}_3})$,  the percentage of uncertain galaxies classified as AGN-like by a particular $\cal{ZE}$$_{\text{x}_1\text{x}_2\text{x}_3}$ diagnostic with respect to the total number of galaxies with uncertain classification.}\label{fig:eta}
\end{figure}

The thirteen $\cal{ZE}$ diagnostics defined in Table~\ref{table:all_params} all rely on a subset of nine line ratios, so that they are not strictly independent from one another. To better understand this connection, we focus our attention on the 2636 (2.5\%) galaxies with an uncertain classification. We introduce the quantity $\eta(\cal{ZE}$$_{\text{x}_1\text{x}_2\text{x}_3})$ as the percentage of uncertain galaxies classified as AGN-like by a particular $\cal{ZE}$$_{\text{x}_1\text{x}_2\text{x}_3}$ diagnostic. The value of $\eta$ for the thirteen $\cal{ZE}$ diagnostics is shown in Figure~\ref{fig:eta}. A low value of $\eta$ indicates a diagnostic which is too lax and will classify most uncertain galaxies as H\,{\sc ii}-like. On the other hand, a high value of $\eta$ indicates a diagnostic which is too tight. In such a case, the majority of the uncertain galaxies are classified as being AGN-like by the diagnostic concerned. All thirteen $\cal{ZE}$ diagnostics have 40\%$<\eta(\cal{ZE}$$_{\text{x}_1\text{x}_2\text{x}_3})<$60\%, an indication that \emph{overall}, they are all consistent with each other (as expected from our iterative procedure to define the $\alpha$, $\beta$, $\gamma$ and $\delta$ coefficients, see Section~\ref{sec:def_line}). The underlying distribution of SDSS galaxies is not uniform, and small ``local'' displacements of the diagnostic lines in the $\cal{ZE}$ diagram can easily give rise to the range of $\eta$ values shown in Figure~\ref{fig:eta}.

\begin{figure}[htb!]
\centerline{\includegraphics[scale=0.3]{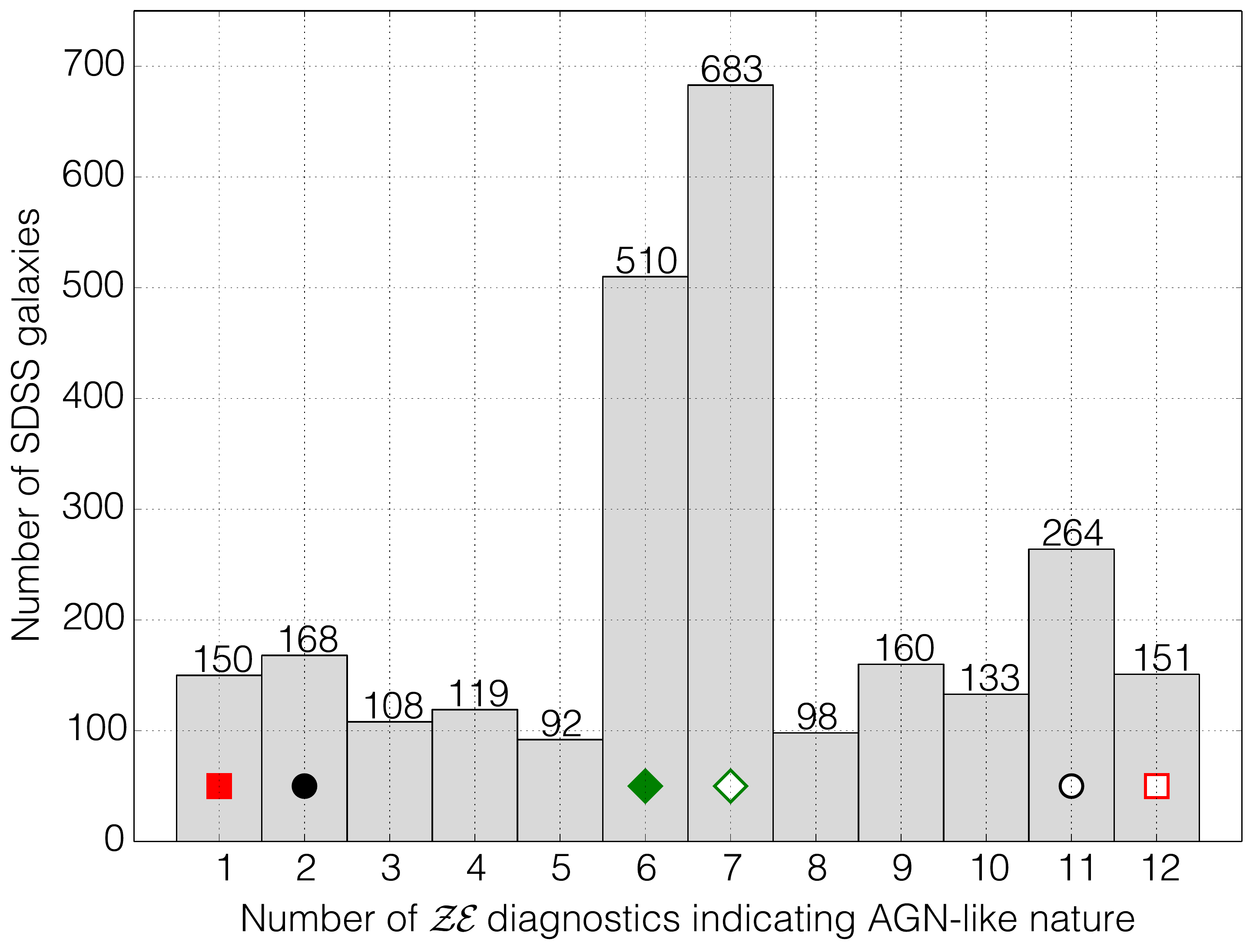}}
\caption{The distribution of the number of uncertain galaxies as a function of the number of $\cal{ZE}$ diagnostics indicating that they are AGN-like. The symbols in the bins '1', '2', '6', '7', '11', and '12' correspond to the different lines in Figure~\ref{fig:min-max} and are shown here for completeness.}\label{fig:hist}
\end{figure}

A complimentary way to look at the SDSS galaxies with uncertain classification is illustrated in Figure~\ref{fig:hist}, where we show the distribution of the number of uncertain galaxies as a function of how many $\cal{ZE}$ diagnostics indicate that they are AGN-like. For example, the bin '1' corresponds to galaxies classified as \ion{H}{2}-like by all {ZE} diagnostics but one, and the bin '12' corresponds to galaxies classified as AGN-like by all $\cal{ZE}$ diagnostics but one. The distribution harbors a sharp peak at $\sim$6.5 and a secondary, minor peak at 11, in addition to a base level of $\sim$130 objects per bin. The base level of uncertain galaxies in all bins is most certainly a consequences of small mismatch between the different diagnostics, as well as observational errors of certain line ratios. 

\begin{figure}[htb!]
\centerline{\includegraphics[scale=0.3]{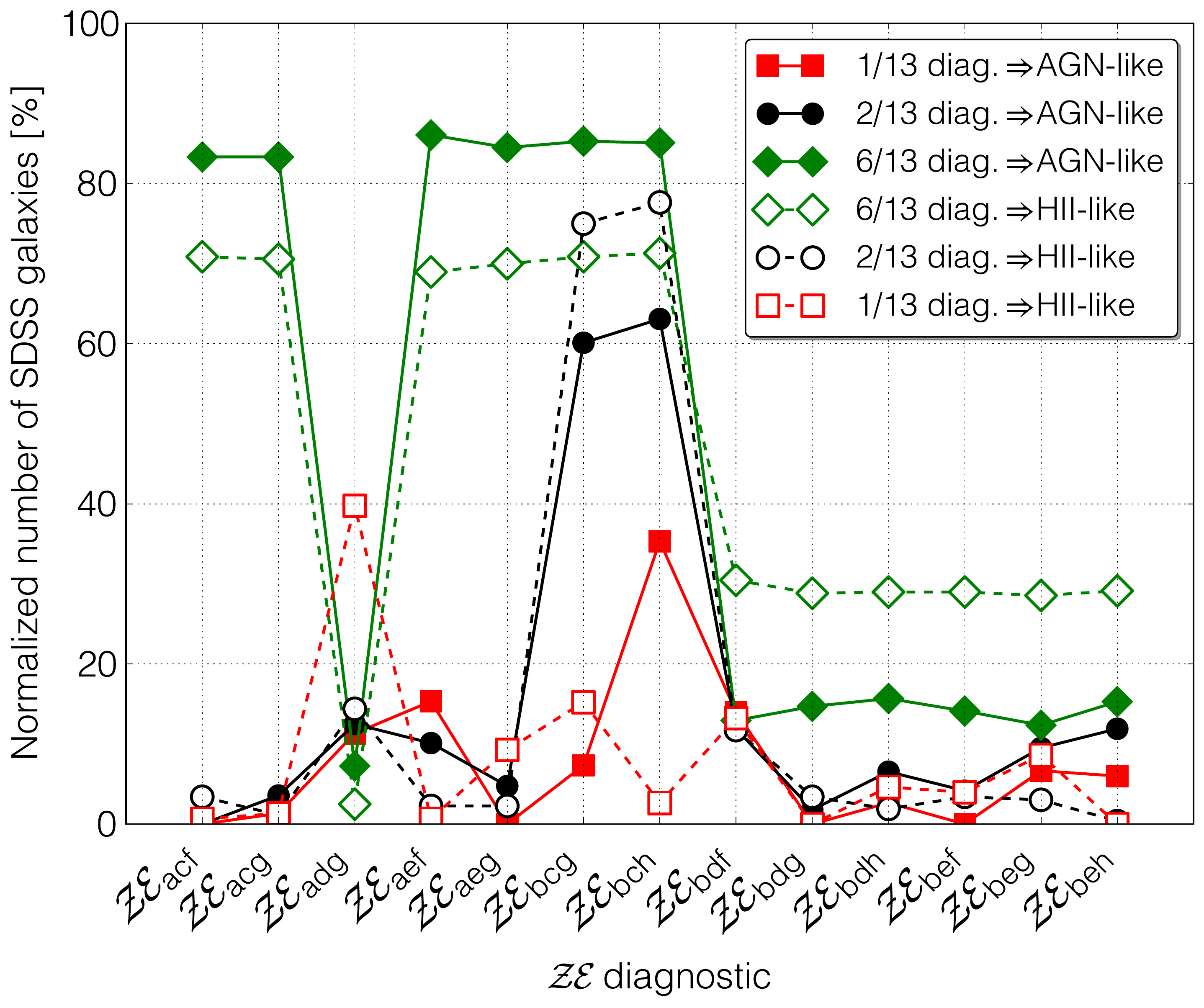}}
\caption{For each $\cal{ZE}$$_{\text{x}_1\text{x}_2\text{x}_3}$ diagnostic, the normalized number of uncertain galaxies for which this $\cal{ZE}$ diagnostic is discordant, when all diagnostics but one agree (squares), all diagnostics but two agree (circles) and only 6 diagnostics agree (diamonds). Full symbols correspond to uncertain objects classified as \ion{H}{2}-like by most $\cal{ZE}$ diagnostics, and empty symbols to objects classified as AGN-like by most $\cal{ZE}$ diagnostics.}\label{fig:min-max}
\end{figure}

The existence of the central and secondary peaks in the histogram shown in Figure~\ref{fig:hist} are perhaps more interesting. To understand their origin, we compute in Figure~\ref{fig:min-max} (for each $\cal{ZE}$ diagnostic) the normalised number of uncertain galaxies for which this $\cal{ZE}$ diagnostic is discordant. In other words, we ask whether some diagnostics are more discordant than the others. For clarity, we restrict ourselves to the data corresponding to the bins 1,2,6,7,11 and 12 in Figure~\ref{fig:hist}, that is objects that have only one (square symbols), only two (circles) and six (diamonds) discordant diagnostics.

First, we note that $\cal{ZE}_\text{bcg}$ and $\cal{ZE}_\text{bch}$ are responsible for $\sim$75\% of galaxies with only two discordant diagnostics, and are thereby mostly responsible for the secondary peak in the bins 11 of Figure~\ref{fig:hist}. This is a first indication of the somewhat lesser quality of these two diagnostics compared to the other ones. 

\begin{figure}[htb!]
\centerline{\includegraphics[scale=0.35]{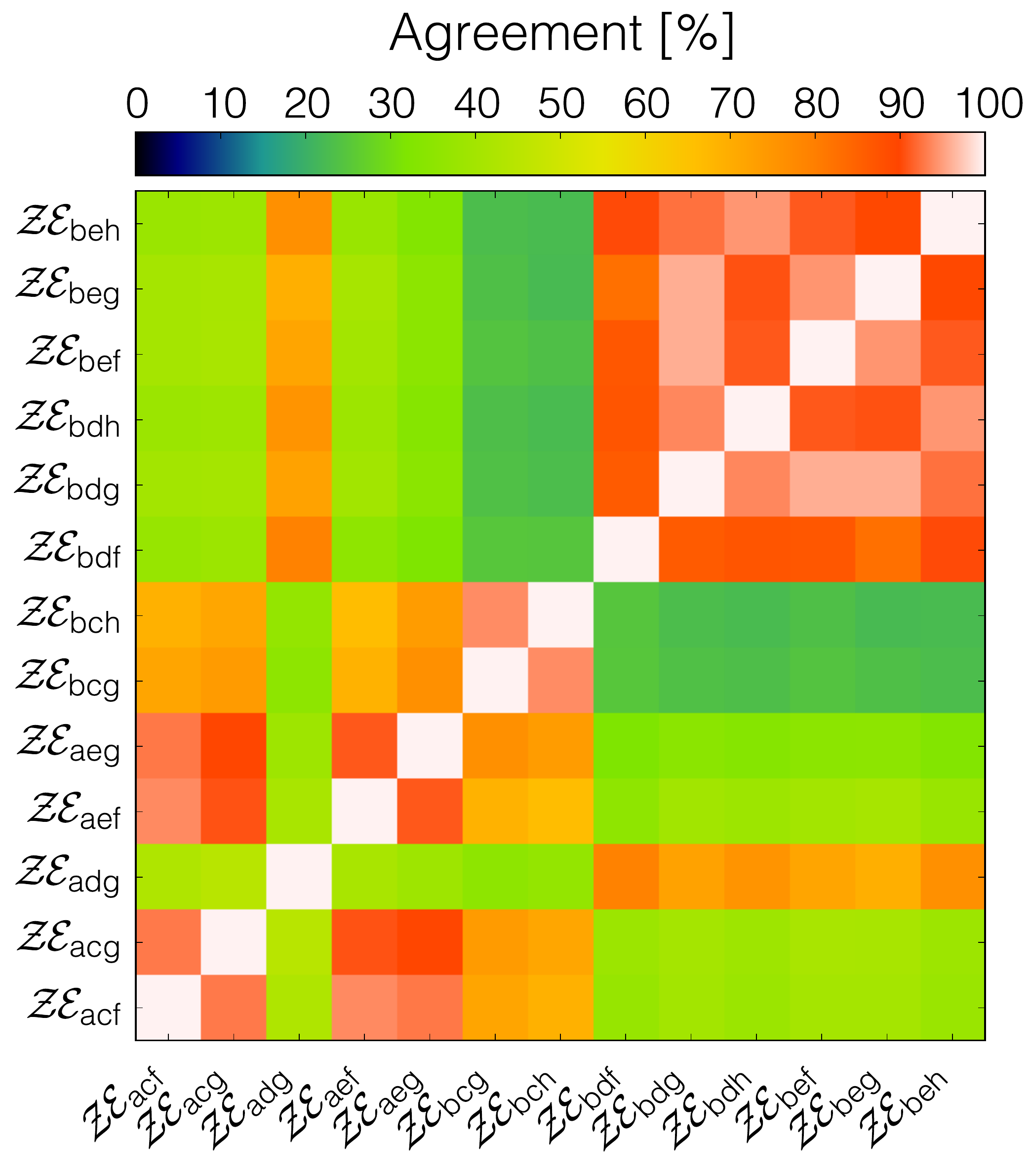}}
\caption{Level of agreement between the different $\cal{ZE}$ diagnostics regarding the classification of uncertain galaxies. }\label{fig:corr}
\end{figure}

Concerning the uncertain galaxies with six discordant $\cal{ZE}$ diagnostics (diamond symbols), we detect a clear dichotomy. On one side, [$\cal{ZE}_\text{acf}$;  $\cal{ZE}_\text{acg}$; $\cal{ZE}_\text{aef}$; $\cal{ZE}_\text{aeg}$; $\cal{ZE}_\text{bcg}$; $\cal{ZE}_\text{bch}$] are each in disagreement with the dominant classification in $\sim$70\%-85\% of the cases. On the other side, [ $\cal{ZE}_\text{bdf}$; $\cal{ZE}_\text{bdg}$; $\cal{ZE}_\text{bdh}$; $\cal{ZE}_\text{bef}$; $\cal{ZE}_\text{beg}$; $\cal{ZE}_\text{beh}$] are concordant with the dominant classification in $\sim$70\%-85\% of the cases.  $\cal{ZE}_\text{adg}$ is almost always ($\sim$95\%) consistent with the dominant classification. The overall consistency between the two groups of $\cal{ZE}$ diagnostics is suggestive of a possible underlying correlation. To explore this possibility, we show in Figure~\ref{fig:corr} the \emph{agreement} matrix between each pair of $\cal{ZE}$ diagnostics. A high value indicates that two diagnostics tend to be concordant in their classification of uncertain galaxies (100\% = always in agreement), while a lower value indicates that the two diagnostics mostly disagree on the classification of uncertain galaxies (0\% = never in agreement).

Indeed, we detect a clear correlation between the two sets of diagnostics
\begin{equation}
[\cal{ZE}_\text{acf};  \cal{ZE}_\text{acg}; \cal{ZE}_\text{aef}; \cal{ZE}_\text{aeg}; \cal{ZE}_\text{bcg}; \cal{ZE}_\text{bch}],\nonumber
\end{equation}
and 
\begin{equation}
[ \cal{ZE}_\text{adg}; \cal{ZE}_\text{bdf}; \cal{ZE}_\text{bdg}; \cal{ZE}_\text{bdh}; \cal{ZE}_\text{bef}; \cal{ZE}_\text{beg}; \cal{ZE}_\text{beh}]. \nonumber
\end{equation}

The level of agreement within each group is high ($\sim$75-95\%), but the agreement between the two groups overall is much poorer, of the order of $\sim$25-45\%. This strong dichotomy is giving rise to the central peak in the distribution of uncertain objects in Figure~\ref{fig:hist}. 

This bi-modal grouping of the $\cal{ZE}$ diagnostics is most certainly no accident, as it separates $\cal{ZE}$ diagnostic involving [\ion{O}{2}] from the others, with $\cal{ZE}_\text{adg}$ the only exception. The fact that diagnostics involving [\ion{O}{2}] are more discordant than other ones is strongly suggestive of reddening errors. Most likely, such reddening correction errors in our sample are responsible for this dichotomy, and responsible for most of the 510+683=1193 galaxies with 6 discordant diagnostics.  As expected and shown in Figure~\ref{fig:me_1},~\ref{fig:me_2} and \ref{fig:me_3} via the $\zeta$ shift, most $\cal{ZE}$ diagnostics involving [\ion{O}{2}] are more sensitive to such reddening corrections. We also note that the overall sensitivity of the SDSS spectrograph is decreasing sharply below $\sim$4000{\AA} \citep{Smee13}. The flux calibration may be less reliable in these spectral regions, and may possibly not be accurately reflected in the associated errors of the [\ion{O}{2}] lines.

In summary, the agreement matrix shown in Figure~\ref{fig:corr} highlights the limitations associated with defining consistent diagnostics relying on observational data, subject to observational and data processing errors. With a total of 2636 (2.5\%) of uncertain galaxies, the thirteen $\cal{ZE}$ diagnostics are nevertheless in very good agreement. For specific purposes, for example if reddening corrections are large and/or uncertain, working with only a subset of $\cal{ZE}$ diagnostics not involving the [\ion{O}{2}] lines could lead to an even better agreement of the combined $\cal{ZE}$ classification.

\section{Discussion}\label{sec:discussion}
\subsection{Comparison of $\cal{ZE}$ diagnostics with the standard optical line ratio diagnostics}\label{sec:comp_bpt}

Having introduced thirteen new composite line ratio diagnostics, and estimated the uncertainties associated with galaxy classification when combining each of these diagrams, it behoves us to investigate how these new diagnostics compare with the three standard optical line ratio diagnostic diagrams;  $\log${\NIIHa} \emph{vs.} $\log${\OIIIHb}, $\log${\SIIHa} \emph{vs.} $\log${\OIIIHb} and $\log${\OIHa} \emph{vs.} $\log${\OIIIHb}. 

\begin{figure*}[htb!]
\centerline{ \includegraphics[scale=0.22]{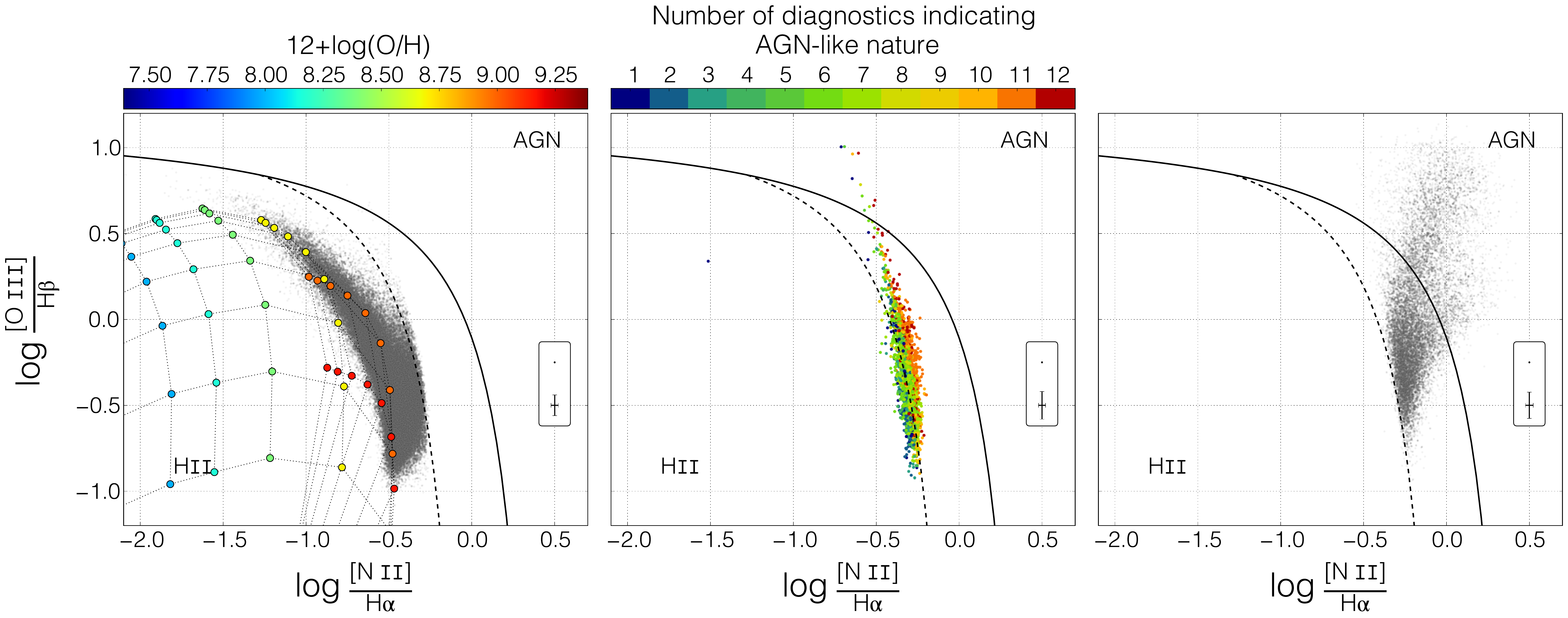}}
\caption{Location of the \ion{H}{2}-like (left), AGN-like (right) and uncertain (middle) SDSS galaxies in the standard $\log${\NIIHa} \emph{vs.} $\log${\OIIIHb}  diagram. Uncertain galaxies are colour-coded as a function of how many diagnostics indicate a non-H\,{\sc ii}-like nature. The diagnostic lines from \cite{Kewley01b} and \cite{Kauffmann03b} are marked with a full and dashed line, respectively. The different sectors (\ion{H}{2}, AGN) are labelled following the nomenclature defined in the Figure~4 of \cite{Kewley06}. We also show the \emph{MAPPINGS IV} model grid for $\kappa=20$ in the left-side diagram, where each grid point is color-coded as a function of $12+\log$(O/H).}
\label{fig:bpt_o3n2}
\end{figure*}

\begin{figure*}[htb!]
\centerline{ \includegraphics[scale=0.22]{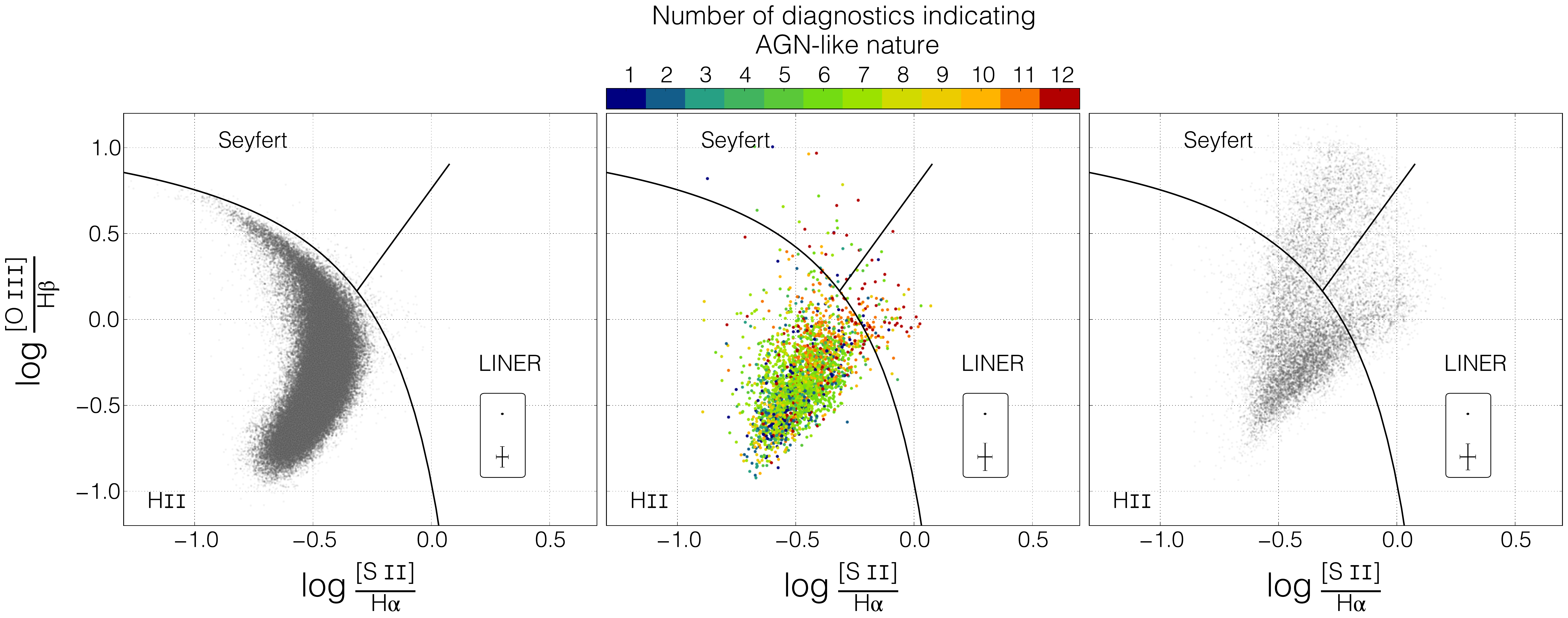}}
\caption{Same as Figure~\ref{fig:bpt_o3n2}, but for the standard $\log${\SIIHa} \emph{vs.} $\log${\OIIIHb} diagram. The maximum starburst line from \cite{Kewley01b} and the Seyfert-LINER line from \cite{Kewley06} are plotted accordingly. }\label{fig:bpt_o3s2}
\end{figure*}

\begin{figure*}[htb!]
\centerline{ \includegraphics[scale=0.22]{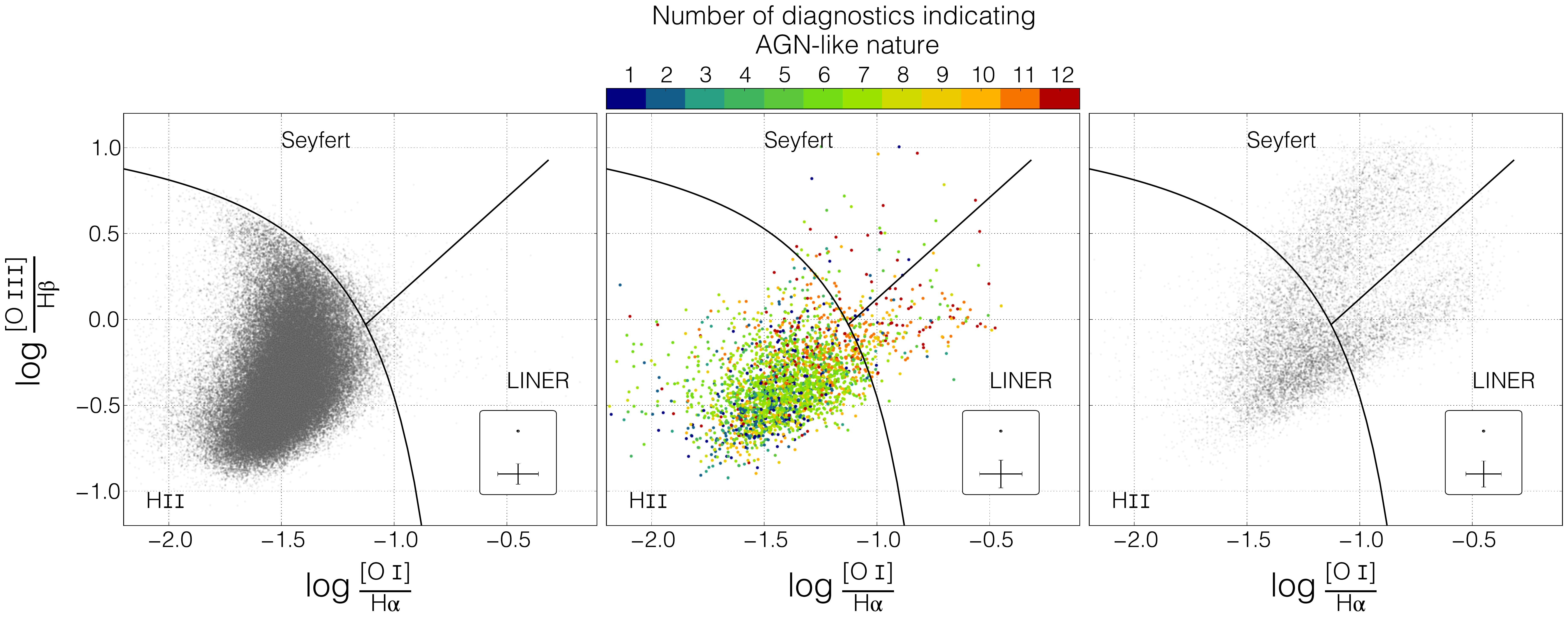}}
\caption{Same as Figure~\ref{fig:bpt_o3n2}, but for the standard $\log${\OIHa} \emph{vs.} $\log${\OIIIHb} diagram.  }\label{fig:bpt_o3o1}
\end{figure*}

For clarity, in Figures ~\ref{fig:bpt_o3n2}, \ref{fig:bpt_o3s2} and \ref{fig:bpt_o3o1}, we have separated out into three distinct diagrams the H\,{\sc ii}-like (\emph{left}), the AGN-like (\emph{right}), and the uncertain (\emph{center}) galaxies, as defined by our new $\cal{ZE}$ diagnostics (see Section~\ref{sec:me_diag}). The advantage of our new classification scheme is that by combining multiple diagnostics, we can assign a probability for all points to be H\,{\sc ii}-like, or not. This probability (i.e. the number of consistent diagnostics) is color-coded and shown in the middle panel for all objects with uncertain classification. The concept of associating a probability to a given classification is reminiscent of the MEx diagram of  \cite{Juneau11}, although our respective methods differ fundamentally in practice.  \cite{Juneau11} rely on a \emph{prior sample of SDSS galaxies} first classified via two of the standard line ratio diagnostic diagrams to quantify the inherent additional confusion between the different classes (star forming, Seyferts, LINERs, composites) in the MEx diagram. On the other hand, with the $\cal{ZE}$ diagnostics, we derive a classification probability for the \emph{prior sample of SDSS galaxies} itself, by combining thirteen new and complementary line ratio diagnostic diagrams. In both cases however, the need to quantify the certainty of a given classification is motivated by the fact that objects located near a given diagnostic line can have an inherently uncertain classification.
 
Clearly, our new composite diagnostics are concordant with the \cite{Kauffmann03b} line in Figure~\ref{fig:bpt_o3n2}, with only 1128 (1.1\%) galaxies classified with certainty by the $\cal{ZE}$ diagnostics as being located on the \emph{wrong} side of the \cite{Kauffmann03b} diagnostic line. This agreement is in itself not surprising, since both approaches rely on an empirical fit of SDSS galaxies. Yet, we stress here that the $\cal{ZE}$ diagnostics have all been defined \emph{before} attempting any comparison with the standard diagnostic diagrams. The agreement between the $\cal{ZE}$ diagnostics and the \cite{Kauffmann03b} diagnostic line is also a confirmation that overall, objects defined as \ion{H}{2}-like based on the $\cal{ZE}$ diagnostics are indeed similar to star-forming galaxies classified using the standard line ratio diagram $\log${\NIIHa} \emph{vs.} $\log${\OIIIHb}. On the other hand, the AGN-like objects (as defined by the $\cal{ZE}$ diagnostics) can be associated with the ``traditional'' Seyfert, LINER and composite classes. 

From the distribution of uncertain galaxies in the middle panel of Figure~\ref{fig:bpt_o3n2}, it is clear that the $\cal{ZE}$ diagnostics are more consistent with the \cite{Kauffmann03b} diagnostic towards the locus of the SDSS galaxies. For $\log${\OIIIHb}$\gtrsim 0$, the distribution of uncertain galaxies, which effectively trace the location of the boundary between \ion{H}{2}-like versus AGN-like objects as defined by the $\cal{ZE}$ diagnostics, is not following the original diagnostic lines. The scarcity of galaxies present in this area make it difficult to objectively decide which set of diagnostics may be more appropriate, and further observations are required (for example of very metal poor galaxies). 

There is a smooth left to right gradient in the number of $\cal{ZE}$ diagnostics classifying the object as AGN-like through the zone occupied by the objects with uncertain classification in the $\log${\NIIHa} \emph{vs.} $\log${\OIIIHb} diagram. This gradient suggests that galaxies with at least 85\% agreement between the different $\cal{ZE}$ diagnostics can be classified accurately. In addition, we also note that if the galaxies with uncertain classifications mostly cluster around the starburst line from \cite{Kauffmann03b}, many galaxies with ``certain'' classification from the $\cal{ZE}$ diagnostics also lie very close to the line. 

The other two standard optical line ratio diagnostic diagrams ( $\log${\SIIHa} \emph{vs.} $\log${\OIIIHb} and $\log${\OIHa} \emph{vs.} $\log${\OIIIHb}) are known to be less efficient at separating star-forming objects from AGNs. The reason, clearly visible in Figure~\ref{fig:bpt_o3s2} and \ref{fig:bpt_o3o1}, is that the Seyfert/LINER branches extend deeply in the star-forming regions \citep{Kewley06,Yuan10}. Consequently, these diagrams are in practice rarely used to identify star-forming galaxies, with the diagnostics solely relying on the $\log${\NIIHa} \emph{vs.} $\log${\OIIIHb} diagram. For completeness, we list in Table~\ref{table:std_uncertain} the percentage of ``standard'' uncertain galaxies that would result from different combination of the three standard line ratio diagrams. To that end, we follow the criteria listed in \cite{Kewley06}, which we repeat here verbatim: ``standard'' star-forming galaxies lie below and to the left-hand side of the \cite{Kauffmann03b} diagnostic in the $\log$[\ion{N}{2}]/H$\alpha$ \emph{vs.} $\log$[\ion{O}{3}]/H$\beta$ diagram, and below and to the left-hand side of the \cite{Kewley01b} diagnostics in the $\log$[\ion{S}{2}]/H$\alpha$ \emph{vs.} $\log$[\ion{O}{3}]/H$\beta$ and  $\log$[\ion{O}{1}]/H$\alpha$ \emph{vs.} $\log$[\ion{O}{3}]/H$\beta$ diagrams. For the sake of comparison with our $\cal{ZE}$ diagnostics, we do not distinguish between composites, Seyfert and LINERS, so that ``standard'' AGN-like objects lie above and to the right-hand side of the \cite{Kauffmann03b} diagnostic in the $\log$[\ion{N}{2}]/H$\alpha$ \emph{vs.} $\log$[\ion{O}{3}]/H$\beta$ diagram, and above and to the right-hand side of the \cite{Kewley01b} diagnostics in the $\log$[\ion{S}{2}]/H$\alpha$ \emph{vs.} $\log$[\ion{O}{3}]/H$\beta$ and $\log$[\ion{O}{1}]/H$\alpha$ \emph{vs.} $\log$[\ion{O}{3}]/H$\beta$ diagrams. In that scheme, ``standard'' uncertain galaxies are those classified as star-forming in some diagrams, and AGN-like in others.

\begin{table}[htb!]
\caption{Number of uncertain classifications resulting from different combination of standard optical line ratio diagnostics and $\cal{ZE}$ diagnostics.}\label{table:std_uncertain}
\center
\vspace{-10pt}
\begin{tabular}{c c c c }
\hline\hline
Diagnostics &\ion{H}{2}-like &AGN-like &uncertain\\
\hline \\ [-1.5ex]
$\cal{ZE}$ & \begin{tabular}{@{}c@{}} 88918 \\ (84.6\%) \end{tabular} & \begin{tabular}{@{}c@{}} 13516 \\ (12.9\%) \end{tabular} & \begin{tabular}{@{}c@{}} 2636 \\ (2.5\%) \end{tabular} \\ [4ex]

[\ion{N}{2}]/H$\alpha$;[\ion{O}{3}]/H$\beta$ &\begin{tabular}{@{}c@{}} 88933 \\ (84.6\%) \end{tabular} & \begin{tabular}{@{}c@{}} 16137 \\ (15.4\%) \end{tabular} & - \\ [4ex]

\begin{tabular}{@{}c@{}} [\ion{N}{2}]/H$\alpha$;[\ion{O}{3}]/H$\beta$ \\ {[\ion{S}{2}]/H$\alpha$;[\ion{O}{3}]/H$\beta$} \end{tabular} & \begin{tabular}{@{}c@{}} 88729 \\ (84.5\%) \end{tabular} & \begin{tabular}{@{}c@{}} 5377 \\ (5.1\%) \end{tabular} & \begin{tabular}{@{}c@{}} 10964 \\ (10.4\%) \end{tabular} \\[4ex]

\begin{tabular}{@{}c@{}} [\ion{N}{2}]/H$\alpha$;[\ion{O}{3}]/H$\beta$ \\ {[\ion{S}{2}]/H$\alpha$;[\ion{O}{3}]/H$\beta$} \\ {[\ion{O}{1}]/H$\alpha$;[\ion{O}{3}]/H$\beta$} \end{tabular} & \begin{tabular}{@{}c@{}} 86942 \\ (82.7\%) \end{tabular} & \begin{tabular}{@{}c@{}} 5208 \\ (5.0\%) \end{tabular} & \begin{tabular}{@{}c@{}} 12920 \\ (12.3\%) \end{tabular} \\[4ex]

\hline
\end{tabular}
\end{table}

We did not apply any S/N cut on the [\ion{O}{1}] line to obtain the standard classification. Indeed, the data presented in Table~\ref{table:std_uncertain} is not intended as a mean of evaluating different classification strategies, but rather to provide a simple comparison of the size of the population of uncertain objects resulting from the combination of different line ratio diagnostics. As discussed previously, the $\cal{ZE}$ diagnostics are very consistent with the standard [\ion{N}{2}]/H$\alpha$ versus [\ion{O}{3}]/H$\beta$ diagram alone. The number of galaxies classified as \ion{H}{2}-like is very comparable (0.1\% change) when combining both $\log$[\ion{N}{2}]/H$\alpha$ \emph{vs.} $\log$[\ion{O}{3}]/H$\beta$ and $\log$[\ion{S}{2}]/H$\alpha$ \emph{vs.} $\log$[\ion{O}{3}]/H$\beta$, as most of the ''standard'' uncertain galaxies are in this case comprised of AGN-like objects. Comparatively, combining all three standard optical line ratios does impact the number of galaxies classified as \ion{H}{2}-like by $\sim$2\%. The large amount of uncertain galaxies ($\sim$10\%) resulting from the combination of different standard line ratio diagnostics is a direct consequence of the AGN branch extending deeply in the \ion{H}{2}-like region in the diagrams involving [\ion{S}{2}] and [\ion{O}{1}]. The number of uncertain galaxies associated with the more numerous $\cal{ZE}$ diagnostics remains comparatively small (2.5\%) by design, as each diagnostic is chosen to cleanly separate the star-forming and AGN branches of SDSS galaxies. 

As already discussed in Section~\ref{sec:consistency}, one disadvantage of the $\cal{ZE}$ diagnostics is that they involve line ratios which are sensitive to reddening corrections. Comparatively, the standard [\ion{N}{2}]/H$\alpha$ versus [\ion{O}{3}]/H$\beta$ is much more immune to this source of errors. We have mentioned in Section~\ref{sec:consistency} the possibility to work with a subset of six $\cal{ZE}$ diagnostics much less prone to reddening errors.Yet, the complete set of $\cal{ZE}$ diagnostics might also act as a way of identifying possible reddening correction issues in a given dataset, which would result in a comparatively large numbers of uncertain $\cal{ZE}$ classifications. 

\subsection{$\cal{ZE}$ diagrams for high redshift objects } \label{sec:high-z}

The limited number of atmospheric windows, and the stretching of the spectrum with redshift makes the observation of the full set of emission lines of high-redshift galaxies difficult. At intermediate redshift (0.5$<$z$<$1.0), the {\SII}, {\NII} and H$\alpha$ lines are redshifted in the near-infrared, while the {\OII}, {\OIII} and H$\beta$ lines are still in the optical range. This observational constraint has motivated several \emph{hybrid} alternatives to the standard line ratio diagnostic diagrams that replace the (unaccessible) redder line ratio with another observable, such as $\log$\OII/H$\beta$ \citep[the ``blue diagram'', see][]{Lamareille04,Lamareille10}, the U-B rest-frame color of the galaxy \citep[the CEx diagram,][]{Yan11} or the galaxy's stellar mass \citep[the MEx diagram,][]{Juneau11,Juneau14}.  Other propositions, such as the DEW diagram \citep{Stasinska06,Marocco11} or the TBT diagram \citep{Trouille11} replaced both line ratios of the original diagnostics in favor of $D_n(4000)$ vs. max(EW\OII,EW[\ion{Ne}{3}]) or the rest frame g-z color vs. $\log$[\ion{Ne}{3}]/\OII, respectively. In the case of $\cal{ZE}$ diagnostics, the complete set relies on the measurement of (at least) one ``red'' line for the Category I (abundance-sensitive) line ratio, and at least one ``blue'' line for the Category II (q-sensitive) line ratio (see Table~\ref{table:keys}). Hence, the $\cal{ZE}$ diagnostics are subject to the same limitation than the original line ratio diagnostics for intermediate redshift objects.

Spectroscopic studies of galaxies at redshifts 1.0$<$z$<$1.7 in the infrared miss observing the [O\,{\sc ii}$]\lambda$3727+$\lambda$3729 lines. While this fact rules out direct access to a certain number of line ratio diagrams, of the thirteen $\cal{ZE}$ diagnostics introduced in this article, six do not rely on [O\,{\sc ii}] (see Table~\ref{table:all_me} and Figure~\ref{fig:me_3}). Thus, while the full set is not available, the remainder should still be useful to ensure that a reliable classification can be made for high redshift objects. We stress that although the $\cal{ZE}$ diagrams can be used, modelling of the evolution of metallicity and ISM conditions are essential for understanding the evolution of ZE diagnostics with redshift. The $\cal{ZE}$ diagnostic lines are in fact subject to the same uncertainty as the standard diagnostic lines of \cite{Kewley01b, Kauffmann03b} and \cite{Kewley06} at high-redshifts \citep[e.g.][]{Liu08, Brinchmann08, Trump13,Kewley13a,Kewley13b, Juneau14}. 

\section{Alternative applications for the $\cal{ZQE}$ 3D line ratio diagrams}\label{sec:AGN}

In Sections~\ref{sec:3d-2d} and \ref{sec:discussion}, we used the $\cal{ZQE}$ 3D optical line ratio diagrams to devise thirteen new diagnostic diagrams that can separate \ion{H}{2}-like and AGN-like objects in a consistent and robust way. A second application for $\cal{ZQE}$ diagrams is to perform a similar analysis specifically targeting the different classes of the AGN family: Seyfert, LINERs, and composites (which belong to the AGN-like group defined by the $\cal{ZE}$ diagnostics). The complex structure of the AGN branch is visible in the interactive $\cal{ZQE}$$_\text{adg}$ diagram in Figure~\ref{fig:3D_example}. Other $\cal{ZQE}$ diagrams, for example those involving the {\OII} line, can provide an even cleaner separation between these different substructures. Hence, $\cal{ZQE}$ diagrams appear as a useful tool to find new view points on the AGN galaxies (in the multi-dimensional line ratio space), and to gain insight on the underlying physics by comparing these with theoretical models. This analysis, outside the scope of this article, will be the subject of a future publication.

As highlighted in Section~\ref{sec:high-z}, observational limitations for intermediate redshift galaxies can hinder the use of the $\cal{ZE}$ diagnostics. One solution to this limitation would be the creation of a series of \emph{hybrid} $\cal{ZQE}$ diagrams, in which the inaccessible lines are replaced by alternative observables, similarly to the MEx diagram \citep{Juneau11} and other similar 2D propositions. Depending on the data available, one could decide to either replace the ``red" lines redshifted outside the optical region, or the ``blue" lines not redshifted enough in the near-IR region. Similarly to the approach applied in this article for the creation of the $\cal{ZE}$ diagrams, the interactive inspection of a \emph{hybrid} $\cal{ZQE}$ diagram might allow for the identification of specific points-of-view reducing the inherent confusion between the different classes of objects associated with the replacement of a line ratio with another observable - a common issue for all the ``hybrid'' 2D diagnostic mentionned in Section~\ref{sec:high-z}.
 
\section{Summary}\label{sec:conclusion}

In this article, we have demonstrated the utility of 3D line ratio diagrams for the classification of galaxies. Unlike standard (and historical) line ratio diagnostics in two dimensions, these new (interactive) diagrams allow for a better understanding of the spatial distribution of galaxies in the multi-dimensional line ratio space. Such diagrams are especially powerful when combing three different line ratios sensitive to a) the gas-phase oxygen abundance $12+\log$(O/H), b) the ionisation parameter $q$, and c) the excitation mechanism. The key advantage of these 3D line ratio diagrams is to allow the identification of specific points-of-view of interest on the spatial distribution of galaxies.

In particular, we identified line-ratio triplets which, when projected onto a specific plane, compact the theoretical \emph{MAPPINGS IV} photoionization grids to be almost completely degenerate along the ionization parameter direction, and in which the different chemical abundances of \ion{H}{2} regions fall along a well-defined curve. From these specific view points, we introduced thirteen new composite line ratio diagnostics diagrams, the $\cal{ZE}$ diagnostic diagrams, which enable us to efficiently separate H\,{\sc ii}-like objects from galaxies excited by an AGNs independently of the standard optical line ratio diagrams.  

We proved that this new set of composite line ratios is very consistent with the standard $\log${\NIIHa} \emph{vs.} $\log${\OIIIHb} diagnostic diagram, and especially with the \cite{Kauffmann03b} diagnostic line. In other words, we have confirmed independently the known ability of the  $\log${\NIIHa} \emph{vs.} $\log${\OIIIHb} diagram to efficiently separate star-forming galaxies from AGN. The $\cal{ZE}$ diagnostics also have the distinct advantage to attribute a probability for each measurements to belong to the \ion{H}{2}-like class (or not) for galaxies close from the diagnostic boundaries. For our sample of 105070 SDSS galaxies, we find that 2.5\% have an uncertain classification when combining the thirteen $\cal{ZE}$ diagnostics. Among our thirteen new diagrams, six do not rely on [O\,{\sc ii}$]\lambda$3727+$\lambda$3729. These specific diagnostics are highly consistent ($\sim$90\%) with one another, less prone to reddening correction issues, and are also suitable for spectroscopic studies of high-redshift objects which may not have access to the [\ion{O}{2}] emission lines.

The notion of multi-dimensional, interactive line ratio diagram opens a new way to look at the classification of extragalactic sources based on their emission line characteristics. Galaxies form a complex structure in this multi-dimensional space, a structure which is intrinsically hard to understand with two-dimensional diagrams. With interactive three-dimensional diagrams, we identified a new and complimentary technique to look at line ratio diagrams. This technique will be of special utility to understand the physical difference between the Seyferts and LINER class of objects, which will be explored in a future paper.

\acknowledgments 
We thank David Nicholls for stimulating discussions, Bill Roberts and the IT team at the Research School of Astronomy and Astrophysics (RSAA) at the Australian National University (ANU) for their support installing and maintaining the PDF3DReportGen software on the school servers, Kristen Anchor and the staff of the Digital Media Center at Johns Hopkins University for their help in developing and testing the 3D-printable model of $\cal{ZE}_\text{adg}$, Nathaniel J. "Niel" Leon of the WSE Advanced Manufacturing Lab at Johns Hopkins University for 3D-printing the high resolution copy of the model, and the anonymous referee for his/her constructive review that helped us improve this article.  

Vogt acknowledges a Fulbright scholarship, and further financial support from an Alex Rodgers Travelling scholarship from the RSAA at the ANU. Vogt is grateful to the Department of Physics and Astronomy at Johns Hopkins University for hosting him during his Fulbright exchange. Dopita acknowledges the support of the Australian Research Council (ARC) through Discovery project DP130103925. This work was funded by the Deanship of Scientific Research (DSR), King Abdulaziz University, under grant No. (5-130/1433 HiCi). The authors, therefore, acknowledge technical and financial support of KAU. Scharw\"achter acknowledges the European Research Council for the Advanced Grant Program Num 267399-Momentum.

This research has made use of NASA's Astrophysics Data System. Funding for SDSS-III has been provided by the Alfred P. Sloan Foundation, the Participating Institutions, the National Science Foundation, and the U.S. Department of Energy Office of Science. The SDSS-III web site is http://www.sdss3.org/. SDSS-III is managed by the Astrophysical Research Consortium for the Participating Institutions of the SDSS-III Collaboration including the University of Arizona, the Brazilian Participation Group, Brookhaven National Laboratory, Carnegie Mellon University, University of Florida, the French Participation Group, the German Participation Group, Harvard University, the Instituto de Astrofisica de Canarias, the Michigan State/Notre Dame/JINA Participation Group, Johns Hopkins University, Lawrence Berkeley National Laboratory, Max Planck Institute for Astrophysics, Max Planck Institute for Extraterrestrial Physics, New Mexico State University, New York University, Ohio State University, Pennsylvania State University, University of Portsmouth, Princeton University, the Spanish Participation Group, University of Tokyo, University of Utah, Vanderbilt University, University of Virginia, University of Washington, and Yale University. 

\appendix

\section{$\cal{ZQE}$ diagrams with larger confusion between the starburst and AGN sequences}

\begin{figure*}[htb!]
\centerline{ \includegraphics[scale=0.27]{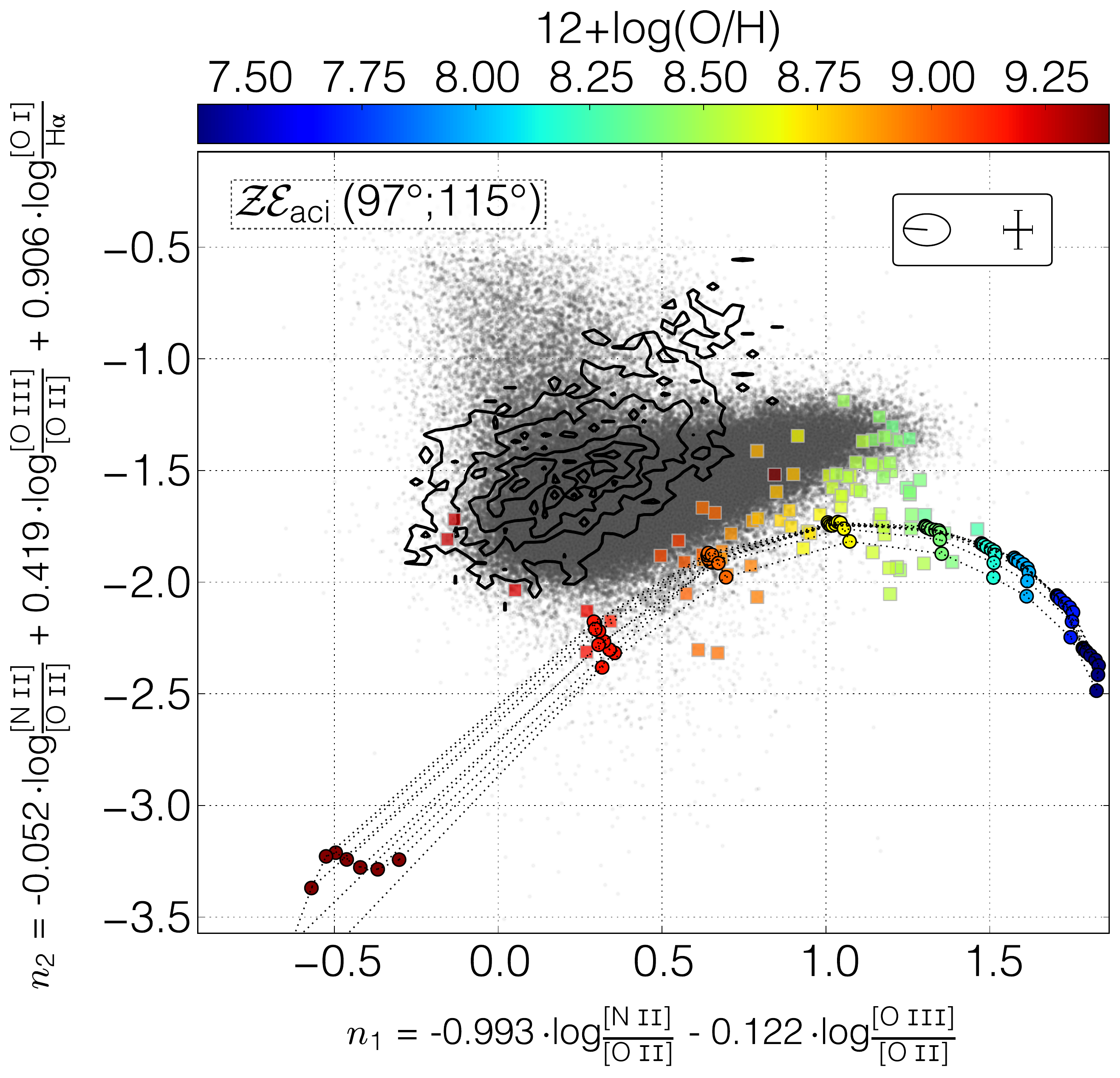}\qquad\qquad \includegraphics[scale=0.27]{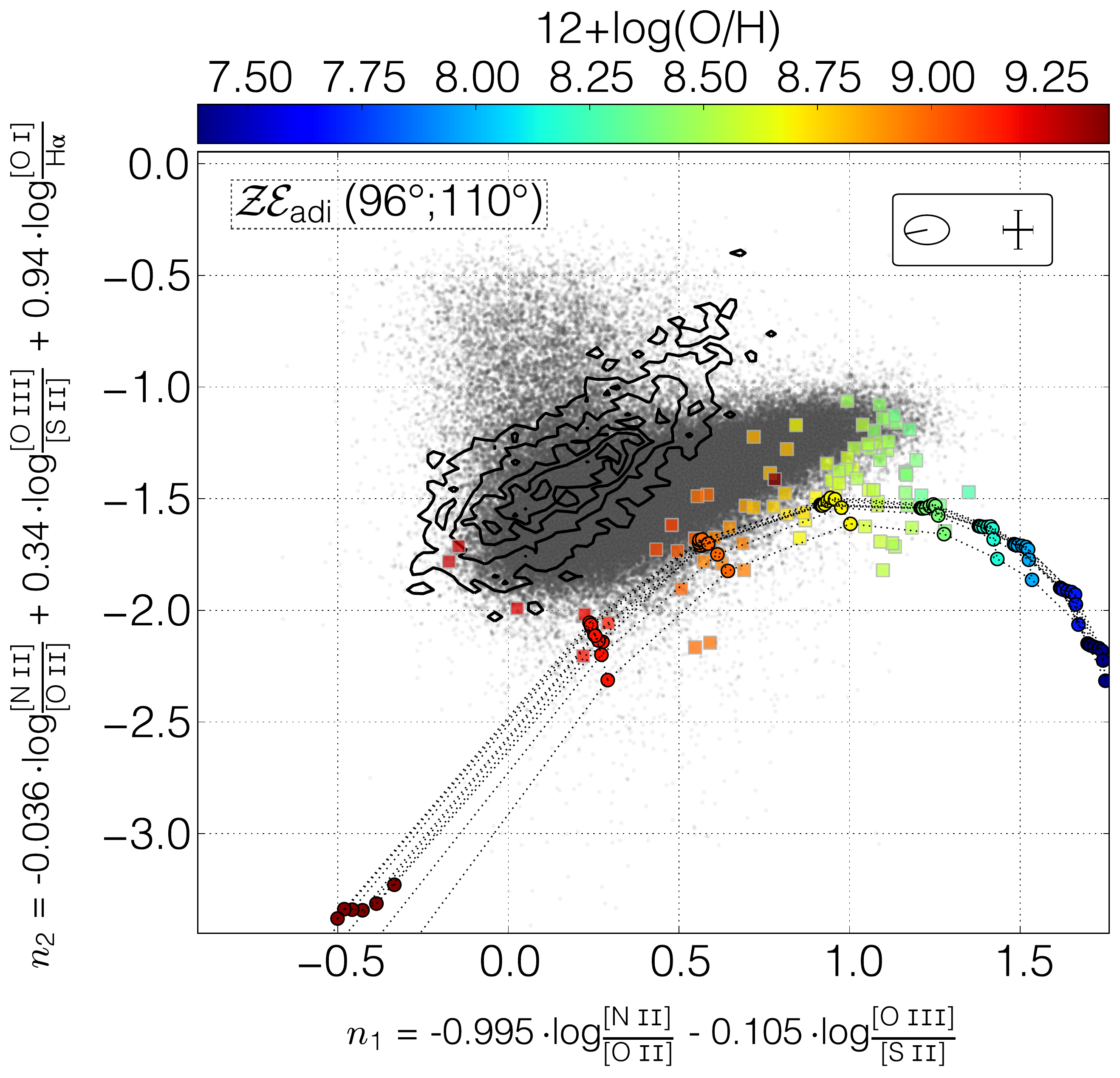}}\centerline{  \includegraphics[scale=0.27]{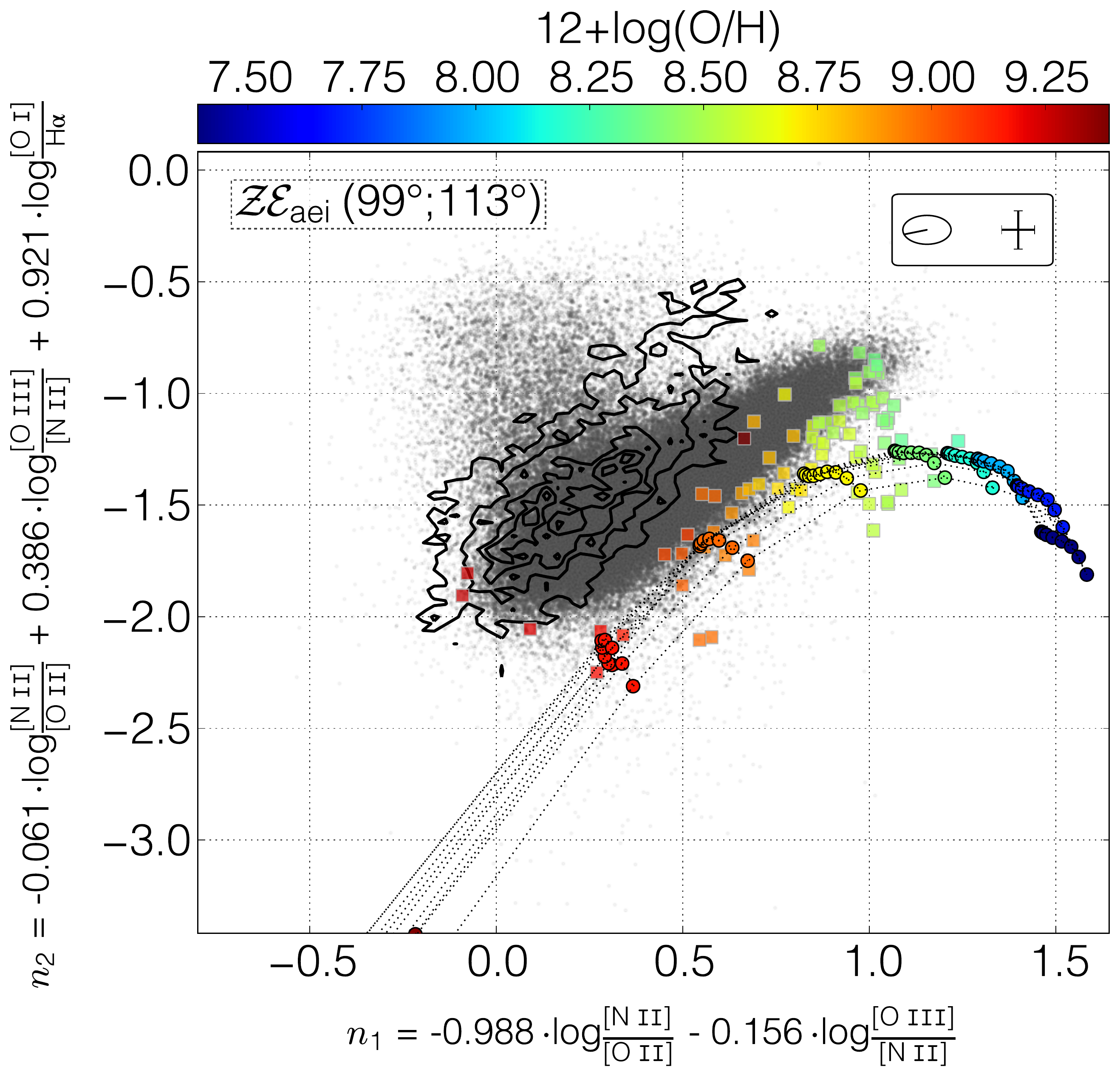}\qquad\qquad \includegraphics[scale=0.27]{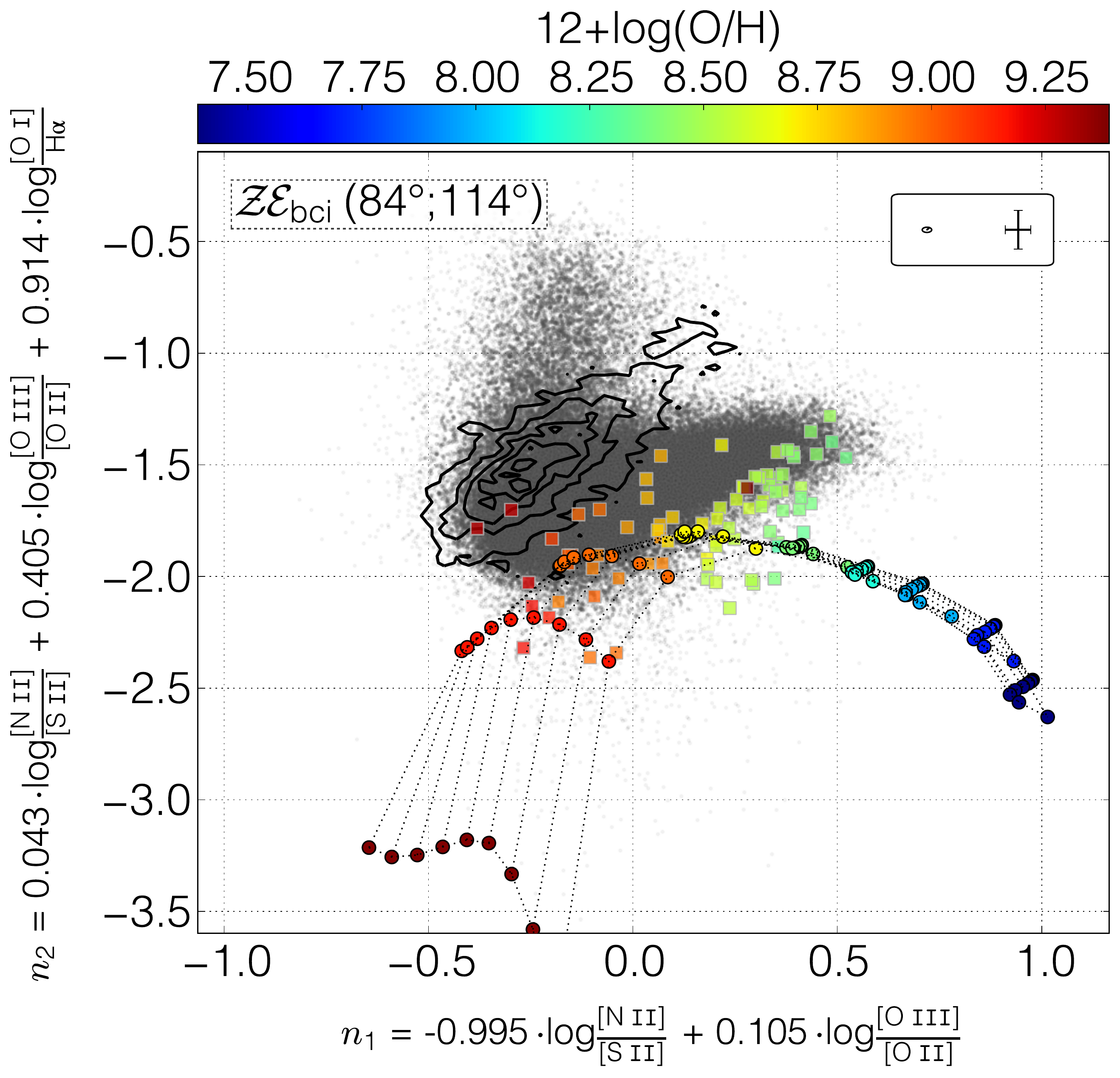}}
\centerline{  \includegraphics[scale=0.27]{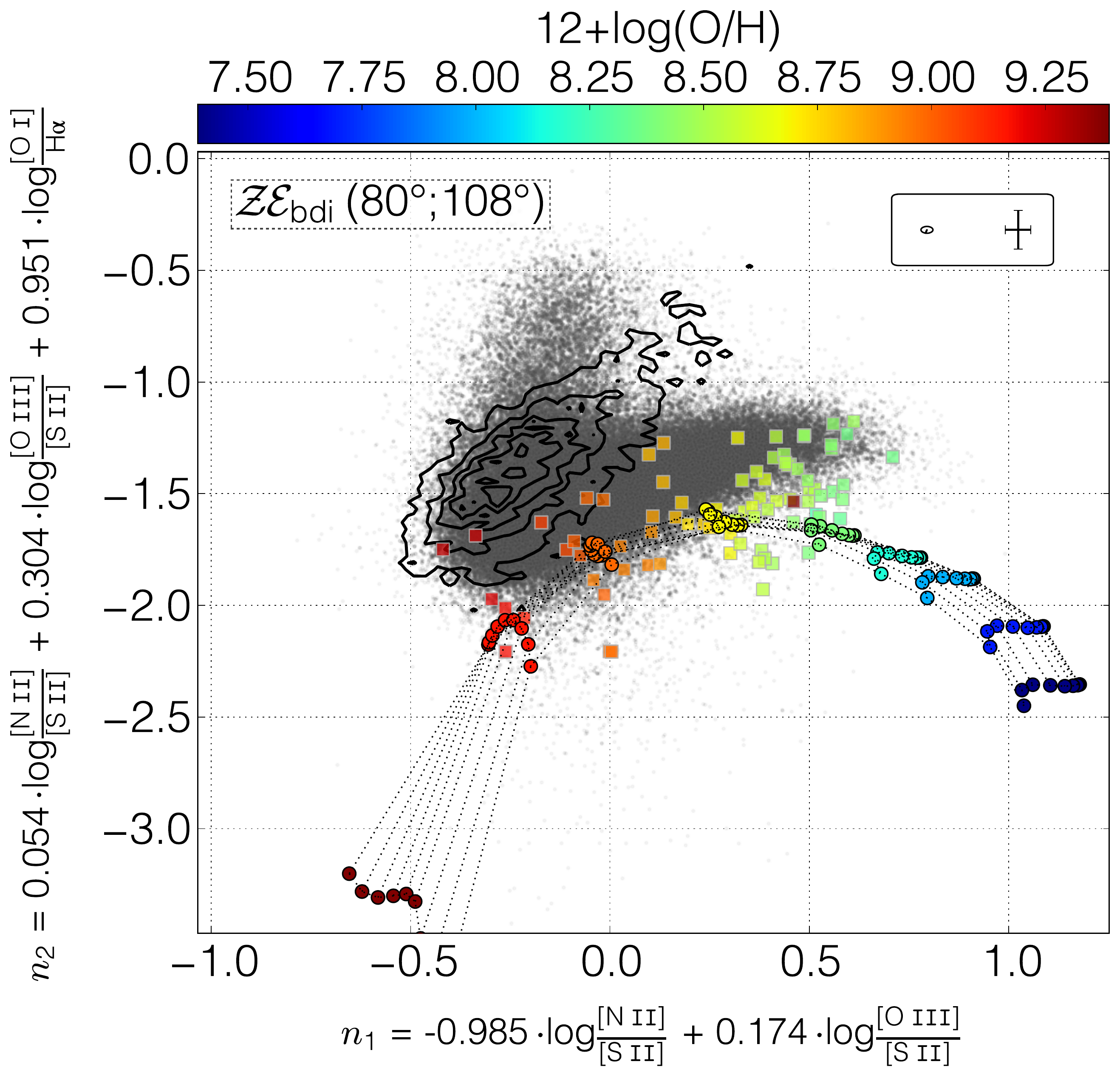}\qquad\qquad \includegraphics[scale=0.27]{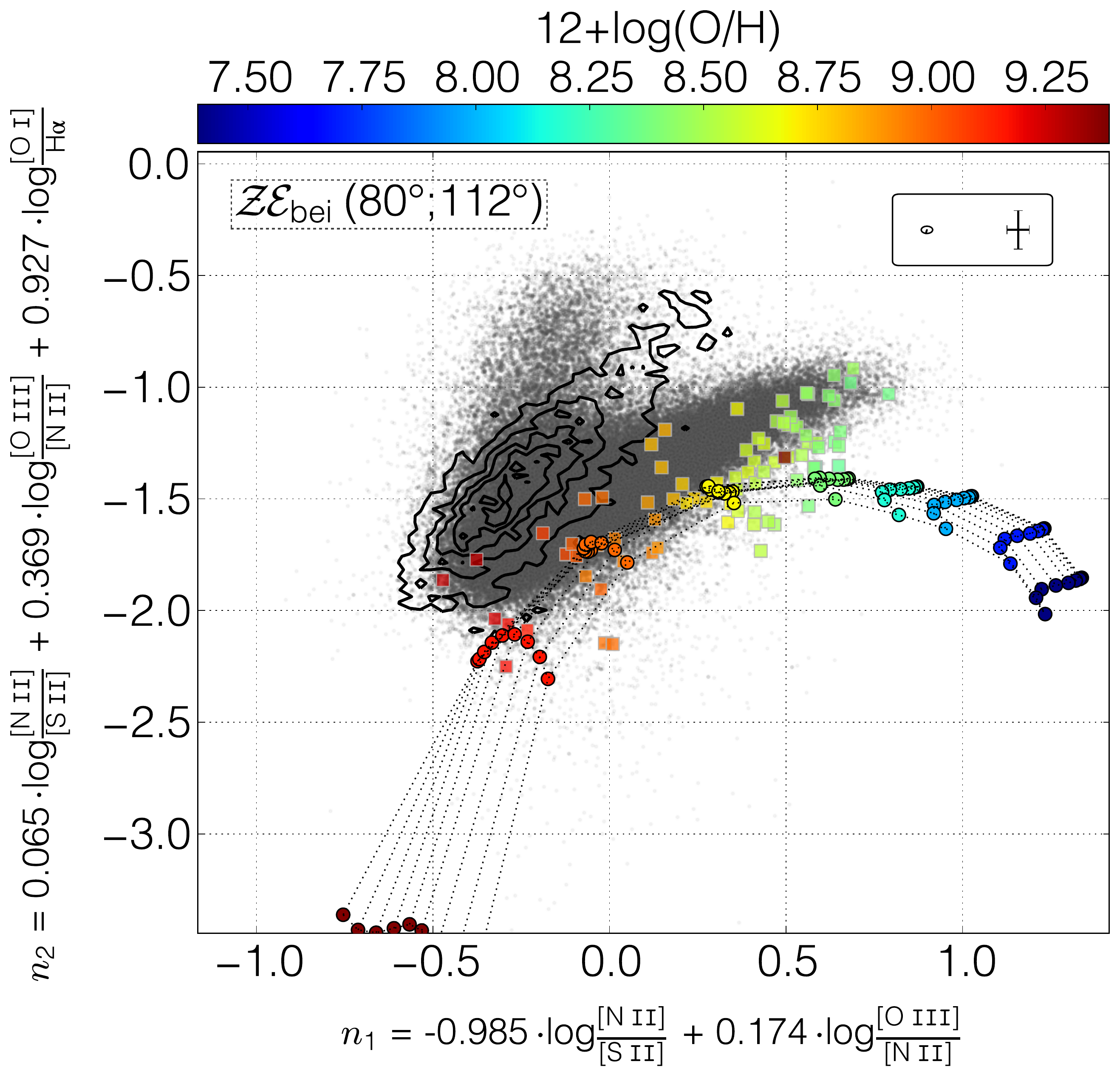}}
\caption{$\cal{ZE}$ diagrams for $\cal{ZQE}$ diagrams involving $\log${\OIHa} with no associated $\cal{ZE}$ diagnostic. The diagram name and associated values of $(\phi,\theta)$ is shown in the top-left corner for completeness. \ion{H}{2}-like and AGN-like SDSS galaxies are in grey. Uncertain galaxies (based on all $\cal{ZE}$ diagnostics) are represented by density contours (5\%, 20\%, 40\% and 80\% of the maximum density). The coloured dots (connected by the dotted lines) correspond to the \emph{MAPPINGS IV} models from \cite{Dopita13a}. The \cite{vanZee98} points are represented by small squares, and the measurements from NGC\,5427 are marked with small triangles.  All measured \ion{H}{2} regions are color-coded according to their oxygen abundance. }\label{fig:ZQE_Oi}
\end{figure*}

\begin{figure*}[htb!]
\centerline{ \includegraphics[scale=0.27]{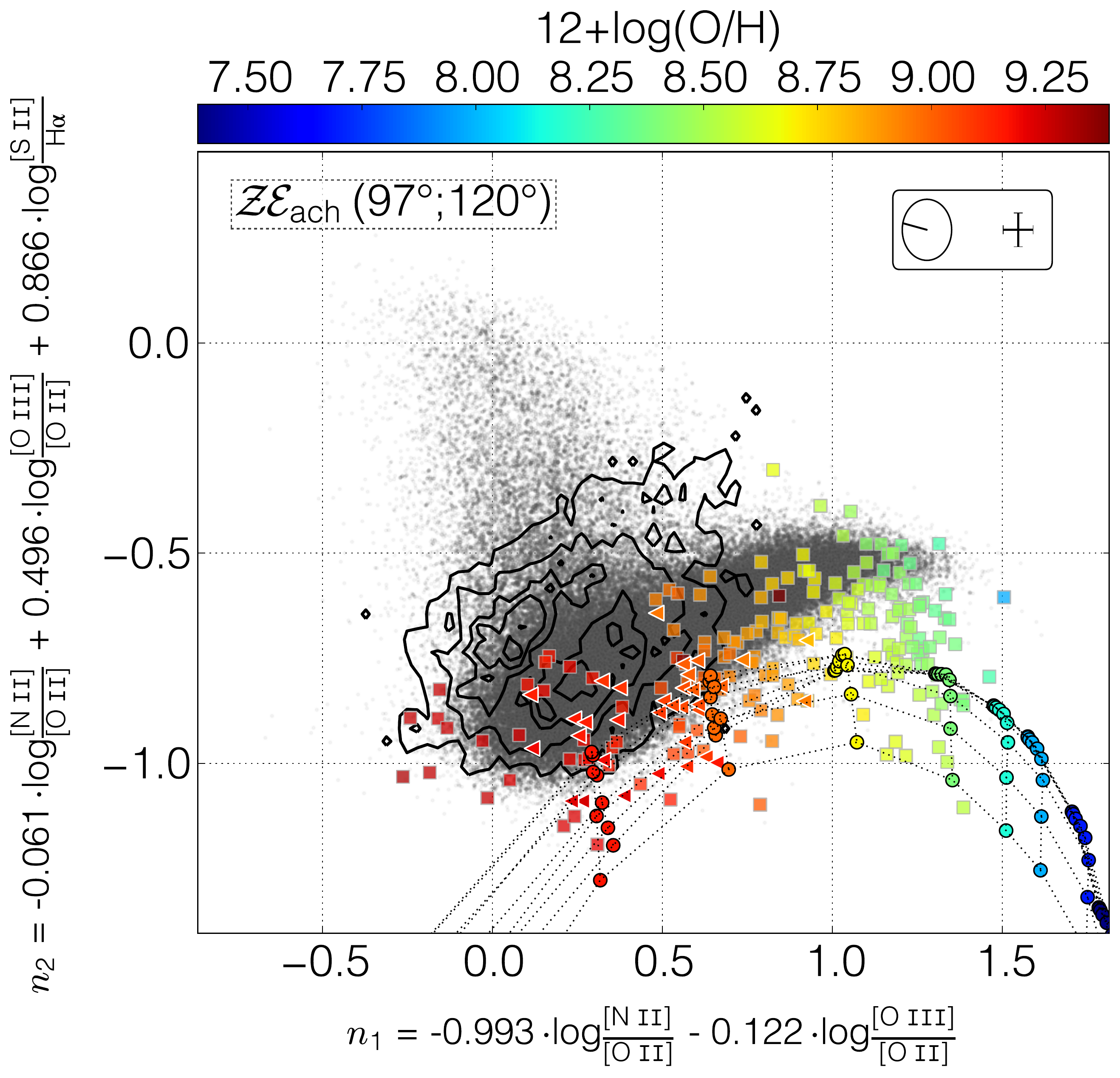}\qquad\qquad \includegraphics[scale=0.27]{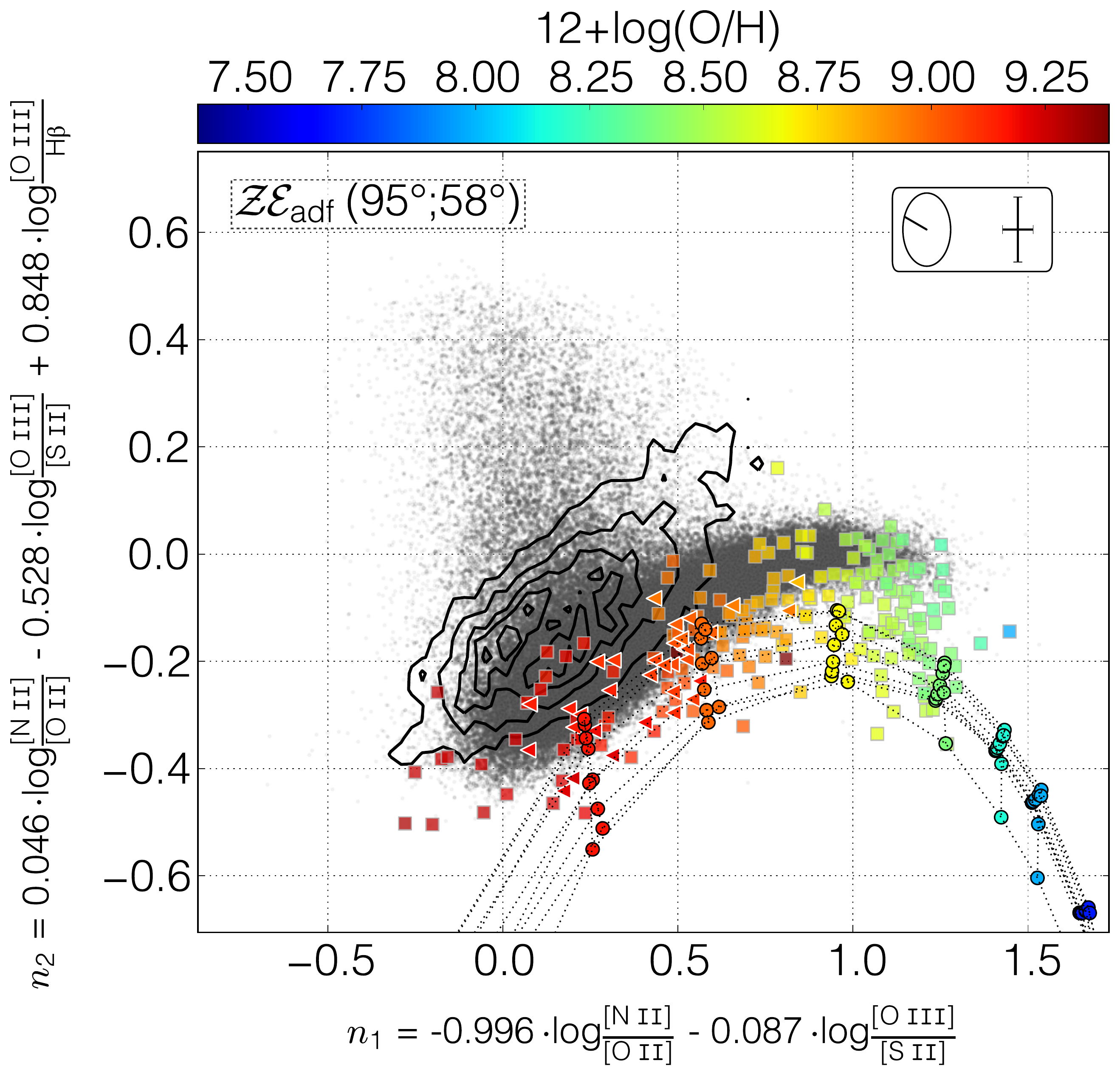}}\centerline{  \includegraphics[scale=0.27]{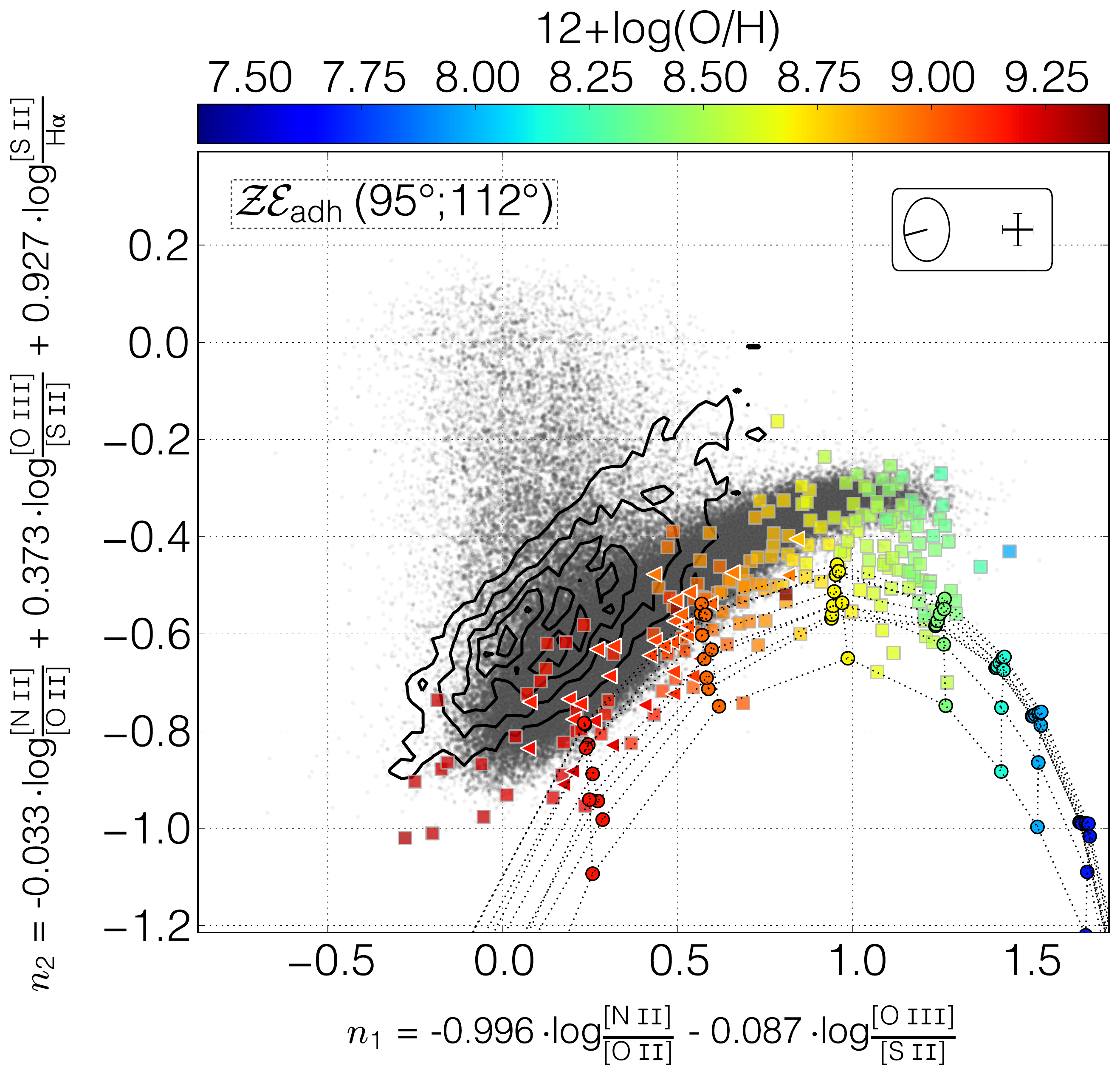}\qquad\qquad \includegraphics[scale=0.27]{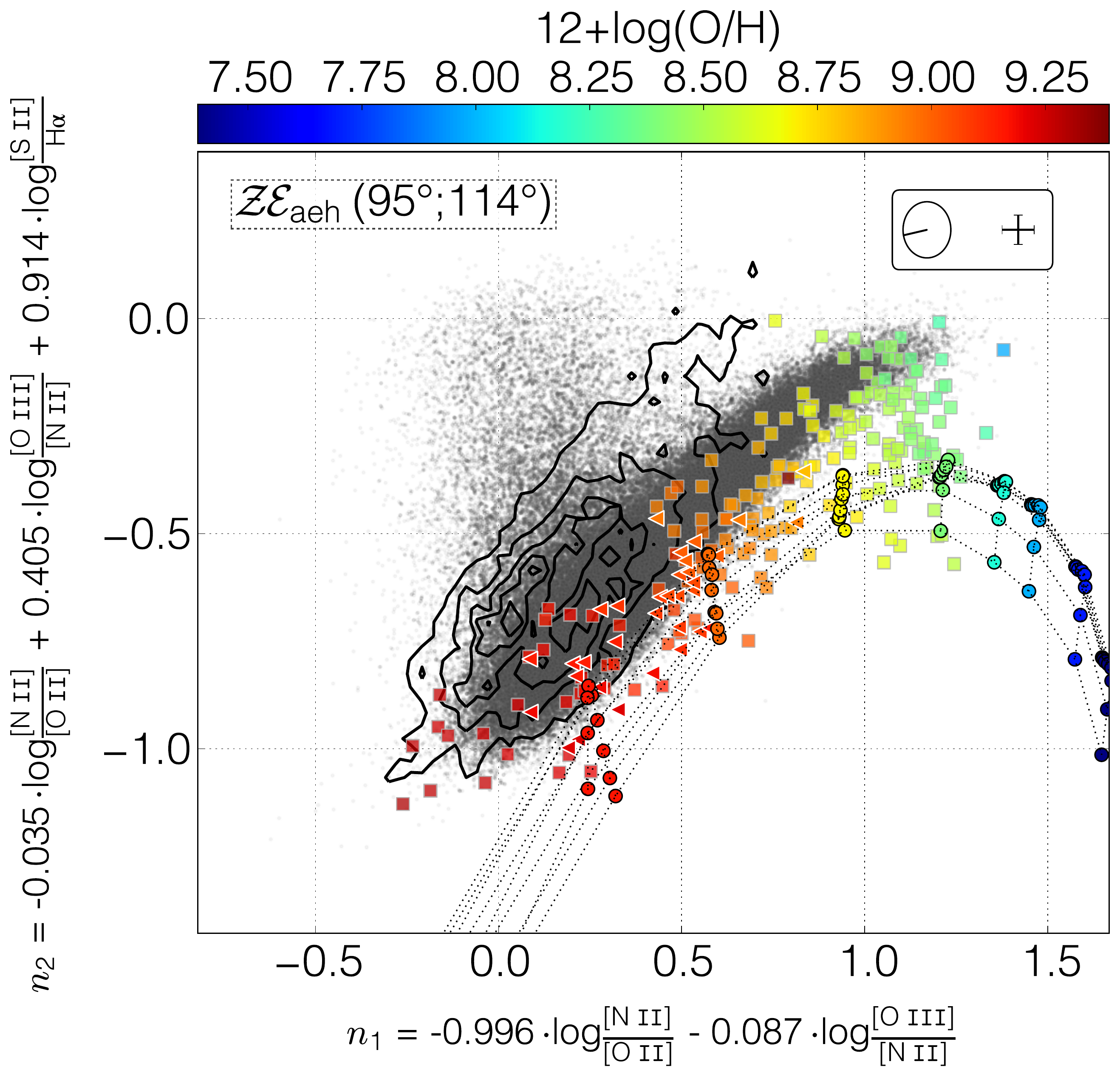}}
\centerline{  \includegraphics[scale=0.27]{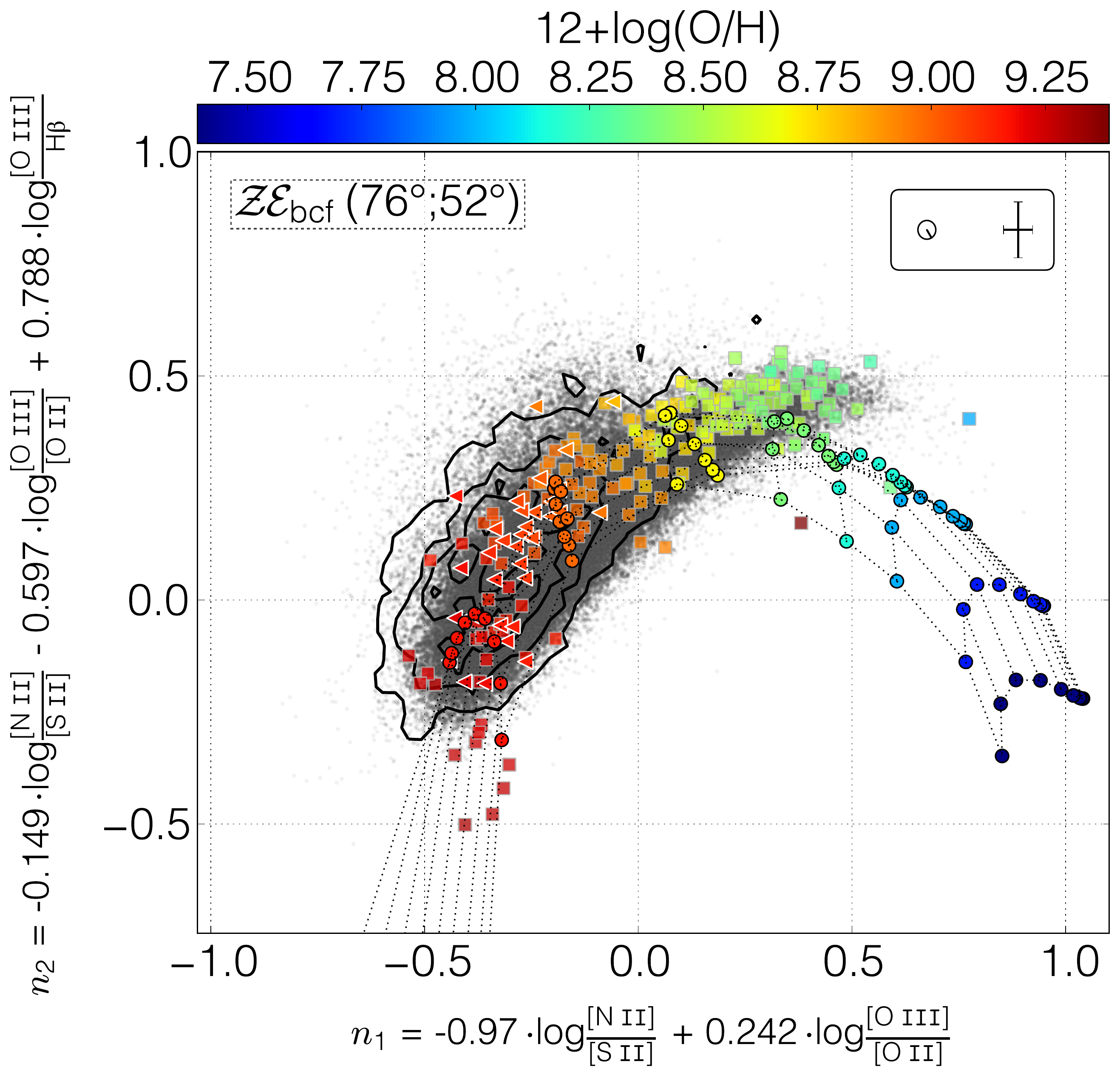}}
\caption{Same as Figure~\ref{fig:ZQE_Oi}, but for $\cal{ZQE}$ diagnostics not involving $\log$ \OIHa, and with no associated $\cal{ZE}$ diagnostic.}\label{fig:ZQE_drop}
\end{figure*}

Of the twenty-four $\cal{ZQE}$ diagrams defined in Section~\ref{sec:3d-diag}, eleven have a comparatively higher confusion between the starburst and AGN branches of the cloud of points of SDSS galaxies, such that there is no view-point from which both branches can be clearly separated. In Figures~\ref{fig:ZQE_Oi} and \ref{fig:ZQE_drop}, we show the $\cal{ZE}$ diagrams, for each eleven ``diagnostic-less'' $\cal{ZQE}$ space, that best collapse the starburst sequence onto itself. \ion{H}{2}-like and AGN-like SDSS galaxies are in grey. Uncertain SDSS galaxies are once again indicated via density contours. In all cases, both the distributions of the uncertain objects and of the starburst sequence are broad. Similarly, while the grid of \ion{H}{2} photoionisation models can be collapsed within $\sim$0.1 dex, the observational measurements of \ion{H}{2} regions are more spread out. These diagrams may be of interest for different applications, such as the study of the AGN branch-itself, although the points-of-views associated with the $\cal{ZE}$ diagrams in Figures~\ref{fig:ZQE_Oi} and \ref{fig:ZQE_drop} do not directly reveal the clear split between the different AGN classes in some of the associated $\cal{ZQE}$ diagrams.

\section{3D printing $\cal{ZQE}$ diagrams}\label{app:3d_print}

\begin{figure*}[htb!]
\centerline{ \includegraphics[scale=0.7]{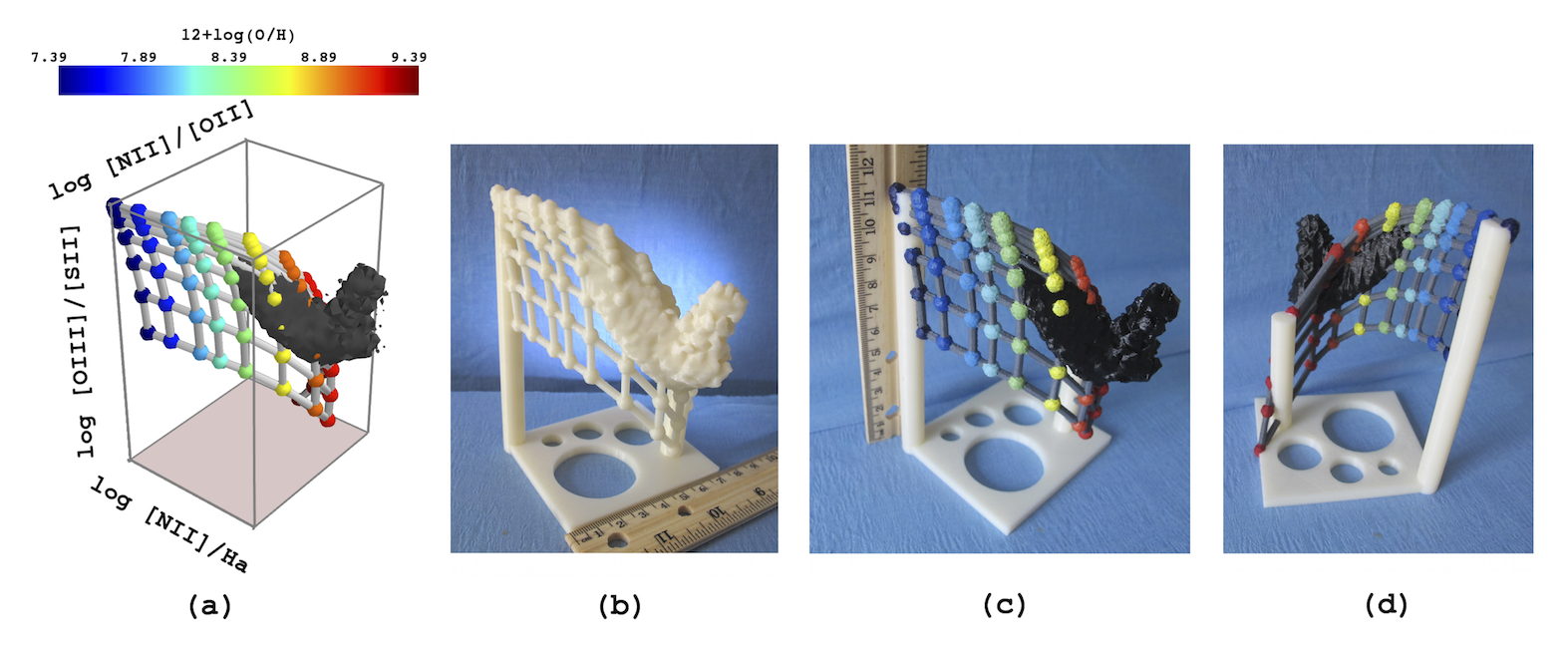}}
\caption{(a) $\cal{ZQE}_\text{adg}$ as shown in Figure~\ref{fig:3D_example}, but from a different point of view. SDSS galaxies are represented by a iso-density surface (black) instead of being shown individually as a cloud of point. (b) 3d printed model of $\cal{ZQE}_\text{adg}$ using ABS plastic The print direction was bottom to top, starting from the base plate. Two columns act as support structure. The model size is approximatively 7cm$\times$7cm$\times$11cm. (c) \& (d) Front and back view of the 3D printed model after applying a layer of acrylic paint manually to reproduce the oxygen abundance color scheme. }\label{fig:3d_print}
\end{figure*}

In an effort to explore new methods for sharing 3D structures such as $\cal{ZQE}$ diagrams, we give the interested reader the possibility to use 3D printing to construct a physical model of $\cal{ZQE}_\text{adg}$ (shown in Figure~\ref{fig:3D_example}). The concept is similar to Steffen et al. (2014, in press) and their 3D printable model of the Homunculus nebula around Eta Carina. A 3D printed model of $\cal{ZQE}_\text{adg}$ is shown in Figure~\ref{fig:3d_print}. The model is comprised of the grid of MAPPINGS IV simulations, of an iso-density surface (black) tracing the position of SDSS galaxies, and of a support structure comprised of two cylindrical columns and a base plate. The entire structure is defined in an STL file attached to this article as a supplementary material. This file format ought to be compatible with most (if not all) of the 3D printers currently on the market. Having access to a monochromatic printer only, we have used acrylic paint to reproduce the color scheme tracing the oxygen abundance of MAPPINGS IV simulations.

\newpage
$ $\\
\newpage 

The STL file has been designed to be ready to print, and should not require any additional modifications before being sent to a 3D printer. While the presence of large gaps in the structure would benefit from the use of a dissolvable support material (and therefore a 3D printer with dual extrusion), the model is also compatible with less advanced, single extrusion devices.

The 3D model was first generated with the \texttt{Mayavi2} module in \texttt{Python}, and saved as a VRML file. We manually added the external support structure using the freely available \texttt{Blender} software, exported the entire structure to the OBJ format, and finally converted it to an STL file with the freely available \texttt {Meshlab} software. We welcome readers with practical questions regarding the 3D printing of the $\cal{ZQE}_\text{adg}$ diagram, the structure design or the file creation process to contact us directly. 

\bibliographystyle{apj}
\bibliography{Vogt_2014}

\end{document}